\documentclass[12pt]{article}
\usepackage{epsfig,graphicx,amssymb,bm,amsmath}

\def \Zbar {\overline{Z}}
\def \be {\begin{equation}}
\def \ee {\end{equation}}
\def \bea {\begin{eqnarray}}
\def \eea {\end{eqnarray}}
\def \nn {\nonumber}

\def \R {{\textsf{I}\kern-.10em \textsf{R}}}
\def \T {{\textsf{T}\kern-.45em \textsf{T}}}
\def \C {{\textsf{C}\kern-.37em \textsf{C}}}
\def \Z {{\textsf{Z}\kern-.35em \textsf{Z}}}
\def \H {{\textsf{I}\kern-.10em \textsf{H}}}
\def \S {{\textsf{S}\kern-.37em \textsf{S}}}

\def \dels {\partial\kern-.5em / \kern.5em}
\def \As {{A\kern-.5em / \kern.5em}}
\def \Ds {D\kern-.7em / \kern.5em}

\newcommand{\ena}{\end{eqnarray}}

\def\bbox{{\,\lower0.9pt\vbox{\hrule \hbox{\vrule height 0.2 cm
\hskip 0.2 cm \vrule height 0.2 cm}\hrule}\,}}
\newcommand{\dsl}{\pa \kern-0.5em /}

\newcommand{\pa}{\partial}

\def \K {{\tt I\kern-.25em K}}

\begin{document}


\begin{titlepage}
\begin{center}

\hfill\parbox{4cm}{
{\normalsize\tt hep-th/0409070}}\\

\vskip 1in

{\LARGE \bf Weyl Card Diagrams and New S-brane Solutions of Gravity}

\vskip 0.3in

{\large Gregory Jones$^a$\footnote{\tt
  jones@physics.harvard.edu} and
John E.\ Wang$^{a,b,c}$\footnote{{\tt
hllywd2@feynman.harvard.edu}} }

\vskip 0.15in

${}^a$ {\it Department of Physics, Harvard University, Cambridge, MA
02138}\\[3pt]
${}^b$ {\it Department of Physics, National Taiwan University \\
Taipei 106, Taiwan}\\
${}^b$ {\it National Center for Theoretical Sciences, Taiwan\\
National Taiwan University \\
Taipei 10617, Taiwan}\\
[0.3in]

{\normalsize September 2004}

\end{center}

\vskip .3in

\begin{abstract}
\normalsize\noindent We construct a new card diagram which
accurately draws Weyl spacetimes and represents their global
spacetime structure, singularities, horizons and null infinity. As
examples we systematically discuss properties of a variety of
solutions including black holes as well as recent and new
time-dependent gravity solutions which fall under the S-brane
class.  The new time-dependent Weyl solutions include S-dihole
universes, infinite arrays and complexified multi-rod solutions.
Among the interesting features of these new solutions is that they
have near horizon scaling limits and describe the decay of
unstable objects.

\end{abstract}

\vfill

\end{titlepage}
\setcounter{footnote}{0}

\pagebreak
\renewcommand{\thepage}{\arabic{page}}
{\baselineskip=5mm\tableofcontents}


\section{A new diagram for spacetime structure}

Spacetimes are typically characterized by a choice of coordinates
and a metric.  If the coordinates are poorly chosen however many
properties of the spacetime such as horizons, causally connected
spacetime points, null infinity and maximal extensions are not
readily apparent.  One way to surmount these difficulties is to
perform conformal transformations leading to Penrose diagrams.

These diagrams are quite useful and successful especially in
understanding causal structure although there are some limitations
to this approach. For instance just knowing the Penrose diagram
for the subextremal $|Q|<M$ Reissner-Nordstr\o m black hole does
not tell us what happens to the spacetime structure in the
chargeless or extremal limits. Also the Penrose diagram for a Kerr
black hole does not clearly describe the ring singularity and the
possibility of crossing through the ring into a second universe.
Recently, analytic continuation has been applied to black hole
solutions to yield bubble-type \cite{Witten:1981gj} or S-brane
\cite{Gutperle:2002ai} solutions. Oftentimes this is done in
Boyer-Lindquist type coordinates which are hard to visualize.
Again we are not left with a clear picture of the resulting
spacetime and the Penrose diagrams are missing important
noncompact spatial directions. For more complicated spacetimes,
Penrose diagrams (which usually assume symmetry) can only draw a
slice of the spacetime.

It would be useful to have an alternative diagram which could also
capture other important features of a spacetime. For this reason
in this paper we expand the notion of drawing spacetimes in Weyl
space \cite{EmparanWK,Myers:rx}.  The idea will be to draw only
Weyl's canonical coordinates (or coordinates related to them via a
conformal transformation) and not Killing coordinates.

In $D=4$ dimensions a Weyl solution in canonical coordinates
\cite{EmparanWK,Weylpaper,EmparanBB} is written as

\be ds^2=-f dt^2+f^{-1}[e^{2\gamma}(d\rho^2+dz^2)+\rho^2 d\phi^2]
\ \ee

\noindent where $f$ and $\gamma$ are functions of $\rho, z$.
Although only solutions with enough symmetry (two orthogonal
commuting Killing fields $\partial_t, \partial_\phi$ in four
dimensions, or $D-2$ fields for general $D$ dimensions) can be
written in these coordinates, it is quite useful to consider the
Weyl type Ansatz as many of the well known solutions can be
written in the above form.  Weyl's symmetry requirement of $D-2$
orthogonal commuting Killing vectors in $D$ dimensions
\cite{Weylpaper,EmparanWK}, is a quite large and interesting class
of gravitational solutions.  We also include the Weyl-Papapetrou
class for 2 commuting Killing vectors in $D=4$ \cite{papapetrou},
and allow charged static solutions in $D\geq 4$ (see the Appendix
to this paper). Furthermore stationary vacuum solutions in $D\geq
4$ are covered with the very recent work of \cite{Harmark:2004rm}
and axisymmetric spacetimes in $D\geq 4$ are discussed in
\cite{Gregory}. In four and five dimensions this general `Weyl'
class includes spinning charged black holes as well as various
arrays\cite{IsraelKhan} of black holes, S-branes, and includes
backgrounds like Melvin fluxbranes\cite{Melvin, zeroduality} and
spinning ergotubes.

One of the main goals of this paper will be to devise a procedure
to obtain a sensible singly covered diagram to exhibit the
features of these spacetimes.  A key point is that Weyl solutions
depend on only two coordinates and so are amenable to the
construction of easy-to-visualize two dimensional diagrams.  These
diagrams will be reminiscent of a pasting-together of playing
cards, and so we call these card diagrams.  Card diagrams are
efficient in the sense that they show only the non-Killing
directions and so it is easy to see the important features of any
Weyl spacetime.

In this paper rather than focusing on one geometry or family of
geometries, we discuss techniques which generate a host of Weyl
spacetimes and briefly discuss their features.   Many new
spacetimes are presented in the current paper and more will be
presented in an ensuing paper \cite{joneswangfuture}.

In Section 2 we construct a card diagram for the Schwarzschild
black hole as an example before giving a general discussion of the
card diagrams properties.  Card diagrams not only allow us to
visualize the spacetime and instantly see much of its structure,
but are also helpful in performing and keeping track of analytic
continuations. The card diagrams for the
S-branes\cite{Gutperle:2002ai} which we will call
S-Schwarzschild\cite{ChenYQ,
KruczenskiAP,sugraSbranes,Nobu-intersect} and S-Kerr are markedly
different from their Schwarzschild and Kerr counterparts. In
comparison the Penrose diagram for these S-branes does not
emphasize a difference and are $90^\circ$ rotations of the black
hole Penrose diagrams. Card diagrams will draw noncompact
hyperbolic $\theta$-directions in S-brane geometries which
provides a different perspective of their properties. Examining
the Witten bubble and S-brane analytic continuations we also find
that a spacetime may have more than one card diagram due to the
different Killing congruences one may choose for a spacetime.

The Weyl Ansatz (and its Wick rotations) depends on two variables
and is well suited for describing the decay of localized unstable
branes. Homogeneous unstable brane decay also depends on two
variables, time and the distance from the object.  Such solutions
in this paper relevant to the decay of unstable objects appear in
Section 3, which includes a discussion of new S-brane solutions
which are analytic continuations\cite{JonesRG} of
diholes\cite{Bonnor, Emparan:1999au}. These solutions are
certainly interesting in their own right, although they also serve
here as good examples of how Weyl cards can simplify the
understanding of a complicated spacetime's global properties.  In
general, card diagrams are indispensable for creating and quickly
examining new Weyl spacetimes.  These solutions describe the
formation and decay of unstable objects including Melvin type
universes and cones. Also noteworthy is that some of these
solutions include near horizon scaling limits, and some are
related to the cosmological solutions of \cite{Cornalba:2002fi}.
Generalizations to larger array solutions \cite{JonesRG} are
discussed in Section 4 and these solutions may play a role in
understanding unstable brane decay\cite{roll}.

In Section 5 we return to discuss many well known black hole and
S-brane solutions including the Reissner-Nordstr\o m black hole,
Kerr, the five dimensional black ring solution of
\cite{EmparanWK}, S0-branes, twisted
S-branes\cite{Wang:2004by,regSbraneQuevedo,Lu:2004ye}, the
C-metric and new S-black ring and two-rod wave solutions.  As
examples, the card diagram for the Reissner-Nordstr\o m black hole
makes more evident how the chargeless and extremal limits take
place. In addition it will be useful to extend the card diagram
past curvature singularities; positive- and negative-mass
universes can be pasted together in a satisfying way.  Also, the
card diagram for the Kerr black hole shows a manifest symmetry
between inner and outer ergospheres, and presents a clearer
picture of the spacetime near the ring singularity as compared to
the Penrose diagram or the Kerr-Schild picture.

Solutions based on two rods are also discussed in Section 6
including a new five dimensional non-nakedly singular S-brane
solution.  We conclude with discussion and an appendix on how the
higher dimensional vacuum Weyl Ansatz can be extended to include
electro-magnetic fields.

\section{Introducing card diagrams: Schwarzschild and its analytic
continuations}

In this section we construct our first card diagrams and list some
of their general properties.

As a useful first example in this section we examine the
Schwarzschild black hole and the construction of its card diagram
before discussing some general properties of card diagrams in
detail.  Up to now if a solution had horizons, then only the
exterior regions outside the horizons in Weyl coordinates have
been drawn. To go through horizons we complexify the Weyl
coordinates. Although asking what is inside a horizon naively
leads to multiply covered coordinate triangles we discuss the
construction of a singly covered extension.

There are several key steps in constructing the diagrams: we will
discuss how to adjoin the cards representing the exterior of a
horizon to its interior, how to extend past ``null lines'' where
the Weyl coordinates are problematic and how to deal with branch
points on the cards. Note that Weyl diagrams are also a way to
picture the space on which we solve the Laplace/d'Alembert
equation to find a metric; the Schwarzschild black hole is a
uniform rod source. These card diagrams in other words will
represent a full accounting of the boundary conditions necessary
to specify the spacetime. Horizontal cards, where the spacetime is
stationary, will be used for regions where we solve for the
Laplace equation while the new time dependent vertical cards will
be where we solve the wave equation.

Then we will describe the effect of analytic continuation on the
card diagrams by examining two known analytic continuations of
Schwarzschild, the Witten bubble of nothing\cite{Witten:1981gj},
and the S0-brane\cite{Gutperle:2002ai} which we also call
S-Schwarzschild\cite{ChenYQ,KruczenskiAP,sugraSbranes}. We
discover that a spacetime may have more than one card diagram
representation and that these solutions related by seemingly
different analytic continuation have a simple and interesting
relationship in Weyl space.

\subsection{Schwarzschild Black Holes}

\begin{figure}[htb]
\begin{center}
\epsfxsize=5in\leavevmode\epsfbox{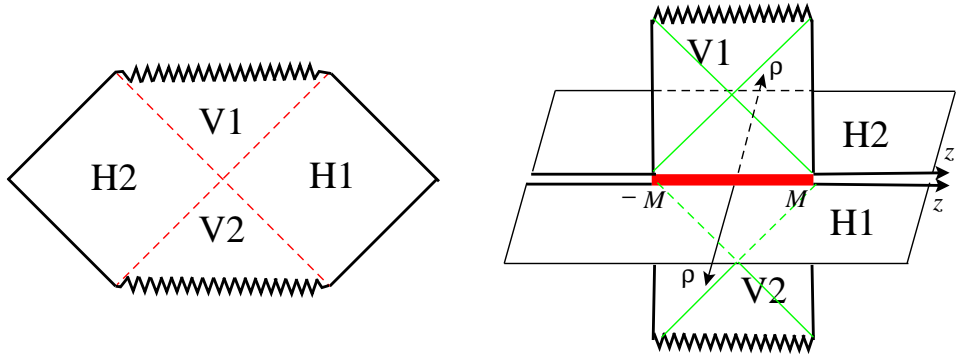}
\caption{The Schwarzschild black hole can be represented in a
Penrose diagram as on the left, or as a Weyl card diagram on the
right. The regions V1 and V2 are vertical in the page while the
regions H1 and H2 are horizontal and are infinitely extended
laterally (half-planes).  The four regions of the card diagram are
joined together like a collection of cards along the black hole
horizon on the $z$-axis.} \label{Penrose-Weyl-comparison}
\end{center}
\end{figure}

Our first example of a concrete card diagram is the Schwarzschild
black hole which is possibly the most well studied four
dimensional gravitational solution.  In this section we will
describe the construction of its Weyl card diagram. The Penrose
diagram and the Weyl card diagram for Schwarzschild are compared
in Fig.~\ref{Penrose-Weyl-comparison}.

\begin{figure}[tp]
\hspace*{2mm}
\begin{minipage}{70mm}
\begin{center}
\includegraphics[width=7cm]{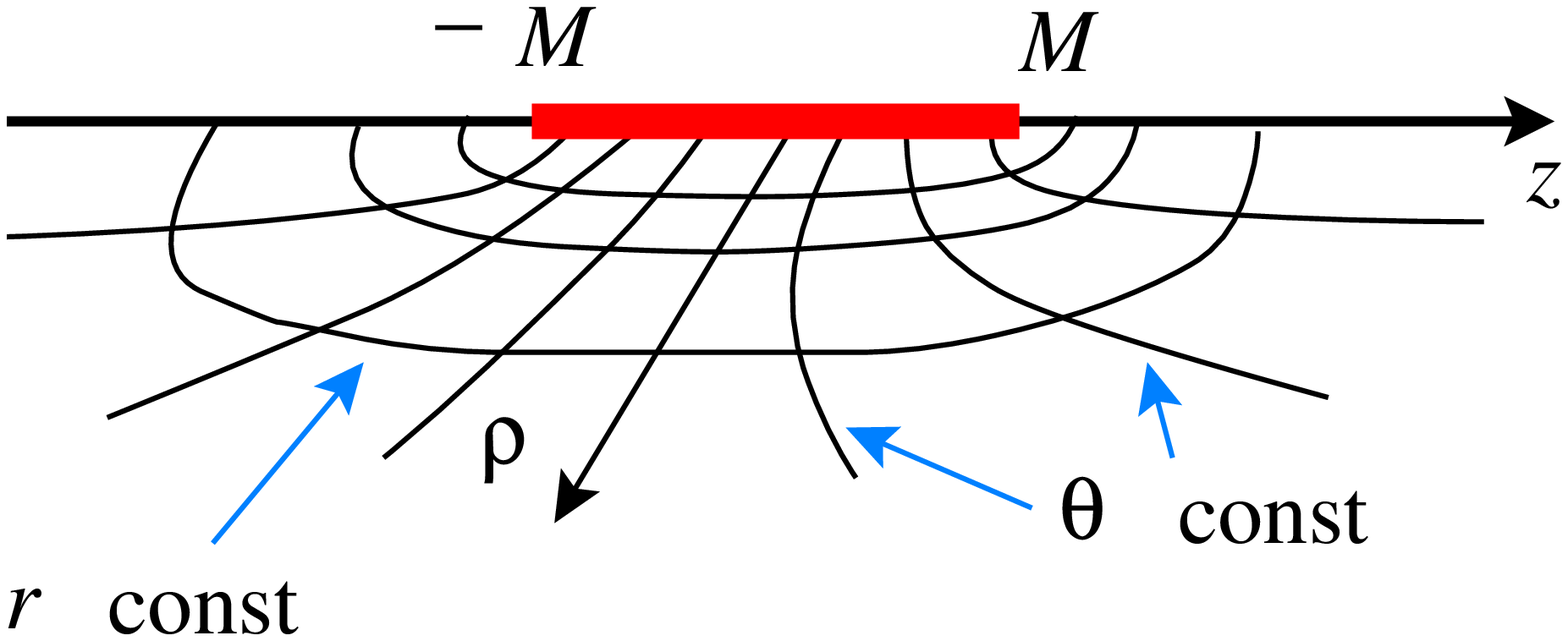}
\caption{In Weyl coordinates the Schwarzschild black hole is
represented as a rod in the $\rho, z$ horizontal half-plane. Lines
of constant Schwarzschild coordinates $r$ and $\theta$ are
outlined.} \label{weylrod}
\end{center}
\end{minipage}
\hspace*{14mm}
\begin{minipage}{70mm}
\begin{center}
\includegraphics[width=7cm]{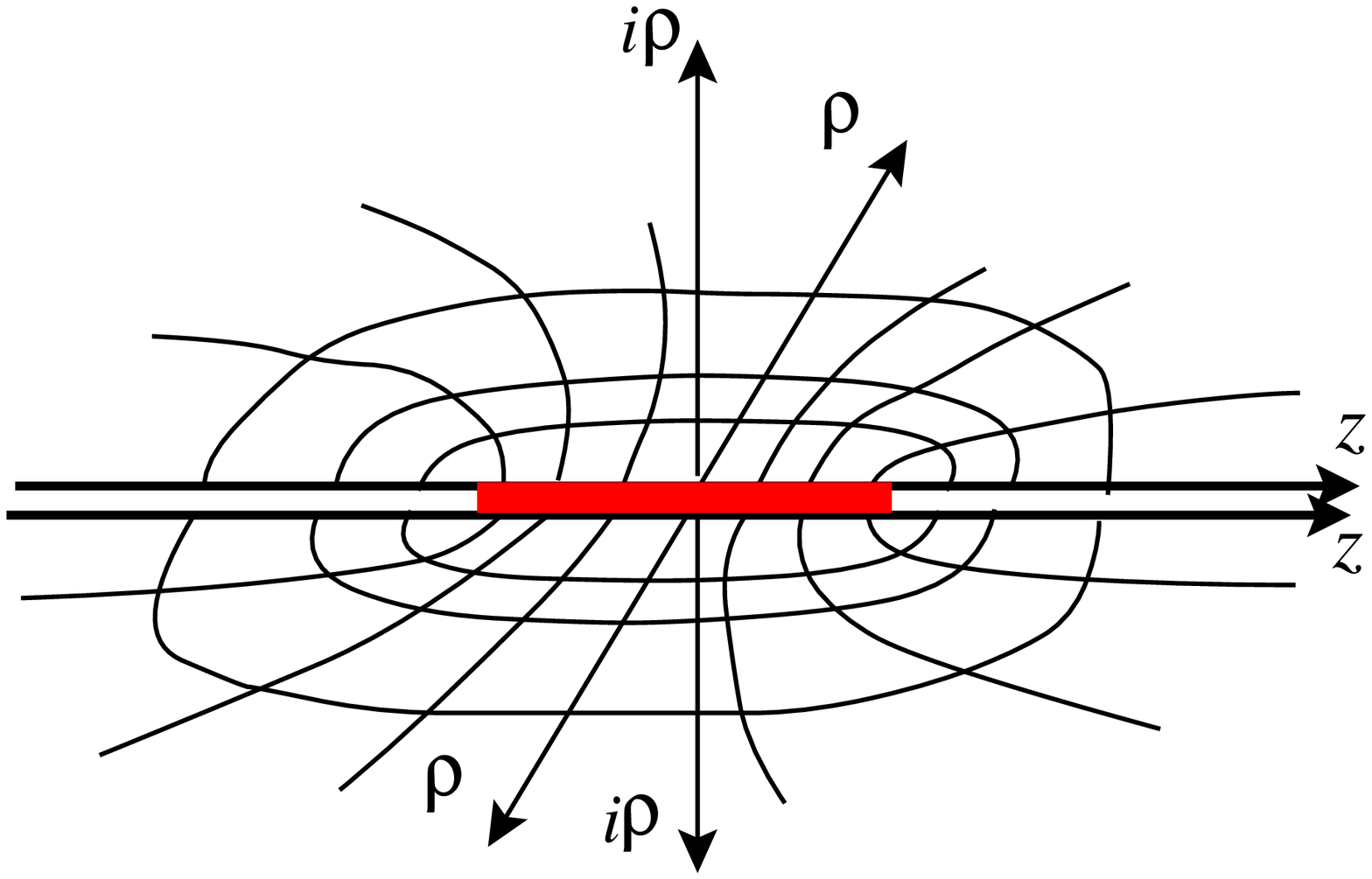}
\caption{The full Weyl representation of Schwarzschild includes
complex $\rho$ values to describe the interior of the horizon and
a second copy of the horizontal half-plane.} \label{buildcard2}
\end{center}
\end{minipage}
\hspace*{5mm}
\end{figure}

The Schwarzschild metric written in the usual spherically
symmetric (Boyer-Lindquist or BL) coordinates is
\begin{equation}\label{schwarz}
ds^2=-(1-2M/r)dt^2+(1-2M/r)^{-1}dr^2+r^2d\theta^2+r^2\sin^2\theta
d\phi^2 \ .
\end{equation}
There is a horizon at $r=2M$ and a curvature singularity at $r=0$.
It is not difficult to also write this solution in Weyl's
canonical coordinates \cite{EmparanWK,Weylpaper,EmparanBB} as
$$ds^2=-f dt^2+f^{-1}(e^{2\gamma}(d\rho^2+dz^2)+\rho^2 d\phi^2)$$
where $f$ and $\gamma$ are functions of the coordinates $\rho$ and
$z$
\begin{eqnarray*}
f&=&{(R_++R_-)^2-4M^2\over (R_++R_-+2M)^2},\\
e^{2\gamma}&=&{(R_++R_-)^2-4M^2\over 4R_+R_-},\\
R_{\pm}&=&\sqrt{\rho^2+(z\pm M)^2}.
\end{eqnarray*}
The half-plane $\rho\geq 0$, $-\infty<z<\infty$, known as Weyl
space, describes the exterior of the Schwarzschild black hole,
whose horizon is represented by a source on the line segment
$\rho=0$, $-M\leq z\leq M$, see Fig.~\ref{weylrod}.  Note that a
solution slice restricted to the two dimensions of Weyl space, is
conformal to the Euclidean $d\rho^2+dz^2$. The coordinate
transformation between BL and Weyl coordinates is
\begin{eqnarray}\label{coordinatetrans}
\rho&=&\sqrt{r^2-2Mr}\,\sin\theta,\\
z&=&(r-M)\cos\theta.\nonumber
\end{eqnarray}

Now we wish to ask how Weyl's coordinates draw the spacetime
inside the horizon.  The BL coordinates (\ref{coordinatetrans})
tell us that for $0<r<2M$, $\rho$ is imaginary and so we set
$\rho'=i\rho$
\begin{eqnarray}\label{coordinatetrans2}
\rho'&=&\sqrt{2Mr-r^2}\,\sin\theta,\\
z&=&(r-M)\cos\theta.\nonumber
\end{eqnarray}
(In general we must perform an analytic continuation of Weyl
coordinates to go through a horizon which are at the zeros of the
Weyl functions $f, e^{2\gamma}$.) The analytic continuation begins
a region with a conformally Minkowskian metric $-d\rho'^2+dz^2$
and we will draw this region as being vertical and attached to the
horizontal card at the horizon $-M\leq z\leq M$.  The vertical
direction is always timelike in card diagrams.

The Penrose diagram in $t,r$ can be divided into four regions that
meet in a $\times$-structure. In Weyl coordinates we also have a
similar structure since we can also perform the analytic
continuation to a second vertical region $\rho'=-i\rho$ and then
to a second horizontal region at negative real $\rho$. So we put
another copy of the horizontal external universe behind the
horizon and another copy of the vertical region below (see
Figure~\ref{buildcard2}) for a total of a four card junction
similar to the Penrose diagram.  The four regions labelled H1, H2,
V1 and V2 in

\begin{figure}[tp]
\begin{minipage}{80mm}
the Penrose diagram map to the similarly labelled region on the
card diagram in Fig.~\ref{Penrose-Weyl-comparison}. Note however
that the Weyl cards we are building represent the $r,\theta$
coordinates of the Schwarzschild solution which is different from
the Penrose diagram.  However the fact that the radial coordinate
$r$ describes four distinct regions, two where $\partial_r$ is
spacelike and two where it is timelike, is still apparent in the
Weyl card diagram.

Let us next examine what the (upper) vertical card looks like.
Looking at an $r$-orbit on the vertical card, we note that $0\leq
\rho'\leq M\pm z$. These bounding lines are where we have a zero
of $R_\pm=\sqrt{-\rho'^2+(z\pm M)^2}$, which we call special null
lines and they are
\end{minipage}
\hspace*{10mm}
\begin{minipage}{70mm}
\begin{center}
\includegraphics[width=7cm]{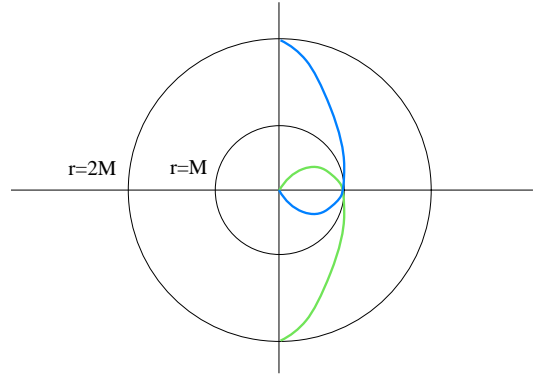}
\caption{The Weyl null lines for the Schwarzschild black hole
correspond to two 3-surfaces whose cross sections are cardioids.}
\label{cardioids}
\end{center}
\end{minipage}
\vspace{-.5cm}
\end{figure}
\begin{figure}[tp]
\begin{minipage}{80mm}
branch of the square root, but as the radicand passes through
zero, we must switch branches of the square root function. This is
just the statement that the function $x\mapsto \sqrt{x^2}$ can be
continued through $x=0$ to negative values.  To obtain a proper
Weyl description of the Schwarzschild solution it is necessary to
extend the spacetime past the null lines. The resolution is that
we may continue to use the coordinates $\rho',z$ on the same
triangle $0\leq\rho'\leq M\pm z$ to describe the interior of the
black hole but with two important details. One is that every time
though we cross a special null line we change the sign of $R_+$ or
$R_-$ depending on which line we cross.  The
\end{minipage}
\hspace*{10mm}
\begin{minipage}{70mm}
\begin{center}
\includegraphics[width=6.5cm]{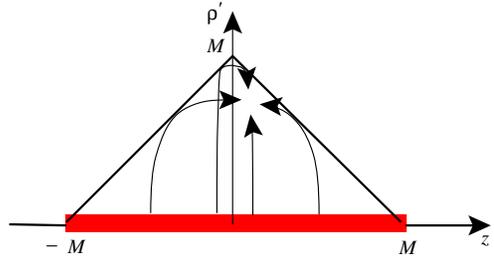}
\caption{The Weyl representation of the interior of the
Schwarzschild black hole naively gives a triangle with base length
$2M$ and height $M$.  The interior is covered four times by orbits
of $r$ at four different values of $\theta$.} \label{4xcoverfig}
\end{center}
\end{minipage}
\vspace{-.5cm}
\end{figure}

%
%
%
%
\noindent a general feature of vertical cards with focal points
(the rod endpoints $z=\pm M$ on the Schwarzschild card). Here the
null lines are the envelope of the $r$-orbits as we vary $\theta$.
Thus when inside the horizon the BL coordinates apparently fill
out a vertical 45-45-90 degree triangle in Weyl coordinates with
hypotenuse length $2M$ as shown in Figure~\ref{4xcoverfig}.

Special null lines play an important role in Weyl card diagrams so
let us explain their significance. Keep in mind that we have
already broken the manifest spherical symmetry when we have
written Schwarzschild in Weyl coordinates, so the existence of
preferred special null lines is relative to this chosen axis. In
Boyer-Lindquist coordinates, we want to look at the vanishing of
\be\label{BLR} R_{\pm}=r-M\pm  M \cos\theta .  \ee As drawn in
Fig.~\ref{cardioids}, the 3-surfaces $R_\pm=0$ have a cardioid
shape. For a given axis there are two, and these surfaces
intersect at $r=M$ and partition the inside-horizon into four
subregions. These regions will correspond to four triangles which
we describe below and the null lines correspond to the above
3-surfaces. At any point in the diagram, two axes may be chosen
and cardioids drawn to bound the future trajectories of null or
timelike curves.

It is clear from (\ref{BLR}) that $R_{\pm}$ is positive outside
the horizon and there is no difficulty going to negative values
inside the horizon. On the other hand in terms of Weyl
coordinates, the functions $R_\pm=\sqrt{-\rho'^2+(z\pm M)^2}$ are
the square root of a positive number when $\rho'<(z\pm M)$ and the
function is imaginary if we naively cross the null line.  As we
will discuss in more detail, the difficulty in Weyl coordinates is
actually that we initially use the positive

\noindent second is that the coordinates cover only a portion of
the interior and we must properly glue these coordinate patches
together to fill out the entire region inside the horizon. So for
example the horizon and the singularity are both described as
being at $\rho'=0$ but the metric near the singularity will have
all instances of $R_{\pm}$ replaced by $-R_{\pm}$.

To understand this resolution let us return to the coordinate map.
An analysis of Eq.~\ref{coordinatetrans2} shows that the BL
coordinates cover the same triangle four times and we depict this
in Figure~\ref{4xcoverfig}.  Of note is the fact that an $r$-orbit
which initially is timelike (upward) on the vertical card turns
seemingly spacelike (leftward) after touching the special null
line, while in fact it should always be timelike. Therefore this
naive extrapolation of Weyl coordinates into the horizon is not
accurate.

\begin{figure}[htb]
\begin{center}
\epsfxsize=5in\leavevmode\epsfbox{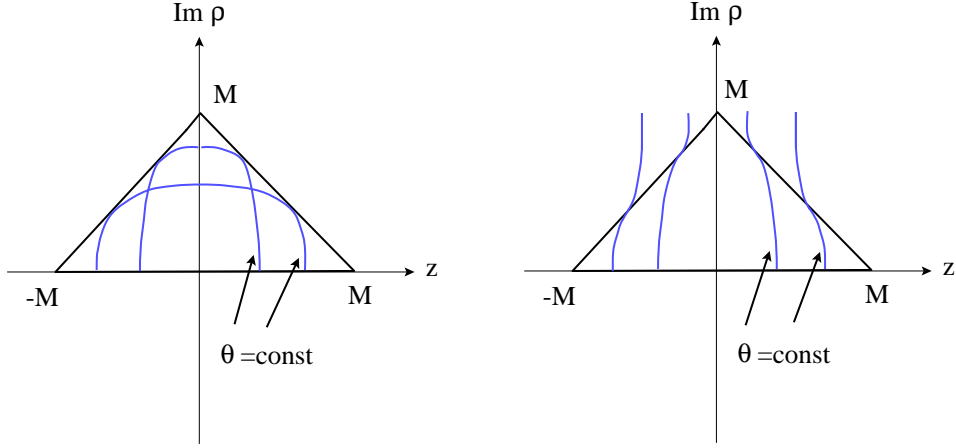} \caption{To
fix the overlap for the region $M\leq r\leq 2M$, we note that each
fixed $\theta$ value tangentially intersects the triangle for two
specific values of the coordinate $r$. Reflecting about the
tangential intersection point on the triangle we fill out a
rectangle of height $M$ and width $2M$. This is half the vertical
card.} \label{buildcard5}
\end{center}
\end{figure}

\begin{figure}[htb]
\begin{center}
\epsfxsize=3in\leavevmode\epsfbox{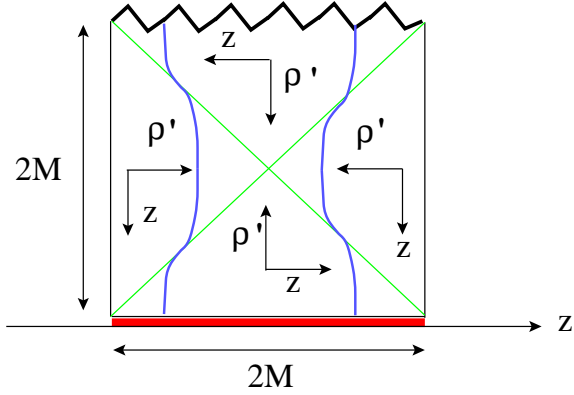}
\caption{Unfolding the four triangles for Schwarzschild the Weyl
space for the range $r=[0,2M]$ produces a square of length $2M$.
The horizon is at the bottom and the singularity at the top.}
\label{buildcard6}
\end{center}
\end{figure}

A very simple and visually appealing solution to this problem
exists.  All difficulties with the coordinates can be fixed by
simply unfolding the four copies of the triangle across the
special null lines to produce a square of length $2M$. For example
this will turn the bottom triangular region on its side as shown
in Fig.~\ref{buildcard6} by the left and right triangles.  In
addition to extending the coordinates, we must also properly
extend the metric across the null lines so the metric in the
different triangles corresponds to different spacetime points. The
first unfolding, drawn down to $r=M$, is shown in
Fig.~\ref{buildcard5}, and the end result with the manifestly
timelike $r$-orbits, drawn all the way to $r=0$, is shown in
Fig.~\ref{buildcard6}.  In the Schwarzschild geometry it is
possible to explicitly see that $e^{2\gamma}$ changes sign when we
change the overall sign of one of the $R_\pm$ in Weyl coordinates.
Changing the sign of $e^{2\gamma}$ as we pass the null lines is
the way to properly unfold the coordinate triangles because it
interchanges which of the the $\rho,z$ coordinates is timelike and
which is spacelike.  Note that $\gamma$ is therefore not
necessarily real and its imaginary part can be either $0$ or
$\pi$.  Also because of the unfolding of the triangles, the
positive $z$-direction on the top triangle horizontal cards points
in the opposite direction compared to on the original horizontal
card.

The vertical cards for the Schwarzschild black hole are thus both
squares of length $2M$.  The bottom of the upper card V1 is the
black hole horizon which connects to three other cards in a four
card junction. The right and left sides of this vertical card at
$\theta=0,\pi$ are boundaries where the $\phi$-circle vanishes.
The top edge of the card at $r=0$ is the black hole curvature
singularity. Fig.~\ref{buildcard6} depicts the region $0\leq r\leq
2M$ on the upper vertical card.  The second vertical card is build
in analogous fashion except the square is built in a downwards
fashion towards negative values of $\rho'$.  Also there is a
second horizontal card plane identical to the first attached to
the same horizon.  We have now described how to build the card
diagram for the Schwarzschild black hole in
Figure~\ref{Penrose-Weyl-comparison}.

\begin{figure}[htb]
\begin{center}
\epsfxsize=4in\leavevmode\epsfbox{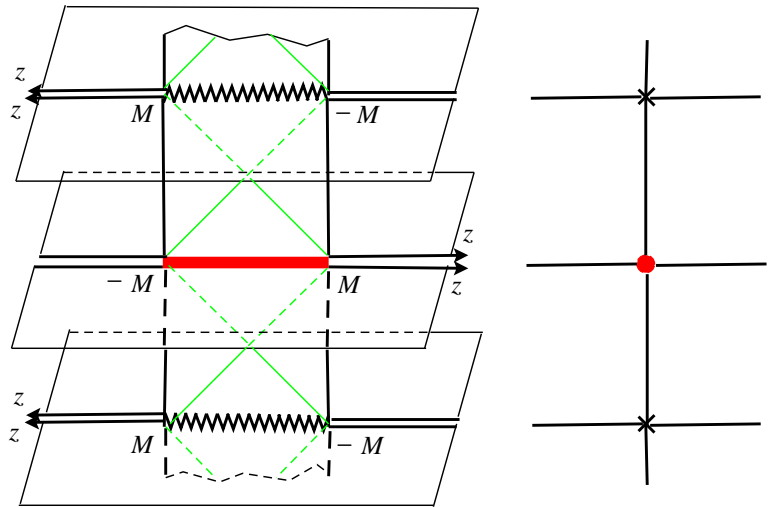}
\caption{The extended card diagram for Schwarzschild includes both
positive and negative mass universes half-planes. Its side-view
for $z=0$ cross-section is drawn on the right.}
\label{FullSchwarzCardfig}
\end{center}
\end{figure}

One would ordinarily stop the construction of Schwarzschild with
the above four regions, but for reasons which become more clear
when we look at the Reissner-Nordstr\o m and Kerr black holes in
Sections~\ref{RNsubsub} and \ref{Kerrsubsec}, we continue the
spacetime past the singularity to attach two horizontal half-plane
corresponding to negative-mass universes, and another vertical
card above. In Schwarzschild coordinates, these negative-mass
universes represent $r<0$ nakedly singular spacetimes. Each
negative mass-universe is one horizontal half-plane card with a
singularity along $-M\leq z\leq M$.  The total card diagram for
Schwarzschild is an infinite stack of cards representing positive
and negative mass universes and the inside-horizon regions, shown
in Fig.~\ref{FullSchwarzCardfig}.

\subsection{General properties of card diagrams}

Having described the construction of the card diagram for
Schwarzschild we now turn to a few general remarks about these
diagrams.  Horizontal cards are conformally Euclidean and
represent stationary regions.  Vertical cards are always
conformally Minkowskian and represent regions with ($D-2$)
spacelike Killing fields.  On a vertical card time is always in
the vertical direction and causal cones lie between the $45^\circ$
angles. Weyl's coordinates certainly go bad at horizons, so these
diagrams are not a full replacement for Penrose diagrams at
understanding causal structure or particle trajectories. However,
it is clear for example from Fig.~\ref{FullSchwarzCardfig} that
two vertical and two horizontal cards attach together in a
$+$-configuration in precisely the sense of the $\times$-like
horizon structure of the Penrose diagram.

The prototypical horizon is
that of the two Rindler and two Milne wedges of flat space; this
spacetime has two horizontal half-plane cards and two vertical
half plane cards that meet along the horizon being the whole $z$-axis.
Zooming in on a non-extremal horizon of any card diagram yields
this Rindler/Milne picture.

A rod endpoint $(\rho,z)=(0,z_i)$ such as $z=\pm M$ for
Schwarzschild is a `focus' for the Weyl diagram and often
represents the end of a black hole or acceleration horizon on a
horizontal card.  Quite generally, multi-black hole Weyl
spacetimes depend only on distances to foci \cite{sibgatullin}.

To understand that it is natural in general to change the branch
of the functions $R_i$ when crossing the special null line that
emanates from the foci at $z=z_i$, take some Weyl spacetime and
imagine moving upwards in a vertical card to meet the special null
line at $R_i=0$ by increasing time $\rho'\geq 0$ for fixed spatial
$z$. Rearranging $R_i=\sqrt{-\rho'^2+(z+z_i)^2}$ as the
semi-ellipse
$$R_i^2+\rho'^2=(z+z_i)^2,$$
we see that a smooth traversal of this semi-ellipse across $R_i=0$
requires a change in the sign of $R_i$.

In many of our solutions, special null lines are used to reflect
vertical triangular cards to create full, rectangular cards.
However, in the Bonnor-transformed S-dihole II geometry of
Sec.~\ref{sdihole-sec} as well as double Killing rotated extremal
geometries and parabolic representations of the bubble and
S-Schwarzschild in Sec.\ref{thirdcard-Poincare}, the special null
lines will serve as conformal null infinity ${\cal I}^\pm$.

Boundaries of cards indicate where Killing circles vanish.  Which
circle vanishes is constant over a connected piece of the
boundary, even when the boundary turns a right angle onto a
vertical card.  Furthermore the periodicity to eliminate conical
singularities is constant along connected parts of the boundary.
For the Schwarzschild, the $\phi$-circle vanishes on both the
connected boundaries and has periodicity $2\pi$.

When a light ray is incident from a horizontal card onto a horizon
(to enter the upper vertical card), it must turn and meet that
horizon perpendicularly.  It then appears on the vertical card,
again perpendicular to the horizon.  Only those light rays which
go from the lower vertical card to the upper vertical card
directly can meet the horizon line at a non-right angle; these
rays would touch the vertex of the $\times$ in a Penrose diagram
such as in the Schwarzschild case. When a light ray on a vertical
card hits a boundary where a spacelike circle vanishes, it bounces
back at the same angle as drawn on the card.

Spacetimes with a symmetry group larger than the minimal Weyl
symmetry can have more than one card diagram representation. This
is due to the fact that there can be more than one way to set up a
Killing congruence on the spacetime manifold which is still
compatible with the Weyl Ansatz. Examples we explicitly discuss in
Sec.~\ref{bubble-sbrane-subsec} are the 4d Witten bubble and the
4d S-Reissner-Nordstr\o m (S-RN) which have three card diagrams
corresponding to the three types of Killing congruences on dS$_2$
or ${\bf H}_2$; Secs.~\ref{5dcard-subsec1} and
\ref{5dcard-subsec2} discuss the 5d cases.  In comparison both the
Schwarzschild and RN black holes have only one card diagram since
on the sphere $S^2$, all Killing congruences are axial rotations
and lead to the same card diagram. The S-Kerr solution of
Sec.~\ref{Kerrsubsec} has symmetry group $U(1)\times {\bf R}$ so
it also has only one representation, which looks like the
`elliptic' representation of S-RN. The multiple representations of
the Witten bubble echo the fact that dS$_2$ can be represented in
different Killing coordinate systems. Using hyperbolic and
trigonometric functions, two types of congruences are easy to
find; one of them is associated with global coordinates, while the
other has patched coordinates. Different representations have
different uses.  For example the patched Witten bubble can be
trivially Wick rotated to give the Schwarzschild black hole or the
S-Schwarzschild, whereas when the bubble is written in global
coordinates more work is required in order to obtain these other
spacetimes in Weyl coordinates (see Sec.~\ref{SRN2sec}).

\subsubsection{Our deck of cards:  The building blocks for Weyl spacetimes}

All spacetimes, new and old, in this paper are built from the following card
types.

Horizontal cards are always half-planes.  They may however have one or
more branch cuts which may be taken to run perpendicular to the $z$-axis.
Undoing one branch cut leads to a strip with two boundaries; multiple
branch cuts lead to some open subset, with boundary, of a Riemann surface.

Vertical cards can be half-planes with a vertical boundary, half-planes
with a horizontal (horizon) boundary; quarter-planes with a vertical and
horizontal
boundary and one special null line; compact squares with two special null
lines; a full plane with two special null lines; a full plane without
special null lines; a noncompact $45^\circ$
wedge with
either a vertical or horizontal boundary, and a special null line serving
as conformal null infinity ${\cal I}^\pm$; a compact $45$-$45$-$90$ wedge
with the
hypotenuse either vertical or horizontal, and short legs being special
null lines serving as ${\cal I}^\pm$; a noncompact $90^\circ$
wedge bounded by two ${\cal I}^\pm$ special null lines; and a compact
$45$-$45$-$90$
wedge with hypotenuse serving as ${\cal I}^\pm$, and legs horizontal and
vertical.

It is satisfying that for a variety of spacetimes, the cards are
always of the above rigid types.

There is one basic procedure which can be performed on vertical
cards and their corresponding Weyl solutions. It is the analytic
continuation $\gamma\to \gamma+ i \pi$, which is allowed since
$\gamma$ is determined by first order PDEs.  This continuation is
equivalent to multiplying the metric by a minus sign and then
analytically continuing the $D-2$ Killing directions.  We call
this a $\gamma$-flip since the way it acts on a card is to
mirror-reflect it about a $45^\circ$ null line (for example, look
at the vertical cards in Figs.~\ref{bubblecard3} and
\ref{S-RN-3}).  Our first example of this procedure as generating
a new solution appears in (\ref{flip-bubble}).

\subsection{Witten bubble and Schwarzschild S-brane}
\label{bubble-sbrane-subsec}

The Schwarzschild black hole can be analytically continued to two
different time dependent geometries, the Witten bubble of
nothing\cite{Witten:1981gj} and
S-Schwarzschild\cite{ChenYQ,KruczenskiAP,sugraSbranes}, and it is
instructive to understand how the card diagram changes.

One of the main new features that arises is that a spacetime can
have multiple card diagram representations.  Both the bubble and
S-Schwarzschild have three different card diagram representations
corresponding to three different ways to select Killing
congruences.  Analyzing the different diagrams we also find that
analytic continuations have a very simple form in Weyl space
allowing us to relate them in a visually satisfying manner.

We will discuss these different representations in detail when
constructing the card diagrams.  To set the stage, however, we
make some preliminary remarks on the symmetries of hyperbolic
spaces.  These three types of Killing congruences can be
understood directly by representing ${\bf H}_2$ as the unit disk
(with its conformal infinity being the unit circle).  The
orientation-preserving isometries of ${\bf H}_2$ are those
M\"obius transformations preserving the disk, $PSL(2,{\bf R})$
\cite{Matsuzaki}. M\"obius transformations $z\mapsto {az+b\over
cz+d}$ have two complex fixed points, counted according to
multiplicity.  In the upper half-plane $z=x+i\sigma$
representation, $a$, $b$, $c$, and $d$ are real, so the fixed
points are roots of a real quadratic. Hence they may be (i)
distinct on the real boundary (hyperbolic), (ii) degenerate on the
real boundary (parabolic), or (iii) nonreal complex conjugate
pairs, one interior to the upper half plane ${\bf H}_2$
(elliptic). Prototypes of Killing fields are (i)
$z\to(1+\epsilon)z$ for the upper half-plane, (ii) $z\to
z+\epsilon$ for the upper half-plane; and (iii) $z\to
e^{i\epsilon}z$ for the disk $|z|<1$. These are the striped,
Poincar\'e, and azimuthal congruences.  In these representations,
the S-Reissner Nordstrom (S-RN) and the Witten bubble each have 0,
1, and 2 Weyl foci.

\subsubsection{Hyperbolic card diagrams} Originally Witten found the
expanding bubble solution by starting from (\ref{schwarz}) and
taking the analytic continuation $\theta\to\pi/2+i\theta$ and
$t\to ix^4$
\begin{equation}
ds^2=(1-2M/r)(dx^4)^2+(1-2M/r)^{-1}dr^2-r^2d\theta^2
+r^2\cosh^2\theta \,d\phi^2.
\end{equation}
Here, $\theta$ plays the role of time and $\theta=0$
is the time where the bubble `has minimum size.'  (This statement
has meaning if we break $SO(2,1)$ symmetry.)  To achieve this in
Weyl's coordinates, we put $z\to i\tau$, $t\to ix^4$; the
resulting space is equivalent to Witten's bubble by the real
coordinate transformation
\begin{eqnarray*}
\rho&=&\sqrt{r^2-2Mr}\,\cosh\theta\\
\tau&=&(r-M)\sinh\theta.
\end{eqnarray*}
Witten's bubble universe is represented in Weyl coordinates as a
vertical half-plane card, $\rho\geq 0$, $-\infty<\tau<\infty$,
where now the $x^4$-circle, and not the $\phi$-circle, vanishes at
$\rho=0$. Note that the vertical card now has Minkowski signature
and is conformal to $-d\tau^2+d\rho^2$. This vertical card does
not have special null lines and is covered only once by the BL
coordinates.
The bubble does have a rod which is along the
imaginary $\tau$ axis and which intersects the card at the $\rho=0$,
$\tau=0$
origin.  The reason there are no special
null lines is that the geometry's foci are at $\tau=\pm iM$.  We
call this the hyperbolic representation of the Witten bubble.

This analytic continuation in Weyl space is precisely the same as
the one used in \cite{JonesRG}, except now the Schwarzschild rod
crosses $z=0$. Solutions which are even-in-$z$ Israel-Khan arrays
where no rod crosses $z=0$ can be Wick rotated to gravitational
wave solutions sourced by rods at imaginary time. Generalizations
of the bubble solution by adding concentric incoming and outgoing
waves include Wick rotations of an Israel-Khan array even about
$z=0$ where one rod does cross $z=0$. As these additional rods are
made to cover more of the $z$-axis and are brought closer and
closer to the principal rod, the deformed Witten bubble solution
hangs longer with a minimum-radius $\phi$-circle.  In the limit
where rods occupy the entire $z$-axis, we get a static flat
solution, which is Minkowski 3-space times a fixed-circumference
$\phi$-circle.

It is well known that sending $t\to ix^4$, $\phi\to i\phi$ gives
another, `elliptic' representation of the Witten bubble (see
Fig.~\ref{bubblecard1}). We will discuss this in the next
subsection. However, sending $t\to ix^4$ and $\phi\to i\phi$ for a
finite Israel-Khan array \cite{EmparanWK,Dowker:2001dg} yields a
solution different than sending $z\to i \tau$; we get black holes
and conical singularities in an expanding Witten bubble.  We see
that to smoothly perturb the Witten bubble by adding more sources
at imaginary time, the use of Weyl's coordinates and $z\to i\tau$
is essential.

The Schwarzschild S-brane vacuum solution of
\cite{ChenYQ,KruczenskiAP,sugraSbranes} \be
ds^2=(1-2M/t)(dx^4)^2-(1-2M/t)^{-1}dt^2+t^2(d\theta^2+\sinh^2\theta
d\phi^2) \label{S-Sch-metric} \ee is gotten from (\ref{schwarz})
by taking $t\to ix^4$, $\theta\to i\theta$, $r\to it$, and $M\to i
M$.  From (\ref{coordinatetrans}) we see that in Weyl's
coordinates we can effect this by sending $t\to ix^4$, $z\to
i\tau$, $M\to i M$, up to a real coordinate transformation. Thus
in Weyl coordinates the only difference between Witten's bubble
and the Schwarzschild S-brane is putting $M\to iM$.  In the
terminology we are using, a hyperbolic representation of the
Witten bubble becomes an elliptic representation of
S-Schwarzschild (see the next section) under the analytic
continuation of the mass.

We can also take the Witten bubble in the hyperbolic
representation and turn the vertical half-plane card on its side
via $\gamma\to \gamma+i\pi$. This yields a hyperbolic
representation of S-Schwarzschild which we will leave for
Sec.~\ref{SRN2sec} because the horizontal card has an interior
branch point which is more technically involved.

Just as we perturbed Witten's bubble solution with an Israel-Khan
array, we can also perturb S-Schwarzschild by adding rods before
analytically continuing.  Here Weyl's coordinates and $z\to i\tau$
are essential. We can choose to analytically continue the mass
parameters of the additional rods or not.  Additionally, we can
displace some rods in the imaginary $z$-direction which affects
the $\tau$-center of their disturbance.  If we do everything in an
even fashion, i.e. we respect ${\rm Im}\,\tau\to-{\rm Im}\,\tau$,
the resulting geometry (at real $\tau$) will be real.  In
particular, rotating a rod at $z>0$ counterclockwise means
rotating its image at $z<0$ clockwise.  We will see in
Sec.~\ref{2rodgravwavesec} in reinvestigating the 2-rod example
\cite{JonesRG} that there may be several choices for branches.

\subsubsection{Elliptic card diagrams}
\label{Witten2sec}

Weyl solutions can have multiple card diagram representations and
we illustrate this by first discussing the elliptic card diagram
for the Witten bubble.  Starting with Schwarzschild and sending
$z\to i\tau$ and $t\to i x^4$ gives the Witten bubble with a
single vertical half-plane card $\rho\geq 0$,
$-\infty<\tau<\infty$. However, one can also take Schwarzschild
and send $t\to i x^4$ and $\phi\to i\phi$ and the metric on the
two sphere becomes two dimensional de Sitter space
$d\theta^2-\sin^2\theta d\phi^2$.  Now $\phi$ is a timelike
coordinate and we leave it noncompact.  At $\theta=0,\pi$ there
are clearly Rindler type horizons about which we analytically
continue $\theta$ and again obtain de Sitter space
$-d\theta^2+\sinh^2\theta d\phi^2$.

\begin{figure}[htb]
\begin{center}
\epsfxsize=3.5in\leavevmode\epsfbox{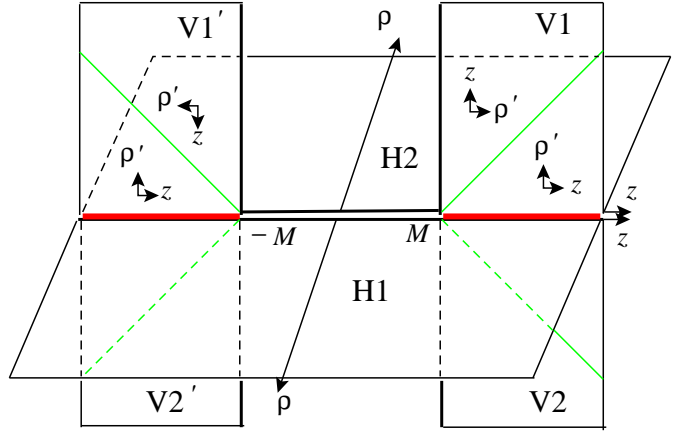} \caption{The
elliptic representation of the Witten bubble contains two horizons
and four special null lines.} \label{bubblecard1}
\end{center}
\end{figure}

This second analytic continuation to the Witten bubble in
Schwarzschild coordinates \be
ds^2=(1-2M/r)(dx^4)^2+(1-2M/r)^{-1}dr^2+r^2(d\theta^2-\sin^2\theta
d\phi^2)  \ee
has a corresponding second analytic continuation in Weyl
coordinates. In Weyl coordinates we find that analytically
continuing the time $t$ turns the horizon of the Schwarzschild
card into a boundary, while continuing the coordinate $\phi$ turns
the boundaries on the horizontal card into noncompact acceleration
horizons along the semi-lines $|z|\geq M$, $\rho=0$; the
coordinate $z$ is not continued.  Setting $\rho'=i\rho$ at the
horizons, we find a vertical 45 degree triangular cards with
special null lines on the semi-infinite rays $\rho'= z-M$ for
$z\geq M$ and $\rho'=-M-z$ for $z\leq -M$.  This region is doubly
covered and so it is necessary to change branches of the function
$R_\pm$ at the null line as in the Schwarzschild case and this
procedure produces a quarter-plane vertical card.  These null
lines are extensions of the null lines of the Schwarzschild
solution onto the bubble card diagram. From the BL coordinate
point of view they lie on $r=M+M \cosh\theta$ which extends
outward to large values of $r$. The two horizontal half-plane
cards are static patches of the solution. This elliptic card
representation of the Witten bubble is shown in
Fig.~\ref{bubblecard1}.

This second card diagram of the Witten bubble is closely related
to the Schwarzschild card we previously constructed.  In the
language of \cite{EmparanWK}, there are actually both $e^{2U_1}$
rods which exist for the coordinate $t$, as well as $e^{2U_2}$
rods for $\phi$. Double Killing rotation (analytic continuation)
of the coordinates $t$ and $\phi$ is equivalent to switching which
Killing coordinate each rod sources. Hence the elliptic
representation of the Witten bubble is sourced by two
semi-infinite (time) rods separated by a ($\phi$) line segment
rod. This is an alternate view of the construction of the elliptic
card diagram for the Witten bubble.

In this representation, one has the freedom to identify the rear
horizontal cards gotten to by going through the left and right
acceleration horizons.  We may also identify after a finite number
of trips around; this gives a discrete set of spacelike
periodicities of dS$_2$.  In the hyperbolic representation (one
vertical card), the azimuth of dS$_2$ was not drawn and we were
free to make any periodicity. On the other hand the elliptic
representation makes clear that someone in the right vertical card
is causally disconnected from someone in the left vertical card.
Different card diagram representations highlight different
features and possibilities.

The reason that more than one card diagram exists for the same
solution can be thought of as different ways to setup a Killing
congruence on our symmetric space.  For the case of the Witten
bubble the symmetric space is dS$_2$ and for S-Schwarzschild it is
${\bf H}_2$.

The symmetry of the bubble solution is large enough so that we may
write the bubble in Weyl coordinates in three distinct ways which
allow us to see different parts of the symmetry structure.  The
two aforementioned hyperbolic and elliptic representations of the
bubble are easily understood by examining the embedding of dS$_2$
into flat space.

To begin with, the unit sphere
\begin{equation}
ds^2=d\theta^2 +\cos^2\theta d\phi^2
\end{equation}
has an $SO(3)$ symmetry and may be embedded into ${\bf R}^3$ as
\begin{equation}
Z=\sin\theta, \ X=\cos\theta\cos\phi, \ Y=\cos\theta\sin\phi,\qquad
Z^2+X^2+Y^2=1.
\end{equation}
The sphere can be analytically continued to global coordinates
covering all of de Sitter space by taking $\theta\rightarrow i
\theta$, which sends $Z\to iZ'$, and so we now have an embedding
into ${\bf R}^{2,1}$
\begin{equation}Z'=\sinh\theta, \ X = \cosh\theta \cos\phi, \
Y=\cosh\theta\sin\phi,\qquad -Z'^2+X^2+Y^2=1
\end{equation}
\begin{equation}
ds^2=-d\theta^2 + \cosh^2\theta d\phi^2 \ .
\end{equation}
The coordinate $\phi$ parametrizes the azimuthal angular direction
whose periodicity we may choose.  This gives the first hyperbolic
representation of the bubble.

It is also possible to obtain by analytic continuation a second
embedding of de Sitter by $\phi\rightarrow i \phi$, so $Y\to iY'$
\begin{equation}
Z=\sin\theta, \  X=\cos\theta\cosh\phi, \
Y'=\cos\theta\sinh\phi,\qquad Z^2+X^2-Y'^2=1
\end{equation}
\begin{equation}
ds^2=d\theta^2- \cos^2\theta d\phi^2 \ .
\end{equation}
In this case the coordinate $\phi$ now parametrizes a noncompact
direction. This embedding gives only a patch $|Z|\leq 1$, $X\geq
0$ of de Sitter.  Our elliptic representation of the bubble comes
from this second analytic continuation.  Both of these two
analytic continuations give de Sitter space with $SO(2,1)$
symmetry but in different parametrizations and this is reflected
in the two card diagrams.

Now let us turn to the `elliptic' construction of S-Schwarzschild
(\ref{S-Sch-metric}), which is a vertical $\rho, \tau$ card, but
the metric is only real for $0\leq \rho\leq \tau-M$ (or for the
equivalent region $0\leq\rho\leq-\tau-M$). The boundary
$\rho=\tau-M$ is a special null line; in BL coordinates the null
lines are $R_{\pm}=t-M\pm M \cosh\theta=0$.   This triangular
region is covered twice and so we flip the triangular region about
the null line to make a quarter-plane vertical card. A null line
appears for the S-brane solution because the focus of
Schwarzschild at $z=M$ continues to $\tau=M$ which is on the real
manifold; special null lines always extend at $45^\circ$ on
vertical cards, from foci on the real manifold. Since we have
analytically continued coordinates, the null line now extends to
large values of time. In Weyl coordinates these null lines are
simply the vanishing of $R_\pm$ which is a function of say
complexified $\rho,\tau$; (S-)Schwarzschild and the Witten bubble
have null lines each being a real section of the same complex
locus.

The vertical boundary is at $\theta=0$ where the $\phi$-circle
vanishes. We then pass down through the lower edge $x^4$-horizon
to a horizontal $\rho', \tau$ half-plane card. The full card
diagram is like the elliptic Witten bubble of
Fig.~\ref{bubblecard1}, except the $\phi$-circle vanishes on the
boundary, and there is a singularity at the intersection of
horizontal and vertical cards on the left side; note that the
coordinate labels $\rho$ and $\rho'$ should be interchanged (as in
Fig.~\ref{chargedSbranecard} which is the charged version of this
solution). Just like for the Schwarzschild black hole card
diagram, past the singularity we have attached negative mass
universes which are represented as quarter plane vertical cards,
for reasons which become more clear when looking at charged
solutions.

Note how the card representation of the S-brane is quite different
from the black hole card diagram while the Penrose diagrams are
related by a simple ninety degree rotation.  This is because the
card diagram shows the compact or noncompact $\theta$ direction,
which shows the distance from the S-brane worldvolume.

There is an alternate way to get (elliptic) S-Schwarzschild from
the elliptic representation of the Witten bubble. Take the
vertical card and perform a $\gamma$-flip by sending $\gamma\to
\gamma + i \pi$. This turns a vertical card on its side,
interchanging the Weyl spacelike and timelike coordinates.
Performing this operation on the Witten bubble's vertical quarter
plane card with its $x^4$-boundary and $\phi$-horizon, results in
the S-Schwarzschild's  vertical quarter plane card with a
$\phi$-boundary and $x^4$-horizon as described above. Continuing
through the four card junction horizon, we fill out the rest of
S-Schwarzschild. In terms of Schwarzschild coordinates this
$\gamma$-flip amounts to starting with the Witten bubble and
changing the signs for $g_{rr}$ and $g_{\theta\theta}$
\begin{eqnarray}
&\ &(1-\frac{2M}{r})(dx^4)^2 +
\frac{dr^2}{1-\frac{2M}{r}}+ r^2(-d\theta^2+\sinh^2\theta
d\phi^2)\nonumber\\
&\ &\qquad \longrightarrow (1-\frac{2M}{r})(dx^4)^2 -
\frac{dr^2}{1-\frac{2M}{r}}+ r^2(d\theta^2+\sinh^2\theta d\phi^2)
\label{flip-bubble}
\end{eqnarray}
to obtain the S-Schwarzschild solution.

\begin{figure}[htb]
\begin{center}
\epsfxsize=3.5in\leavevmode\epsfbox{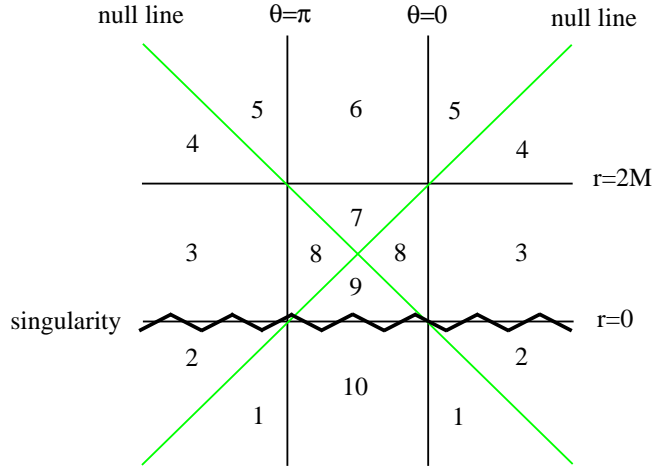} \caption{The
Schwarzschild black hole, Witten bubble and S-Schwarzschild are
all related by $\gamma$-flips and so their card diagrams may be
combined into this one extended diagram.} \label{fullsSchwarz}
\end{center}
\end{figure}

The elliptic form of the card diagrams show that Schwarzschild
S-brane, Witten bubble and Schwarzschild solutions have similar
structures and in fact they are all related by $\gamma$-flips and
trivial Killing continuations. Solutions which are related in this
manner may be conveniently drawn together in one diagram which
simultaneously displays all of their card diagrams.  For example
in the diagram in Fig.~\ref{fullsSchwarz} the S-Schwarzschild
solution comprises regions $1,2,3,4,5$, the Witten bubble is
regions $4,5,6$ and the Schwarzschild black hole is $6,7,8,9,10$.
Regions $1,2,10$ correspond to a singular Witten bubble of
negative `mass.'  For example on the Schwarzschild card the
horizontal card is region 6 and the vertical card square is
regions 7,8 and 9, while region 10 is the negative mass horizontal
card; note that this diagram represents all the different types of
card which appear in the full Schwarzschild card although not
every card in the infinite extended diagram.   In this diagram we
also see that the special null lines extend through the individual
solutions and so are a feature of the complexified spacetime.

\subsubsection{The parabolic diagram: Poincar\'e time}
\label{thirdcard-Poincare}

\begin{figure}[tp]
\hspace*{5mm}
\begin{minipage}{70mm}
\begin{center}
\includegraphics[width=7cm]{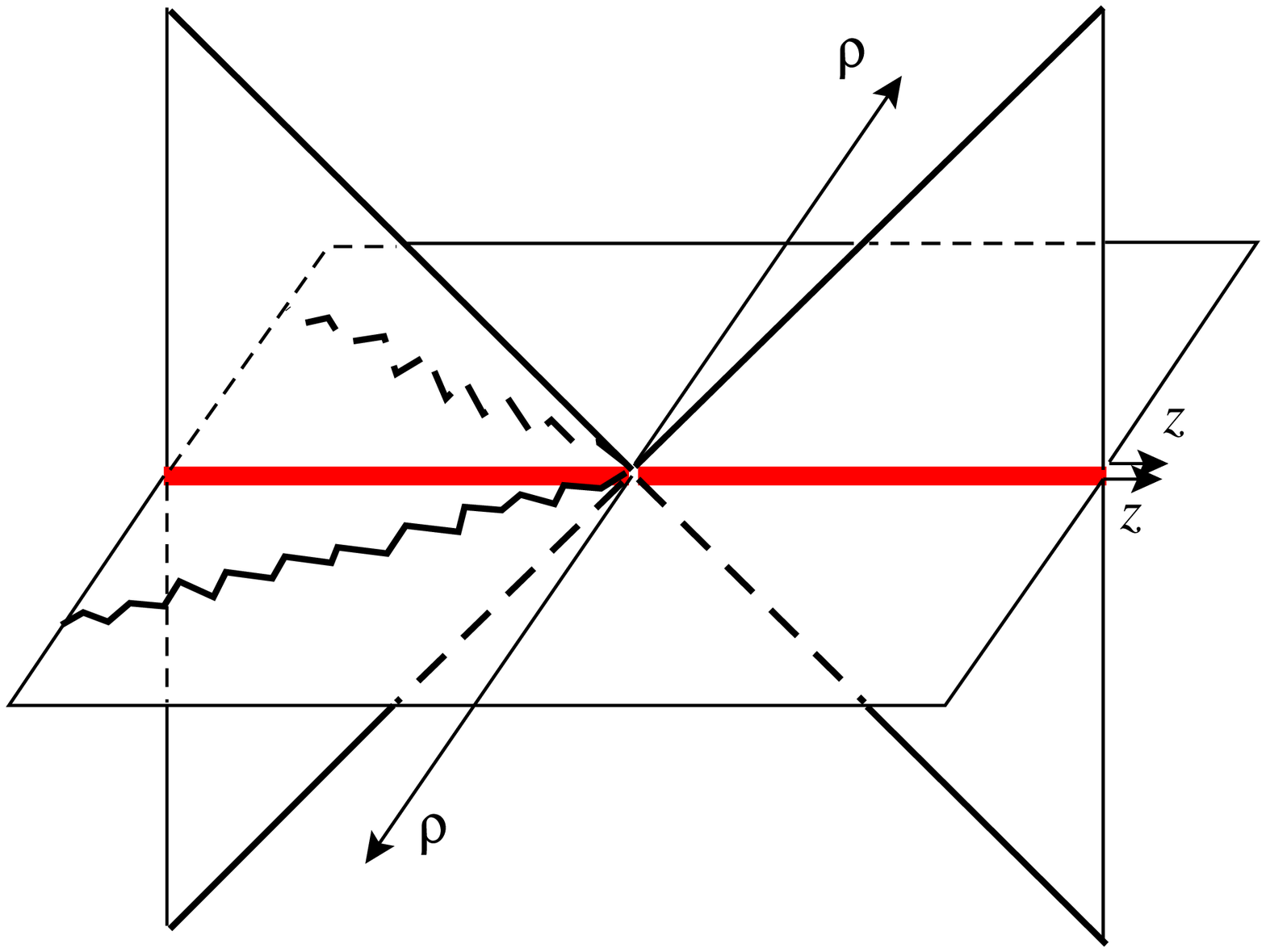}
\caption{The parabolic card diagram representation for S-Reissner
Nordstrom.  The forty five degree vertical lines represent null
infinity.} \label{bubblecard3}
\end{center}
\end{minipage}
\hspace*{15mm}
\begin{minipage}{70mm}
\begin{center}
\includegraphics[width=6cm]{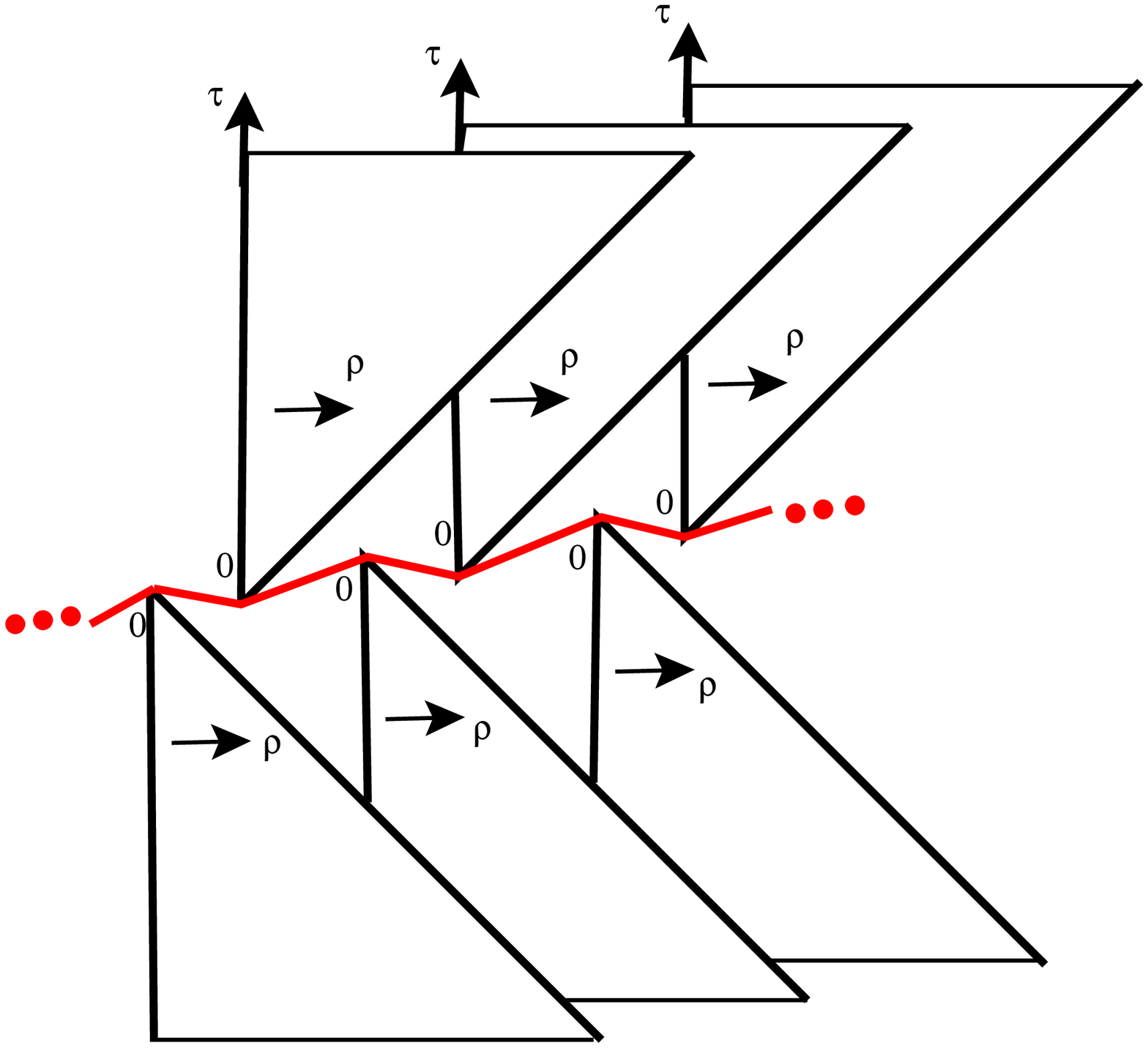}
\caption{The parabolic representation of the Witten bubble
contains an infinite number of wedge vertical cards pointing up
and down with each wedge joined to two others at the tip in a
dS$_2$ fashion.} \label{S-RN-3}
\end{center}
\end{minipage}
\hspace*{5mm}
\end{figure}

There is a third way to put a Killing congruence on ${\bf H}_2$ or
dS$_2$ using Poincar\'e coordinates.  Parametrizing hyperbolic
space (which is just Euclideanized AdS$_2$) as $ds^2=\sigma^2
dx^2+{d\sigma^2\over \sigma^2}$, and keeping the BL coordinate $t$
and the usual $x^4$ we get a Poincar\'e Weyl representation of the
S-Reissner-Nordstr\o m (S-RN)
spacetime\cite{Gutperle:2002ai,ChenYQ,KruczenskiAP}. This is a
charged version of the S-Schwarzschild where the $t$ dependence is
governed by $t^2-2Mt-Q^2$ with zeroes $t_+>0$, $t_-<0$, and a
singularity at $t=0$.  The charged S-brane is
\begin{eqnarray*}
ds^2&=&f(dx^4)^2-f^{-1}(e^{2\gamma}(-d\rho'^2+dz^2)+\rho'^2dx^2),\\
A&=&Qdx^4/t,\\
f&=&(1-2M/t-Q^2/t^2),\\
e^{2\gamma}&=&{t^2-2Mt-Q^2\over \sigma^2(M^2+Q^2)},\\
\rho'&=&\sigma\sqrt{t^2-2Mt-Q^2},\\
z&=&\sigma(t-M).
\end{eqnarray*}
In this Weyl representation, $\rho'$ is timelike on a $t>t_+$
vertical card which is a noncompact $45^\circ$ wedge, $\rho'<|z|$.
This connects along $z>0$ to a $t_-<t<t_+$ horizontal card; a
$t<t_-$ vertical card attaches to $z<0$. So this is similar to the
representation of S-RN (or S-Schwarzschild) that we have seen,
except the line segment $-M<z<M$ has collapsed and the special
null lines are now conformal null infinity (Fig.~\ref{S-RN-3}).
The singularity on
the horizontal card is particularly easy to describe in these
coordinates; it is on a ray $z/\rho=-M/Q$.  For
comparison, the elliptic and hyperbolic representations are in
Figs.~\ref{chargedSbranecard} and \ref{SRN2fig}.

If we take the $t>t_+$ (or $t<t_-$) $45^\circ$ wedges and turn
them on their sides via $\gamma\to \gamma+i\pi$, we get the
parabolic version of the $r\geq r_+$  charged Witten bubble. The
line which used to be the horizon in the S-brane card diagram
becomes a boundary which is the minimum volume sphere, at
$\rho=0$. Time is now purely along the $\tau$ direction as in the
hyperbolic bubble card representation. The special null line is
still ${\cal I}^\pm$ since $\rho=|\tau|$ corresponds to
$r\to\infty$.  The vertex of the card is not the end of the
spacetime.  These wedge cards only represent $\sigma>0$ and so the
card diagram should be extended to negative times.  The card
diagram is an infinite row of $45^\circ$ wedge cards pointing up
and an infinite number pointing down.  The vertex of each upward
card is attached to its nearest two downward neighbors (one to the
left and one to the right), in the usual dS$_2$ fashion as shown
in Figure~\ref{bubblecard3}. One can identify cards so only needs
one upward and one downward card with two attachments. Although
this card diagram is not the most obvious representation of the
Witten bubble, it will be useful in understanding the more
complicated S-dihole II universes of Sec.~\ref{sdihole-sec}.

\subsection{Gravitational wave fall-off}

The Schwarzschild S-brane (\ref{S-Sch-metric})
has a warp factor for the brane direction of $1-2M/t$, which
decays to unity as $1/t$.  This is the expected behavior at fixed
spatial position for a wave produced by a $(2+1)$-codimension
source. To really consider the wave equation, we should move to
conformally Minkowskian coordinates in two dimensions, Weyl
coordinates $(\rho,\tau)$, and follow the wave as it propagates
outward at the speed of light. As we have seen, the
S-Schwarzschild geometry includes such a vertical card region with
some outgoing wave activity centered about $\rho=\tau+M$.  As
$t\to\infty$ along this null line, $\rho\sim t^2/M$, and so our
wave falls off as $1/\tau^{1/2}$. This is the same fall-off as the
warp factor for the S-dihole of \cite{JonesRG}.

We can understand this power-law decay with the following
simplified model, using $x,y,z,\tau$ coordinates in Minkowski
4-space.  A codimension $3+1$ source for the wave equation gives
precisely causal wave fronts, and a codimension $2+1$ source can
be gotten by dimensional reduction.  Hence we may write a wave as
$$\Psi(x,y,\tau)=\int_{-\infty}^\infty\Big({dz\over
x^2+y^2+z^2-\tau^2+i\epsilon}+{\rm c.c.}\Big)$$ where $\epsilon$
is a regulator.  Along $x^2+y^2-\tau^2=0$, $\Psi={1\over
\sqrt{\epsilon}} \int_{-\infty}^\infty {\zeta^2 d\zeta\over
\zeta^4+1}$.  If the wave is regulated by displacing the source
into imaginary time, $\tau\to \tau-ia$, we find that
$\epsilon\propto\tau$.  Therefore the behavior for such
time-displaced wave sources is $\Psi\sim 1/\tau^{1/2}$ and this is
precisely the case for the S-dihole of \cite{JonesRG}. In the
S-Schwarzschild case, the card diagram is more complicated and the
singularity is stationary, but we expect a similar fall off of the
metric along the vertical card.

\section{New S-branes from Diholes} \label{sdihole-sec}

In this section we discuss new time dependent solutions which
arise by analytically continuing dihole solutions.  The use of
Weyl coordinates and the construction of card diagrams will be
helpful in understanding their interesting global structure. We
first review the dihole solution as well as the S-dihole solution
of \cite{JonesRG} before examining new S-dihole solutions.

The extremal black dihole of
\cite{Bonnor,Chandrasekhar:ds,Emparan:1999au} was analytically
continued \cite{JonesRG} to obtain a smooth time dependent
solution free of curvature singularities, closed timelike curves
(CTCs) and horizons.  Let us refer to that solution as S-dihole I.
In the previous section, we discussed how in Weyl coordinates the
Schwarzschild solution may be analytically continued to the bubble
of nothing and if we additionally send $M\to iM$ this gives us the
S0-brane solution which we called S-Schwarzschild.

Many gravity solutions can be analytically continued to obtain new
time dependent spacetimes and it is not uncommon to have two or
more different analytic continuations.  Sending the mass parameter
$M\to iM$ in Weyl coordinates for the S-dihole I solution produces
a substantially different S-dihole II. One of the main differences
for spacetime structure, is that the Weyl foci and the special
null lines are at real values and so will play an important role.
These new solutions are related to the decay of unstable objects,
cosmological solutions \cite{Cornalba:2002fi} and posses near
horizon scaling limits.

\subsection{Diholes, S-diholes and Bonnor transformations}

The black magnetic dihole metric in Boyer-Lindquist coordinates
(assuming $a\neq 0$) is \be ds^2= (1-\frac{2Mr}{\Sigma})^2 [ -dt^2
+ \frac{\Sigma^4}{(\Delta+
(a^2+M^2)\sin^2\theta)^3}(\frac{dr^2}{\Delta} + d\theta^2)] +
\frac{\Delta \sin^2\theta}{(1-\frac{2Mr}{\Sigma})^2} d\phi^2 \ee
\begin{equation}
A= \frac{2aMr\sin^2\theta}{\Delta + a^2 \sin^2\theta} d\phi
\end{equation}
\be \Delta= r^2-2Mr-a^2 ,\hspace{.3in} \Sigma= r^2 -a^2
\cos^2\theta \ . \ee The dihole represents two oppositely charged
magnetic black holes separated  by the coordinate distance
$2\sqrt{M^2+a^2}$ along the $z$-axis.

The dihole was first found using a solution generating technique
called the Bonnor transformation starting from the Kerr black hole
\cite{Bonnor,Herlt}.
In Weyl-Papapetrou coordinates this Bonnor transformation takes a
stationary vacuum solution
$$ds^2=-f(dt-\omega
d\phi)^2+f^{-1}(e^{2\gamma}(d\rho^2+dz^2)+\rho^2d\phi^2),$$ and
produces a static charged solution
\begin{eqnarray}
ds^2&=&-f^2dt^2+f^{-2}(e^{8\gamma}(d\rho^2+dz^2)+\rho^2d\phi^2),
\label{bonnortrans}\\
A&=&B(\rho,z)d\phi,\nonumber
\end{eqnarray}
where $\omega=iB$ and $\omega$ is proportional to a parameter (the
angular momentum $a$ in the case of Kerr) which must be
analytically continued to make $B$ real.

Analytically continuing the coordinates
\begin{equation}
\theta \rightarrow \  \frac{\pi}{2} + i \theta, \ t \rightarrow i
x^4
\end{equation}
gives us the S-dihole I solution of \cite{JonesRG}
\begin{equation}\label{sdihole1}
\Delta \rightarrow r^2 - 2Mr - a^2 =\Delta, \hspace{.3in}
\Sigma\rightarrow r^2+a^2\sinh^2\theta =\Sigma,
\end{equation}
\vspace{-.2in}
\begin{equation}
ds^2 = (1-\frac{2Mr}{\Sigma})^2 [ (dx^4)^2 +
\frac{\Sigma^4}{(\Delta+
(a^2+M^2)\cosh^2\theta)^3}(\frac{dr^2}{\Delta} - d\theta^2)] +
\frac{\Delta \cosh^2\theta}{(1-\frac{2Mr}{\Sigma})^2} d\phi^2 \ee
\be A= \frac{2aMr\cosh^2\theta}{\Delta + a^2 \cosh^2\theta} d\phi
\ .
\end{equation}
This analytic continuation is the same as the one performed on
Schwarzschild to obtain the Witten bubble of nothing in the first
global representation.  The dihole and S-dihole I do not have
closed timelike curves in $\phi$, since $r=r_{+,-}$ ($\Delta=0$)
is like the origin of polar coordinates and closes off the
spacetime.  The fact that spacetime stops at $\Delta=0$ and
periodically identifying $\phi\simeq \phi+2\pi{a^4\over
(M^2+a^2)^2}$, ensures that the metric and vector potential are
well behaved and smooth everywhere.

Bonnor transformations and Wick rotations are both sensible ways
to act on the Kerr solution, and it is interesting to consider the
set of solutions which can be obtained from their combined
effects.  Since the dihole is the Bonnor dual of the Kerr black
hole, this means that S-dihole I is the Bonnor dual of the Kerr
bubble\cite{Dowker:1995gb, clean}.  The Kerr bubble is obtained
from Kerr by sending $z\to i\tau$ and $a\to ia$.  Although the
Kerr bubble needs a twisting to close a circle at the boundary
$\rho=0$, we will still consider S-dihole I and the Kerr bubble to
be Bonnor dual. S-dihole I was originally defined to be the
spacetime where $r\geq r_+$.  In addition to the spacetime
connected to large value of the radius, in these BL coordinates it
is clear that S-dihole I actually contains a second non-singular
universe where $r\leq r_-$; the singularities (surrounding the
negative-mass black holes) do not intersect $z=0$ by symmetry.  In
fact the Kerr bubble is similarly well defined and non-singular
for negative values of the radius $r\leq r_-$.

Recently the S-Kerr or twisted S-brane solution has been found
\cite{Wang:2004by,regSbraneQuevedo,Lu:2004ye} and we now examine
its Bonnor dual which we call S-dihole II.  To obtain S-dihole II
we analytically continue the dihole in the following alternative
method \be \label{sd2-wick} r\rightarrow it, \hspace{.3in} t
\rightarrow i x^4, \hspace{.3in}\theta \rightarrow i \theta;
\hspace{.3in}M\rightarrow i M.  \ee The angular momentum is not
Wick rotated; it has already been Wick rotated in the Bonnor
transformation of Kerr. The S-dihole  II solution is \be \Delta
\rightarrow -t^2 + 2Mt - a^2 =-\Delta, \hspace{.3in}
\Sigma\rightarrow -t^2-a^2\cosh^2\theta =-\Sigma, \ee
\be\label{sdihole} ds^2 =(1-\frac{2Mt}{\Sigma})^2 [ (dx^4)^2 +
\frac{\Sigma^4}{(\Delta+
(a^2-M^2)\sinh^2\theta)^3}(-\frac{dt^2}{\Delta} + d\theta^2)] +
\frac{\Delta \sinh^2\theta}{(1-\frac{2Mt}{\Sigma})^2} d\phi^2 \ee
$$A= \frac{2aMt\sinh^2\theta}{\Delta + a^2 \sinh^2\theta} d\phi.$$
The factor $\Sigma$ is always positive.  Remembering that elliptic
S-Schwarzschild is more complicated than the hyperbolic Witten
bubble, it might be expected that S-dihole II is more complicated
than S-dihole I.  In fact S-dihole II does have several new and
unexpected features.  We discuss the properties of this new
solution on a case-by-case basis, discuss relevant regions and
then discuss how certain regions must be pieced together.

\subsection{Extremal case}

Let us first examine the extremal case $|M|=|a|$ for S-dihole II.
The metric near the origin is
$$ds^2={T^4\over M^4}(dx^4)^2-{M^4\over T^4} dT^2 + {M^4\over
T^2}(d\theta^2+\sinh^2\theta d\phi^2)$$C where $T=t-M$.
We do not find a de Sitter space limit but a singular metric
This extremal case is not related to de Sitter space because the
Bonnor transformation has shifted the powers of the coordinate $T$
in the metric components.

\subsection{Superextremal case}
\label{supersdiholeii}

For the superextremal case $|M|<|a|$, $\Delta$ is never zero and
the solution is smooth.  The coordinate $\theta$ is noncompact and
spacelike.  The $\phi$-circle vanishes along $\theta=0$ around which the
metric
has the expansion
$$ds^2\supset {(t^2+a^2)^2\over \Delta}(d\theta^2+\theta^2d\phi^2) \ .$$
The possible conical singularity can be simply removed by taking
$\phi\simeq\phi+2\pi$; this is the same periodicity for the black
dihole on the axis outside the black holes.

The superextremal S-dihole II solution is the smooth magnetic
Bonnor dual of the superextremal S-Kerr \cite{Wang:2004by}. Both
superextremal solutions can be represented by one vertical
half-plane card where the edge of the card is a vertical boundary.

\subsection{Subextremal case}

For the parameter range $|a|<|M|$, S-dihole II has several
interesting surprises.  This solution can be divided into three
separate nonsingular spacetimes and three separate spacetimes with
singularities behind acceleration horizons.  Some of these
spacetimes have near horizon scaling limits and are free of closed
timelike curves in four dimensions.  To begin with we outline some
of their most important features before giving a detailed
analysis.

The metric has three coordinate singularities. First, the quantity
$(\Delta+(a^2-M^2)\sinh^2\theta)^3$ in the denominator vanishes at
$t^{\rm DH}_\pm=M\pm\cosh\theta\sqrt{M^2-a^2}$.  This is an
extended analog of the extremal dihole horizons and it is a
special null line which serves as null infinity. Second, the
factor $\Delta=0$ vanishes when $t^{\rm
KH}_\pm=M\pm\sqrt{M^2-a^2}$. In analogy to the Kerr horizon this
will turn out to be a Weyl card boundary or horizon. Third, the
quantity $\Sigma-2Mt=\Delta+a^2\sinh^2\theta$ vanishes for small
enough $\theta$ $t^{\rm
ergo}_\pm=M\pm\sqrt{M^2-a^2\cosh^2\theta}$; this is the analog of
the Kerr ergosphere. In the case of Kerr, the ergosphere lies
between the two Kerr horizons and can be smoothly traversed. For
the S-dihole II this Bonnor dual analogue of the ergosphere turns
out to be the location of a charged object and is a true
singularity.
We will continue to use the terminology `ergosphere' for this
singular locus.

\begin{figure}[tp]
\hspace*{5mm}
\begin{minipage}{70mm}
\begin{center}
\includegraphics[width=4cm]{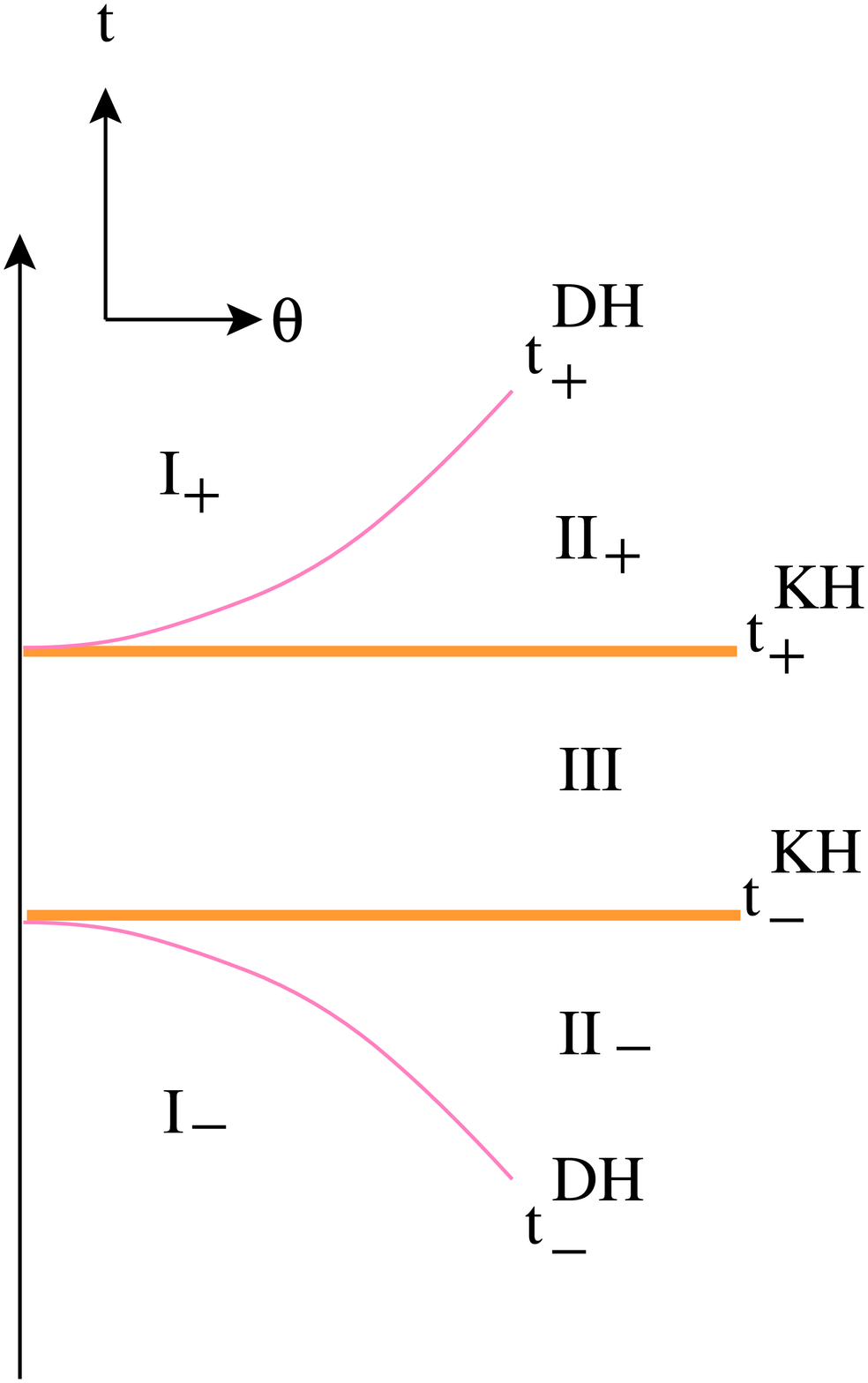}
\caption{S-dihole II is actually several different solutions
including regions I$_\pm$, II$_\pm$ and III.  Regions I$_\pm$ and
II$_\pm$ are separated by null infinity at $t_\pm^{\rm DH}$ while
region III is a signature $+---$ solution separated by a boundary at
$t_\pm^{\rm KH}$.} \label{Sdihole2a}
\end{center}
\end{minipage}
\hspace*{10mm}
\begin{minipage}{70mm}
\begin{center}
\includegraphics[width=6cm]{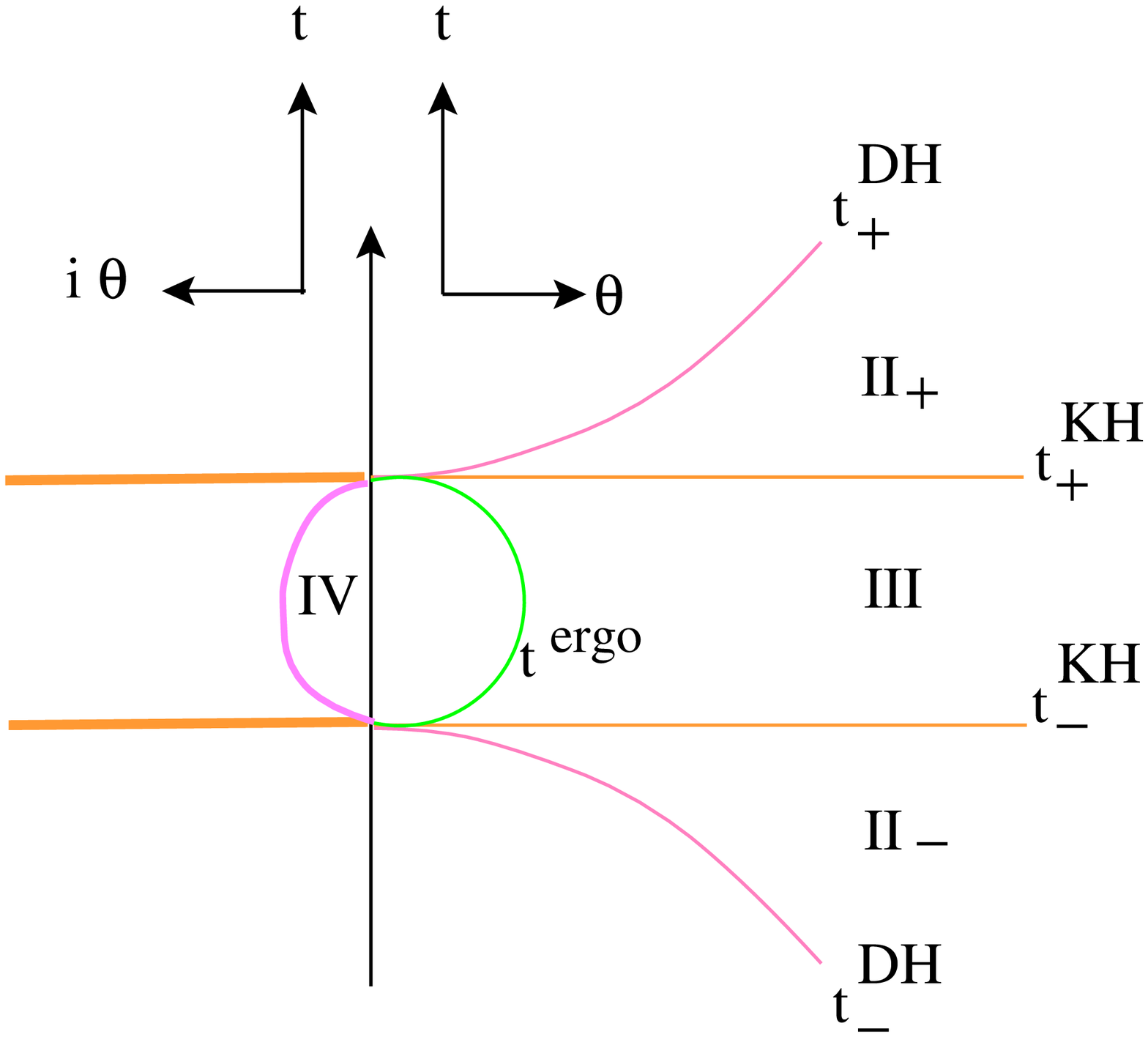}
\caption{Applying a $\gamma$ flip to region III we find a $+++-$ S-brane
solution where now $t_\pm^{\rm KH}$ are horizons. An ergosphere
singularity lies in region III and $\theta=0$ is a horizon which
leads to a new region IV.  A complete picture of S-dihole II is in
Fig.~\ref{fullSdihole2}.} \label{Sdihole2b}
\end{center}
\end{minipage}
\hspace*{5mm}
\end{figure}

Due to the three coordinate singularities, it is natural at first
glance to consider S-dihole II to be a single solution subdivided
into five regions I$_\pm$, II$_\pm$, and III as indicated on
Figure~\ref{Sdihole2a}.

In regions I$_+$ and I$_-$, the coordinate $\theta$ is spacelike,
$t$ is timelike and $\theta=0$ is a boundary.  This spacetime is
smooth and does not have a conical singularity at the boundary if
$\phi\simeq\phi+2\pi$.

In regions II$_+$ and II$_-$, the coordinate $\theta$ timelike and
$t$ spacelike and $t_\pm^{\rm KH}$ is thus a boundary where the
$\phi$-circle closes.  Near $t^{\rm KH}_+$ the metric looks like
$$ds^2\supset {2\Sigma^2\over \sinh^2\theta}\left[{a^4\over
(M^2-a^2)^{7/2}}d\epsilon^2+\epsilon^2{(M^2-a^2)^{1/2}\over
a^4}d\phi^2\right]\ . $$  where we have made the change of
variables $T=t-t^{\rm KH}_+=\epsilon^2$.  To avoid a conical
singularity the $\phi$ coordinate has periodicity
\begin{equation}\label{altperiod}
\phi\simeq\phi+2\pi a^4/(M^2-a^2)^2 \ .
\end{equation}
This is the same periodicity in the black dihole which removes the
strut from between the black holes
\cite{Emparan:1999au}\cite{JonesRG}, with the mass $M$
analytically continued.

Region III is therefore separated from regions II$_\pm$ and has
signature $(1,3)$ with $t$, $\theta$, and $\phi$ all being
timelike coordinates.  To interpret this as a solution to the
Einstein-Maxwell equations we perform the $\gamma\to \gamma+i\pi$
flip so the coordinates $t$ and $\theta$ become spacelike leaving
just $\phi$ as time;\footnote{
A $\gamma$-flip on a horizontal card will change signature (1,3)
to (3,1) and vice-versa.  A signature (1,3) charged S-brane is
equivalent upon $g_{\mu\nu}\to -g_{\mu\nu}$ to a signature (3,1)
charged E-brane solution.}
we still label the region after the $\gamma$
flip as region III.  See Figure~\ref{Sdihole2b}.  Although this is
therefore strictly speaking a new solution we will continue to
refer to it as part of S-dihole II.  There is a singularity at
$t^{\rm ergo}$ which divides this region into an internal small
$\theta$ region and an external large $\theta$ part.  At this
ergosphere, the electric gauge field is infinite and so this is
the location of a brane object.  The region outside the ergosphere
has two horizons, at $t_\pm^{\rm KH}$, with Euclidean periodicity
(\ref{altperiod}). Applying the $\gamma$-flip to region III, has
turned $t_\pm^{\rm KH}$ into horizons which can be traversed.  The
horizons connect to regions II$_\pm$ which have also been turned
on their side via $\gamma\to \gamma+i\pi$; we will continue to
call these region II$_\pm$.  The internal part has a horizon at
$\theta=0$ with Euclidean periodicity $2\pi$. Past this horizon at
$\theta=0$ is the new region IV as shown in Fig.~\ref{Sdihole2b}.
Region IV is compact and bounded in these BL coordinates.  In fact
this boundary is the continuation of $t_\pm^{\rm DH}$ and will
again represent null infinity!

In the next subsection we provide a full description of these
spacetimes and further general discussion of these different
regions of the S-dihole II appears in Sec.~\ref{analogy-sec}.

\subsubsection{Piecing together the spacetimes}

\begin{figure}[htb]
\begin{center}
\epsfxsize=3.5in\leavevmode\epsfbox{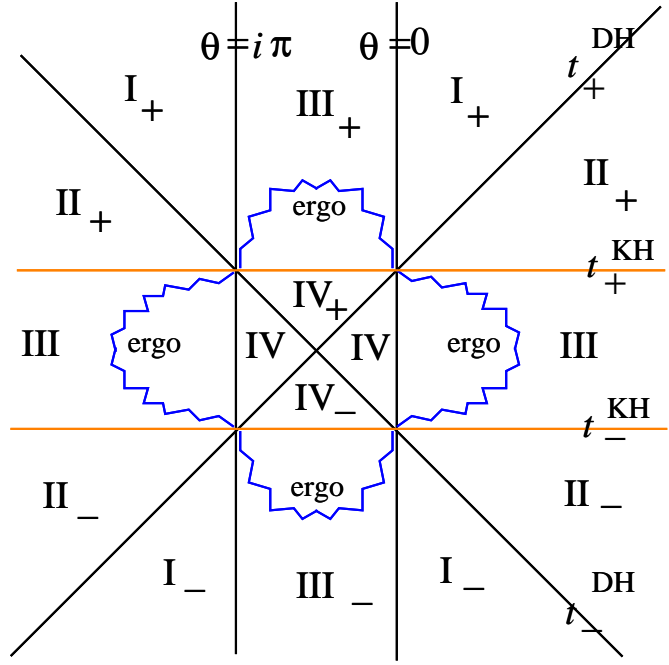} \caption{The
regions of S-dihole II} \label{fullSdihole2}
\end{center}
\end{figure}

The six aforementioned regions are not the end of the spacetimes
and their maximal extensions will include new regions III$_\pm$
and IV$_\pm$ (see Fig.~\ref{fullSdihole2}).  In total S-dihole II
actually contains six different maximally extended universes, and
so six card diagrams, which we describe here in Weyl language.
Three of the spacetimes, which we label $\cal{E}, \cal{E}_\pm$,
are singular while $\cal{U}, \cal{U}_\pm$ are non-singular.
Further details in Weyl coordinates will also appear in
Sec.~\ref{sdihole-weylcoors-subsec}.

\begin{figure}[tp]
\hspace*{5mm}
\begin{minipage}{60mm}
\begin{center}
\includegraphics[width=7cm]{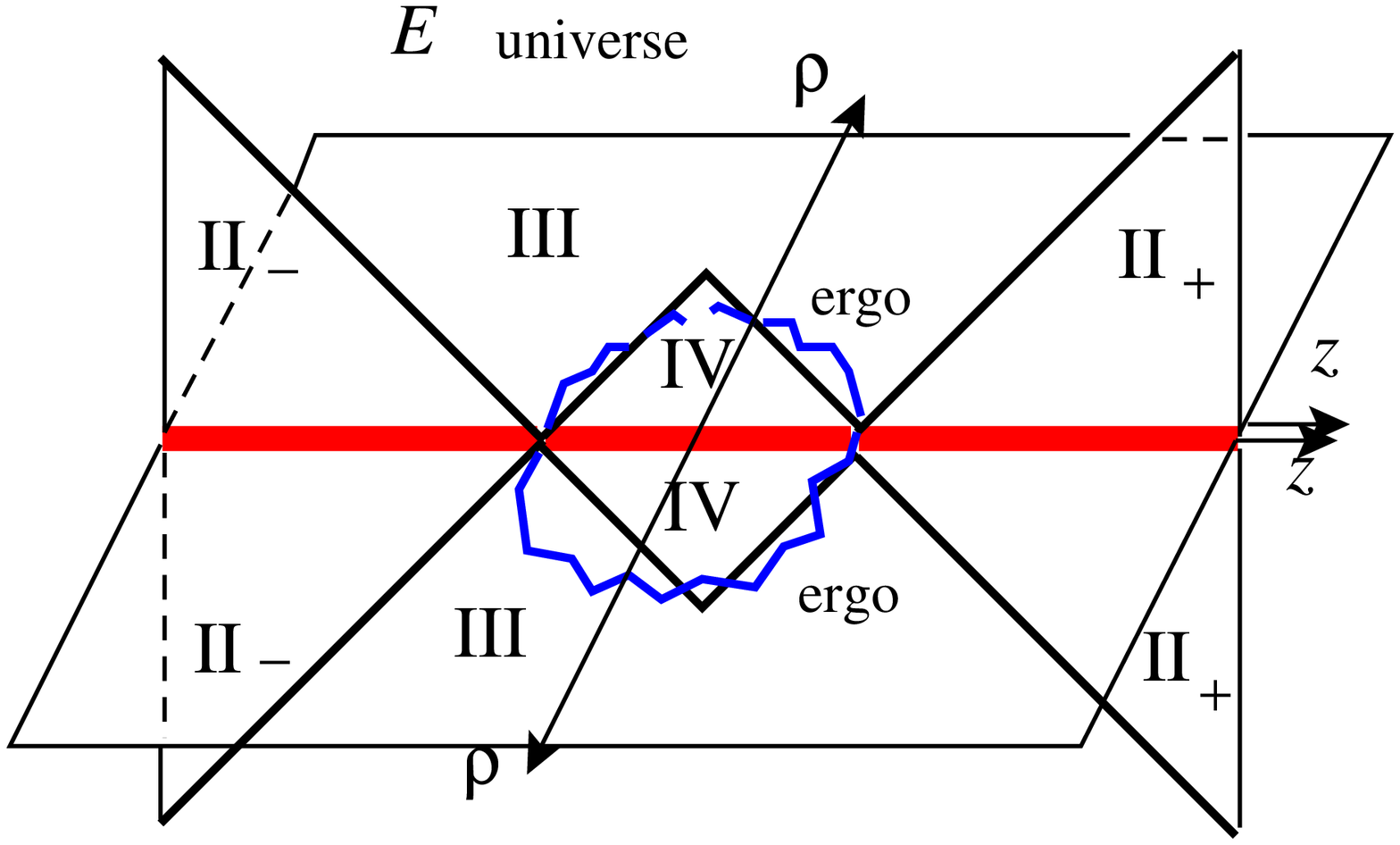}
\caption{The $\cal{E}$ card diagram consists of eight cards and a
singular ergosphere on the horizontal region III.} \label{Euniv}
\end{center}
\end{minipage}
\hspace*{20mm}
\begin{minipage}{60mm}
\begin{center}
\includegraphics[width=7cm]{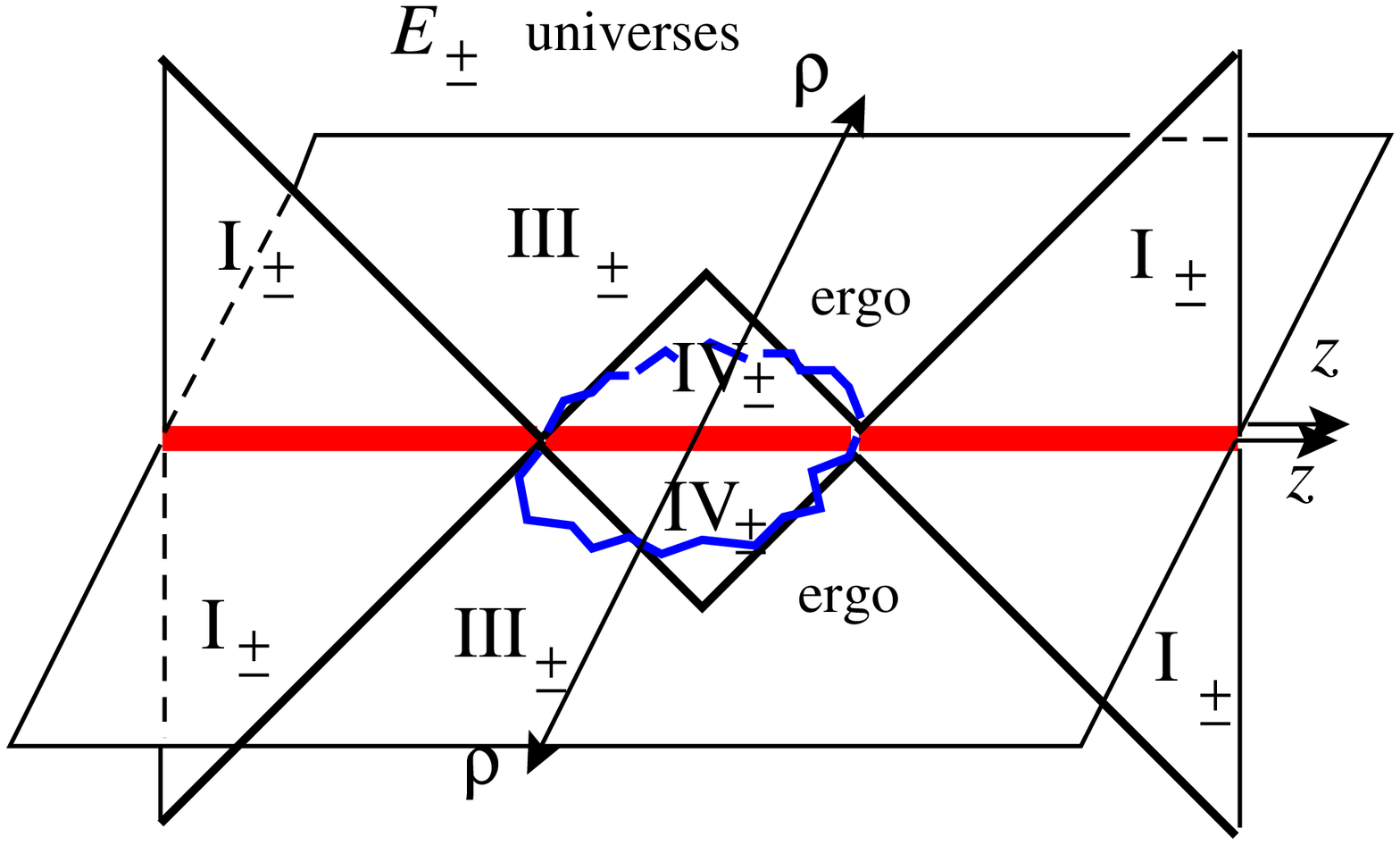}
\caption{The $\cal{E}_{\pm}$ cards are similar to $\cal{E}$ and
each other.  The ${\cal E}_-$ universe has a ring singularity
atop the ergosphere at $z=0$ (not pictured).} \label{Epmuniv}
\end{center}
\end{minipage}
\hspace*{5mm}
\end{figure}

The three ${\cal E}$ card diagrams are drawn in Figs.~\ref{Euniv}
and \ref{Epmuniv}.  Each diagram comprises 8 cards, 2 singular
ergospheres, and is time symmetric. All of these card diagrams
actually describes two universes, one on each side of the
ergosphere.  The ergosphere is located on the horizontal cards so
the singularity is not naked from the point of view of any of the
vertical cards.

The universe ${\cal E}$, comprising $t_-^{\rm DH}<t<t_+^{\rm DH}$,
can be obtained by starting with the region III, $t_-^{\rm KH}\leq
t \leq t_+^{\rm KH}$, and performing $\gamma\to \gamma+i\pi$. This
universe is composed of two copies of regions II$_\pm$, III and
IV.  Region III is represented as a horizontal card with foci at
$z=\pm\sqrt{M^2-a^2}$ and where the coordinate $t$ is conveniently
parametrized as $t=M+\sqrt{M^2-a^2}\cos\zeta$. On this card the
ergosphere is the curve $(M^2-a^2)\sin^2\zeta=a^2\sinh^2\theta$
which ends on the two foci and extends outward to touch
$(\rho,z)=((M^2-a^2)/a,0)$. Along $z>\sqrt{M^2-a^2}$ region III
attaches to region II$_+$ which is a $45^\circ$ wedge card
$\rho'<z-\sqrt{M^2-a^2}$.  These wedges are bounded by special
null lines $\rho'=z-\sqrt{M^2-a^2}$ which are the Weyl coordinate
representation of $t^{DH}$.  We can further attach two more cards
to form a four card junction; region III is connected to two
II$_+$ cards, one pointing up and one pointing down, and another
horizontal region III card in the back. Similarly along
$z<-\sqrt{M^2-a^2}$ we have a four card junction, including two
II$_-$ cards. Surprisingly the line
$-\sqrt{M^2-a^2}<z<\sqrt{M^2-a^2}$ is in fact also a horizon. Here
we again have four cards including two copies of region IV which
is a compact vertical wedge card $\rho'<\sqrt{M^2-a^2}\pm z$.  The
card diagram is shown in Fig.~\ref{Euniv}.  To understand this
card diagram it is helpful to remember that the special null lines
can represent conformal null infinity which is infinitely far
away, and that the focal points themselves lie down throats and
are infinitely far away.  In Sec.~\ref{scriminussec} we explicitly
demonstrate that the null lines here correspond to null infinity.
As additional evidence one may check that the Riemann tensor
squared vanishes along the null lines.

The universe ${\cal E}_+$ has $t>t_+^{\rm DH}$.  This universe is
composed of four copies of region I$_+$, and two copies of regions
III$_+$ and IV$_+$.  It is obtained by taking region I$_+$, which
is a $45^\circ$ wedge noncompact vertical card in Weyl coordinates
and turning it on its side via $\gamma\to \gamma+i\pi$. Instead of
being a boundary, $\theta=0$ is now a horizon which we can go
through to reach III$_+$.  In Weyl coordinates this is one of the
two horizons $|z|>\sqrt{M^2-a^2}$.  Each horizon is a four card
junction which attach to two copies of region I$_+$ and III$_+$ in
the back.  Region III$_+$ is qualitatively similar to region III
and has foci at $z=\pm\sqrt{M^2-a^2}$ and an ergosphere along the
line $(M^2-a^2)\sinh^2\zeta=a^2\sin^2\theta$ running in the
horizontal card.   There is a further horizon $|z|<\sqrt{M^2-a^2}$
which attaches to a region IV$_+$ qualitatively similar to IV; it
is a compact 45-45-90 wedge vertical card.  In Weyl coordinates
this region IV$_+$ can be found by starting with region IV and
passing through the special null line $R_-=0$, where $R_-$ is the
distance to the $t_+$ focus. The card diagram for this universe
and the $\cal{E}_-$ universes, which we describe next, are shown
in Fig.~\ref{Epmuniv}.

A third universe ${\cal E}_-$ has $t<t_-^{\rm DH}$. It is similar
to ${\cal E}_+$ but it involves regions I$_-$, III$_-$, and
IV$_-$.   However since S-dihole II is not `time symmetric' these
are truly different universes.  The ergosphere is specified by the
same locus as for III$_+$. Region IV$_-$ is Weyl-adjacent to IV
and is gotten by passing the special null line $R_+=0$ where $R_+$
is the distance to the $t_-$ focus.

\begin{figure}[htb]
\begin{center}
\epsfxsize=3.5in\leavevmode\epsfbox{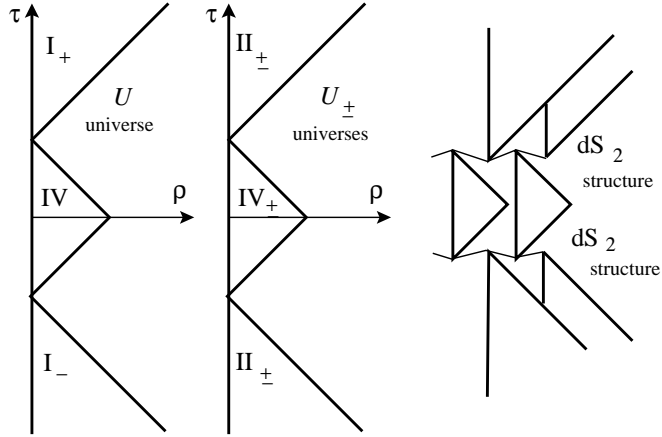} \caption{The $\cal{U}$
universes. Each pointlike intersection actually connects to two
similar spacetimes as illustrated on the right.}
\label{U-universes}
\end{center}
\end{figure}

As for the three non-singular $\cal{U}$ universes, they consist of
three vertical cards starting with one vertical $45^\circ$ wedge
card joined at its tip to a compact $45$-$45$-$90$ card which is
joined at its other tip to another vertical $45^\circ$ wedge card
(see Fig.~\ref{U-universes}).   The cards are joined pointwise at
their tips by a de Sitter type near horizon scaling limit
$\cal{W}$. Although the ${\cal U}$, ${\cal U}_\pm$ card diagram
structures are time symmetric, the $\cal{U}$ universe itself is
not time symmetric.

The universe ${\cal U}$ is composed of regions I$_-$, I$_+$, and
IV. Take region I$_+$ which is the vertical wedge $0\leq
\rho<\tau-\sqrt{M^2-a^2}$; in BL coordinates $t$ is time. At the
corner of the wedge, we will see in Sec.~\ref{scriminussec} that
the metric has a near-horizon scaling limit ${\cal W}$ which
includes dS$_2$ in the coordinates $x^4$, and $\sigma$, where
$\sigma$ loosely is the distance into the Weyl wedge from the
corner.  From our discussion of the third parabolic card
representation for the Witten bubble, we know then that such a
wedge card can connect to two regions `below the $t_+$ tip' and
they are precisely IV. Region I$_+$ is connected to region IV by
crossing a horizon and performing an analytic continuation of
$\theta$ and $\zeta$. Although this connection is unusual, if
I$_+$ did not connect to any other region, the universe would have
to begin at the tip of the wedge since $t$, or $\tau$ in Weyl
coordinates, is a timelike variable. Similarly at the bottom
region of IV, the $t_-$ tip attaches via ${\cal W}$ to two I$_-$
wedges. This universe is not time symmetric.  An interesting
feature of these card diagrams is that we can see that region IV
has both future and past null infinities, ${\cal I}^\pm$. They
meet along the point at spacelike infinity.  There is no easy
Penrose diagram of the ${\cal U}$ universes.  We must envision it
as the card diagram with a dS$_2$ Penrose structure near each Weyl
vertex.

The universe ${\cal U}_+$ is gotten from II$_+$, where $\theta$ is
time.  At the corner, we find ${\cal W}$ and this attaches to two
copies of IV$_+$.  At each other tip of IV$_+$, which is still at
$t_+$, we attach two copies of II$_+$.  This universe is time
symmetric.  Lastly ${\cal U}_-$ is gotten from II$_-$ and IV$_-$.
This universe is time symmetric.

It is important that the I$_\pm$, II$_\pm$, IV, IV$_\pm$ wedges
are free of ergosphere singularities, and that the `ring
singularity' $\Sigma=0$ occurs only as a point in III$_-$, and sits
atop the ergosphere.

As described earlier in Sec.~\ref{Witten2sec} the Schwarzschild
black hole, Witten bubble and S-Schwarzschild are all related by
$\gamma$-flips and may be combined into one diagram.  This diagram
contains the information for all of their individual card
diagrams. Similarly the combined multitude of S-dihole II regions
is shown collectively in Fig.~\ref{fullSdihole2}.

\subsubsection{Other ways to get the regions and analogy with
Kerr}\label{analogy-sec}

To complete our general description of the global structure of
S-dihole II, we detail other ways to find the various regions in
BL coordinates and the analogy between the (S-)dihole and (S-)Kerr
geometries. To avoid confusion, we always use the name that
applies in the Kerr geometry.  For example even though the Kerr
ergosphere is mapped by a Bonnor transformation to a singularity
in S-dihole II, we still call it the ergosphere.

A simple way to obtain region III$_+$ is to begin with the $r\geq
r_+$ portion of the dihole and analytically continue \be
\label{wb1} t\to ix^4, \hspace{.3in} \phi\to i\phi, \hspace{.3in}
a\to ia \ .\ee Region III$_-$ can similarly be obtained from the
$r\leq r_-$ portion of the dihole and continuing in the same way.
If we start with a separated pair of one negative mass black hole
and one positive mass one black hole instead of the dihole, we can
similarly analytically continue and get region III. In Weyl
coordinates this is achieved by sending $R_-\to -R_-$; in BL
coordinates $\theta\to i\theta$ and $\zeta\to i\zeta$.  We note
that in the case of the Schwarzschild black hole, the analytic
continuation in (\ref{wb1}) and \be \label{wb2} t\to
ix^4,\hspace{.3in} \theta\to \pi/2 +i\theta\ee both gave the
Witten bubble just in different parametrizations.  For the dihole,
however, (\ref{wb1}) does not give S-dihole I.  This is clear
since the analytic continuation in (\ref{wb1}) leads to
$\cal{E}_+$ universes which are singular, while S-dihole I is
non-singular. The fact that these two analytic continuations give
different spacetimes in the case of the dihole, is a result of its
reduced $SO(2)$ rotational symmetry as compared to $SO(3)$ for
Schwarzschild.

Regions III$_\pm$ are Bonnor duals to horizontal cards in new
S-Kerr universes ${\cal K}_+$ and ${\cal K}_-$ obtained by taking
$t\to ix^4$, $\phi\to i\phi$, $a\to ia$ in Kerr and looking at
$r\geq r_+$ and $r\leq r_-$ respectively.  For these cases we must
twist the $x^4$ and $\phi$ directions before compactifying
$\tilde{x}^4$.  These two spacetimes are different from the S-Kerr
universe of \cite{Wang:2004by} which in this notation can be
called ${\cal K}$.  All three ${\cal K}$ type universes have the
same card diagram structure as elliptic (two Weyl foci)
S-Schwarzschild.  In the language of cards, the two nonidentical
vertical quarter-plane cards of $\cal{K}$ can be turned on their
sides and continued to yield ${\cal K}_+$ and ${\cal K}_-$.
\footnote{Using the same card diagram techniques, the Taub-NUT
metric \cite{Stephani} also yields nonsingular analogs of ${\cal
K}$, ${\cal K}_\pm$, and the Kerr bubble \cite{joneswangfuture}.}

For Kerr geometries, switching the roles of time and space Killing
vectors means that the boundaries of the ergospheres and regions
with closed timelike curves are interchanged.  For Bonnor dual
dihole geometries, the ergosphere $\Sigma-2Mr=0$ becomes a
singularity and the CTC boundary does not map to anything
immediately relevant.

The `ring singularity' $\Sigma=0$ can persist as a singularity in
some S-Kerr and dihole geometries. It may or may not overlap with
the ergosphere singularity.  For the dihole, there is no
ergosphere for either $r\geq r_+$ or $r\leq r_-$, but there is a
ring singularity which represents the actual black hole
singularities, on the $r\leq r_-$ (both negative mass) horizontal
card.

A magnetic dihole with magnetic potential $A_\phi$ becomes
electrically charged after a double Killing rotation. Electric
flux $*dA$ must then be computed with a metric, which is not well
defined at say the ergosphere singularity.  A dual magnetic
potential $A_4$ is however well behaved at the singularity and the
flux passes through the singularity on the card as if it were not
there. For the black dihole, $A={2Ma\cos\theta
dt/(r^2-a^2\cos^2\theta)}$ does blow up at the ring singularity,
$\Sigma=r^2-a^2\cos^2\theta=0$.

\subsection{Scaling limits}

\subsubsection{The $\theta$, $\zeta$ coordinates, ${\cal I}^-$, and
the near-horizon limit }
\label{scriminussec}

It is convenient to hyperbolically parametrize the region
$t>t_+^{\rm KH}$ by $t=M+\sqrt{M^2-a^2}\cosh\zeta$.  In this case
the subextremal, $M>a$, S-dihole II solution is given by
\begin{eqnarray*}
ds^2&=&\Big({\Sigma-2Mt\over
\Sigma}\Big)^2\Big((dx^4)^2+{\Sigma^4(-d\zeta^2+d\theta^2)
\over (M^2-a^2)^3(\sinh^2\zeta-\sinh^2\theta)^3}\Big)\\
&\ &+{(M^2-a^2)\sinh^2\zeta\sinh^2\theta d\phi^2\over \big({\Sigma-2Mt\over
\Sigma}\big)^2},\\
A&=&{2aMt\sinh^2\theta\over \Sigma-2Mt}d\phi,
\end{eqnarray*}
where $\Sigma-2Mt=(M^2-a^2)\sinh^2\zeta+a^2\sinh^2\theta$
and $\Sigma=t^2+a^2\cosh^2\theta$.  The locus $t_+^{\rm DH}$ is
now simply $\theta=\zeta$ and $\zeta\geq \theta$ gives region
I$_+$.

To see that the line $\theta=\zeta$ serves as ${\cal I}^-$, the
relevant non-Killing part of the metric is
$${-d\zeta^2+d\theta^2\over (\sinh^2\zeta-\sinh^2\theta)^3}.$$
Let us change variables so $U={\zeta+\theta\over 2}$,
$V={\zeta-\theta\over 2}$ where $U\geq V>0$.  For small $V$ and
staying away from small $U$, we have $ds^2\sim -dU dV/V^3$.  Next
define $v=-1/V^2$, $u=-1/U^2$, so the metric is $ds^2\sim -du dv$
for $v\leq u<0$.  From these coordinate transformations it is
clear that region $I_+$ extends infinitely far into the negative
$v$ direction. The $uv$ chart has its own ${\cal I}^+$, $u=0$.

Next we examine the corner of the wedge where $\theta$ and $\zeta$
are small and find a near horizon dS$_2$ type scaling limit.
Define $\theta=\sqrt{\sigma}\sinh\eta$,
$\zeta=\sqrt{\sigma}\cosh\eta$ and scale the coordinates as
$\sigma\to \sigma/\lambda$, $x^4\to \lambda x^4$,
$\lambda\to\infty$.  This gives the $\sigma>0$ half of a universe
which we call ${\cal W}$
\begin{eqnarray*}
ds^2&=&((M^2-a^2)\cosh^2\eta+a^2\sinh^2\eta)^2\Big({\sigma^2(dx^4)^2\over
\Sigma_+^2}
+{\Sigma_+^2\over(M^2-a^2)^3}(-d\sigma^2/4\sigma^2+d\eta^2)\Big)\\
&\ &+{(M^2-a^2)\Sigma_+^2\cosh^2\eta\sinh^2\eta d\phi^2\over
((M^2-a^2)\cosh^2\eta+a^2\sinh^2\eta)^2},\\
A&=&{2aM t_+^{\rm KH}d\phi\over a^2+(M^2-a^2)\coth^2\eta}
\end{eqnarray*}
where the constant $\Sigma_+=(t_+^{\rm KH})^2+a^2$. This universe
is a fibering of dS$_2$ over the $\eta$ direction for $\eta\geq0$.
The magnetic field points along the $x^4$ direction and is finite
and bounded.  Although the ergosphere singularity apparently
intersects the corner of the wedge, the singularity is smoothed
out as we approach the corner by the shrinking of the space for
small $\sigma$.

This is the first appearance of dS$_2$ as a nontrivial
near-horizon limit in a non-spinning Einstein-Maxwell geometry. It
might be interesting to search for a similar scaling limit in the
extremal limit.  We stress that this is not an E-brane solution.

\subsubsection{Asymptotics: flat space}

To better understand the S-dihole I and II, it is useful to study
the spatial asymptotics.  This can be done by taking a
large-$\theta$ or large-$\zeta$ (or $r$ or $t$ in BL coordinates)
scaling limit, whichever variable represents space.  All large
time limits give flat space with no conical singularity.

For the $r\geq r_+$ universe of S-dihole I, the large spatial
limits are along the radial $r$ direction.  Scaling the metric and
coordinates $r\to\lambda r$, $x^4\to \lambda x^4$, $g_{\mu\nu}\to
g_{\mu\nu}/\lambda^2$, and the vector potential $A\to A/\lambda$,
we find that the vector potential scales to zero and we get the
vacuum solution
$$ds^2=(dx^4)^2+dr^2-r^2d\theta^2+r^2\cosh^2\theta d\phi^2.$$
The constraint $r\geq r_+$ scales to $r\geq 0$, and we have a
Rindler wedge of flat space.  Changing to Weyl coordinates, the
metric is
$$ds^2=(dx^4)^2+d\rho^2-d\tau^2+\rho^2 d\phi^2 \ .$$  Although this looks
like
flat
space there is an asymptotic conical singularity as we earlier
identified $\phi\simeq \phi+2\pi{a^4\over (M^2+a^2)^2}$ to avoid a
conical singularity near the origin.  By an asymptotic conical
singularity we mean that only asymptotically is the spacetime a
cone.  This solution is therefore interpreted as the creation of
an S0-brane with energy per unit length equal to the deficit angle
over $8\pi$, so $E/L={1\over 4}(1-(1+M^2/a^2)^{-2})$
\cite{Vilenkin}. The $r\leq r_-$ S-dihole I universe gives the
same result.

For S-dihole II superextremal, the $\theta$ coordinate is
spacelike and we scale $e^\theta\to \lambda e^\theta$, $x^4\to
\lambda x^4$, $g_{\mu\nu}\to g_{\mu\nu}/\lambda^2$, $A\to
A/\lambda$. In this limit the solution again simplifies to a
vacuum solution
\begin{equation}\label{dsiholeequation101}
ds^2=(dx^4)^2+{a^8\over (a^2-M^2)^3}(-R^2 d\zeta^2
+dR^2)+(a^2-M^2)R^2\cosh^2\zeta d\phi^2,
\end{equation}
where $t-M=\sqrt{a^2-M^2}\sinh\zeta$ and $-\infty<\zeta<\infty$
parametrizes a Rindler wedge.  Changing to dimensionless Weyl
coordinates, the metric becomes
\begin{equation}\label{sdiholeequation102}
ds^2=(dx^4)^2+{a^8\over (a^2-M^2)^3}(-d\tau^2+d\rho^2)+(a^2-M^2)\rho^2
d\phi^2.
\end{equation}
The angular $\phi$ was previously periodically identified with
$\phi\simeq\phi+2\pi$ to avoid a conical singularity at the origin
so S-dihole II has an asymptotic conical singularity.  We have
created an S0-brane with $E/L={1\over 4}(1-(1-M^2/a^2)^2)$.

Next let us examine the large-$\theta$ limit of the S-dihole II
subextremal solutions.  If we attempt to take the large-$\theta$
limit for S-dihole II subextremal in region I$_+$ of the ${\cal
U}$ universe, we move into region II$_+$.  So large spatial limits
cannot be taken in the same way here.  However if we $\gamma$-flip
region I$_+$ so it is part of the ${\cal E}_+$ universe, then a
large spatial limit gives the Milne wedge.
The large spatial limits of the horizontal cards III,
III$_\pm$ are also just Rindler wedges; there are no conical
singularities, or Rindler or Milne orbifolds.

Finally the compact wedges IV, IV$_\pm$ interpolate from one
${\cal W}$ to the same or another one, either $\Sigma_-\to
\Sigma_+$, $\Sigma_+\to \Sigma_+$, or $\Sigma_-\to\Sigma_-$. If we
regard them as part of the ${\cal U}$ universes, then IV$_\pm$ are
time-symmetric, while IV is not.  While these regions have ${\cal
W}$ as timelike scaling limits, it is not clear if they have
spacelike scaling limits.

\subsubsection{Melvin scaling and KK CTCs}

While the above two scaling limits are new, there is also a known
scaling limit of the dihole \cite{Emparan:2001gm} and S-dihole I
\cite{JonesRG} which gives the Melvin universe
\cite{Melvin,zeroduality}. The Melvin universe has an axially
symmetric magnetic field which decays to zero in the transverse
direction. The quantity $\Sigma-2Mr$, whose zero locus yields the
ergosphere singularity, is the quantity of interest yielding the
nontrivial spatial dependence; both the parameters $M$, $a$, and
$\tilde\theta=\theta-\pi/2$, $\zeta$ must be scaled such that
$\tilde\theta\sim\zeta\to 0$ and $M\zeta\sim a$.  In this limit we
have $M\gg a$.

The dihole and S-dihole I are related by analytic continuation,
and the Melvin universes which come from the dihole and S-dihole I
are actually from the same neighborhood of their complexified
4-manifolds.  Since the $r\geq r_+$ dihole can be continued to
region III$_+$, and IV$_+$ is directly adjacent near $\rho=0$,
$z=0$, we must also have a Melvin scaling limit in IV$_+$. For
$r\leq r_-$, similar remarks apply to III$_-$ and IV$_-$. As part
of the ${\cal U}_\pm$ universes, IV$_\pm$ scale to
\begin{eqnarray}
ds^2&=&\Big({a^2+\rho^2\over 4M^2}\Big)^2\Big({4M^2\over
a^2}\Big)^4\big((dx^4)^2-d\tau^2+d\rho^2\big)
+\big({4M^2\over a^2+\rho^2}\big)^2\rho^2 d\phi^2 \label{melvinlimit}\\
A&=&-a\tau dx^4/2M^2\nonumber.
\end{eqnarray}
As part of the ${\cal E}_\pm$ universes, we must turn (\ref{melvinlimit})
on its side to yield a 4-card S-Melvin scaling limit, with an ergosphere
singularity on the horizontal cards.

There is no corresponding Melvin scaling limit for regions III,
IV, or for the superextremal S-dihole II.

The ergosphere singularity of a dilatonized version of S-Melvin
was found and discussed in \cite{Cornalba:2002fi}.  Just as
dilatonized Melvin can be obtained by twisting a completely flat
KK direction with an azimuthal angle, S-Melvin can be obtained by
twisting a completely flat KK direction with a boost parameter.
The ergosphere singularity is then where the twisted KK direction
becomes null.  On one side of the ergosphere singularity (small
$\rho$ on the horizontal card), the twisted KK direction is
spacelike whereas on the other side (large $\rho$ on the
horizontal card) it is timelike yielding a KK CTC.

Actually, this is a general feature of ergosphere singularities.
One may wonder about the following connection:  The `ergosphere,'
where a timelike Killing direction of say Kerr becomes null and
switches to spacelike, maps via the Bonnor transformation to an
ergosphere singularity of say the ${\cal E}$ S-dihole II, where a
dilatonized version has a KK circle changing signature. The
precise connection is that the Bonnor transformation can be
understood from the KK perspective in reducing from five to four
dimensions. If we take a dihole-type solution (\ref{bonnortrans})
and dilatonize it
with $\alpha=\sqrt{3}$ \cite{EmparanBB} we get
\begin{eqnarray*}
ds_{\rm dil}^2&=&-f^{1/2}dt^2
+f^{-1/2}\big(e^{2\gamma}(d\rho^2+dz^2)+\rho^2 d\phi^2\big)\\
A_{\rm dil}&=&{1\over 2}Bd\phi\\
e^{2\phi}&=&f^{\sqrt{3}/2}.
\end{eqnarray*}
Lifting to 5 dimensions \cite{Dowker:1995gb}\cite{JonesRG}, we get
$$ds_{\rm 5d}^2=f(dx^5+B
d\phi)^2-dt^2+f^{-1}\big(e^{2\gamma}(d\rho^2+dz^2)+\rho^2 d\phi^2\big),$$
and the Killing $t_{\rm dihole}$ becomes completely flat.  It may be
dropped and the resulting 4d solution is a Kerr-type instanton.
Upon Wick rotating $x^5\to i t_{\rm Kerr}$ and $B\to -i\omega$,
the $x^5$ direction becomes Kerr time.  Hence $x^5$ and $t_{\rm
Kerr}$ change signature on the same complexified locus, the
`ergosphere.'

\subsection{Weyl Coordinates} \label{sdihole-weylcoors-subsec}

Here we give the Weyl description of S-dihole II and discuss
some technical details involving branches.

In Weyl coordinates for the dihole
$$\rho=\sqrt{r^2-2Mr-a^2}\sin\theta,\qquad z=(r-M)\cos\theta$$
the Wick rotation (\ref{sd2-wick}) is the same as \be\label{sd2c}
t\rightarrow i x^4, \hspace{.3in} z\rightarrow i \tau,\hspace{.3in} M\to iM.
\ee
This means that
S-dihole II can be easily obtained by sending $M\to iM$ in
S-dihole I of \cite{JonesRG}. This also makes it clear that the
foci for S-dihole II are at $z=\pm\sqrt{M^2+a^2}\to
\tau=\pm\sqrt{M^2-a^2}$ in Weyl space and therefore we get the
special null lines for $|M|>|a|$ and not for $|M|<|a|$.

Due to the square roots which appear in Weyl functions, however,
there is a branch choice to make for $R_\pm$ and this will
sometimes generate an additional choices for signs in the
solution.  Previously we have noted that $R_\pm$ will change sign
across a special null line and in fact this is precisely what will
account for the branch choices in subextremal S-dihole II.

The magnetic dihole in Weyl coordinates is
$ds^2=-fdt^2+f^{-1}(e^{2\gamma}(d\rho^2+dz^2)+\rho^2 d\phi^2)$ with
\begin{eqnarray}\label{magdiholeweyl}
f&=&\left[{(R_++R_-)^2-4M^2-{a^2\over M^2+a^2}(R_+-R_-)^2
\over (R_++R_-+2M)^2-{a^2\over M^2+a^2}(R_+-R_-)^2}\right]^2\\
e^{2\gamma}&=&\left[ {(R_++R_-)^2-4M^2-{a^2\over
M^2+a^2}(R_+-R_-)^2
\over 4 R_+ R_-}\right]^4\nonumber\\
A&=&{aM(R_++R_-+2M)(1-{(R_+-R_-)^2\over 4(M^2+a^2)}) \over {1\over
4}(R_++R_-)^2-M^2-a^2{(R_+-R_-)^2\over
4(M^2+a^2)})}d\phi \nonumber,\\
R_\pm&=&\sqrt{\rho^2+(z\pm\sqrt{M^2+a^2})^2}.\label{Rpmform}
\end{eqnarray}
Analytically continuing $t\to i x^4$, $z\to i\tau$ and $M\to i M$,
gives S-dihole II.  However, in the superextremal case, the quantity
$R_++R_-+2M$ appearing in $f$ is no longer real, since $R_\pm$ are
complex conjugates. The problem is that a
transformation equation like $2(r-M)=R_++R_-$ no longer makes
sense when doing the analytic continuations in both coordinate
systems.  The problem is remedied by replacing all instances of
$R_-$  by $-R_-$ in (\ref{magdiholeweyl}); this is a choice of
branch.  This sign change is common
in Weyl solutions; if we replace $R_-\to -R_-$ in
(\ref{magdiholeweyl}), we get the solution for the interior of the
black hole at $R_-=0$ (and also the universe where we have a
negative mass object `centered' at $R_-=0$ and an extremal black
hole at $R_+=0$).  In particular changing this sign maintains the
status of these functions as solutions to the Einstein-Maxwell
system.

To demonstrate the necessity of changing $R_-\to -R_-$ in the
superextremal case, we begin by simplifying the transformation
formulas for the dihole by introducing the coordinate $\zeta$
through the relation $r-M=\sqrt{a^2+M^2}\cosh\zeta$ so in this new
variable we have $\rho=\sqrt{a^2+M^2}\sinh\zeta\sin\theta$,
$z=\sqrt{a^2+M^2}\cosh\zeta\cos\theta$ and
$R_\pm=\sqrt{a^2+M^2}(\cosh\zeta\pm\cos\theta)$.  Now perform the
analytic continuation of (\ref{sd2-wick}).  To effect the
analytic continuation $r\to it$ we must shift $\zeta\to
\zeta+i\pi/2$, so $\cosh\zeta\to i\sinh\zeta$.  In this case we
get $t-M=\sqrt{a^2-M^2}\sinh\zeta$ and
$R_\pm=\sqrt{a^2-M^2}(i\sinh\zeta\pm\cosh\theta)$.  The sum
$R_++R_-$ is purely imaginary, as required to ensure that the
metric in (\ref{magdiholeweyl}) is real.

For the subextremal case, we do not shift $\zeta$.  The solution is
real, with any branch choice, to the left of both special null lines
on the vertical card.  With
$$\rho=\sqrt{t^2-2Mt+a^2}\sinh\theta,\qquad
\tau=(t-M)\cosh\theta$$ the $t>t^{\rm KH}_+$ region can be covered
by $t-M=\cosh\zeta\sqrt{M^2-a^2}$ as
$$\rho=\sqrt{M^2-a^2}\sinh\zeta\sinh\theta,\qquad
\tau=\sqrt{M^2-a^2}\cosh\zeta\cosh\theta.$$ Since
$\sinh\zeta\sinh\theta+1\leq \cosh\zeta\cosh\theta$, we see this
covers $\tau\geq\rho+\sqrt{M^2-a^2}$ twice.  Both $\theta=0$
($\zeta>0$) and $\zeta=0$ ($\theta>0$) lie on the $\tau\geq
\sqrt{M^2-a^2}$ semiaxis; this is where the $\phi$-circle vanishes
in a smooth or conical singularity fashion.  The envelopes of
$\theta$- or $\zeta$-orbits are the line
$\tau=\rho+\sqrt{M^2-a^2}$, which is precisely $t=t^{\rm DH}_+$
and is a coordinate singularity in BL or Weyl coordinates.  This
is a special null line separating region I$_+$ from I$_-$ and we
have previously demonstrated that it serves as ${\cal I}^-$.

Since $R_\pm=i\sqrt{M^2-a^2}(\cosh\zeta\pm\cosh\theta)$, we see
that interchanging $\theta\leftrightarrow\zeta$ sends $R_-\to
-R_-$ in (\ref{magdiholeweyl}).  If we take the convention that
$R_\pm=\pm i\tau+i\sqrt{M^2-a^2}$ at $\rho=0$, then the large-$t$
wedge involves $R_+-R_-+2iM$ and the large-$\theta$ wedge involves
$R_++R_-+2iM$.  The former branch (regions I$_\pm$) can be identified
with the superextremal S-dihole II upon making $M^2-a^2$ negative,
again explaining the superextremal solution's branch.

So far we have built the vertical card for the subextremal
S-dihole II.  To obtain the region $t^{\rm KH}_-<t<t^{\rm KH}_+$
in the $\cal{E}$ universes we analytically continue $\rho\to
i\rho'$ and use
$$\rho'=\sqrt{M^2-a^2-(t-M)^2}\sinh\theta,\qquad \tau=(t-M)\cosh\theta.$$
Changing to prolate angle
$0\leq\eta\leq\pi$,  we get
$$\rho'=\sqrt{M^2-a^2}\sin\eta\sinh\theta,\qquad
\tau=\sqrt{M^2-a^2}\cos\eta\cosh\theta.$$ This is then the usual
horizontal spacelike Weyl half-plane, except there is an
ergosphere singularity.  The ergosphere is a line segment which is
symmetric in the coordinate $\tau\to -\tau$, and touches the
points $(\rho',\tau)=(0,\pm\sqrt{M^2-a^2})$ on the horizon and
$(\rho',\tau)=((M^2-a^2)/a,0)$ in the interior of the card.  The
$\phi$ direction is timelike on the horizontal card and has
horizons along $\rho'=0$, $\tau>\sqrt{M^2-a^2}$ or
$\tau<-\sqrt{M^2-a^2}$ with Euclidean periodicity
(\ref{altperiod}), and another horizon inside the ergosphere
singularity with Euclidean periodicity $2\pi$ for
$-\sqrt{M^2-a^2}<z<\sqrt{M^2-a^2}$. For the region inside the
ergosphere we can continue to two vertical cards IV at non-primed
$\rho$, much like the Schwarzschild card diagram. However in this
case the spacetimes ends at the special null lines which serve as
${\cal I}^\pm$.

The Weyl foci are an infinite distance from the interior of the
III card from either inside or outside the ergosphere.

Finally we discuss how the three regions of each ${\cal U}$
universes are pieced together on the same Weyl vertical $\rho$,
$\tau$ card (see Figure~\ref{U-universes}).  Label the three
pieces $W_1$, $W_2$, and $W_3$ in chronological order so as time
passes we proceed from $W_1$ to $W_2$ to $W_3$.  These three
wedges are bounded by two lines $0\leq\rho\leq
|\tau\pm\sqrt{M^2-a^2}|$. The key point is that these are null
lines so when we cross from one wedge to another, we must change
branches of the Weyl functions $R_\pm$. Label the null lines so
$S_1$ refers to the special null lines from $\tau=-\sqrt{M^2-a^2}$
and $S_2$ refers to the special null lines from
$\tau=\sqrt{M^2-a^2}$. Label the two different branches for $S_1$
as $\pm$ and similarly for $S_2$.  A spacetime region is then
specified by the four possible branches relative to the two null
lines.
Region $W_1$ can then be labelled by the branches $(-,-)$, while
$W_2$ is then $(+,-)$ since we have passed $S_1$, and $W_3$ is
$(+,+)$ since we have also passed $S_2$. These three regions are
pieced together to give ${\cal U}$, with its wedges I$_-$, IV, and
I$_+$. Similarly, the ${\cal U}_+$ has three wedges labelled by
$(-,+)$, $(+,+)$, $(+,-)$, and the three wedges of ${\cal U}_-$
are $(+,-)$, $(-,-)$ and $(-,+)$.

\subsection{S-Charge}

For an S-brane solution with electric field, the magnetic S-charge
\cite{Gutperle:2002ai}-\cite{Maloney:2003ck} is defined as the
integral of $F$ over a two dimensional surface ${\cal S}$ which is
spacelike and transverse to the brane direction.  In the absence
of sources or singularities and with sufficient decay of fields at
infinity, the S-charge is conserved in the sense that it does not
depend on ${\cal S}$.

In \cite{JonesRG} the S-charge of S-dihole I for $r\geq r_+$ was
computed in Weyl coordinates over a constant-$\tau$ slice to be
$Q_s={M\over a}(M+\sqrt{M^2+a^2})$ and was shown to be conserved.
The S-charge along a constant-$t$ slice in BL coordinates can be
shown to give the same result.  The region $r\leq r_-$ S-dihole I
has a constant S-charge which has a value equal to putting $M\to
-M$ in the above formula.

The S-dihole II (\ref{sdihole}) has a vector potential
$$A={2aMt\sinh^2\theta \over t^2-2Mt+a^2\cosh^2\theta}d\phi.$$
The superextremal $|a|>|M|$ spacetime is globally simple in that
it is free of horizons, singularities and special null lines. To
compute the S-charge on a BL slice, we fix $t$ and integrate
$F_{\theta\phi}$
$$Q_s={1\over 4\pi}\int_0^{2\pi}d\phi\int_0^\infty d\theta{\partial\over
\partial\theta}
\Big({2aMt\sinh^2\theta\over
t^2-2Mt+a^2\cosh^2\theta}\Big)=Mt/a.$$

This is not conserved, and is due to the fact that $F_{t\phi}$
does not decay fast enough; as $\theta\to\infty$ the $dtd\phi$
flux integral is
$$d\Phi_{\infty}={4\pi M\over a}dt.$$
Trying to get this result directly from our knowledge of S-dihole
I, whose S-charge is $Q_s={M\over a}(M+\sqrt{M^2+a^2})={M\over
a}(M+{R_++R_-\over 2})$ evaluated at $r=r_+$.  This Wick rotates
in our prescription to $Q_s=-{M\over a}(M+{R_+-R_-\over
2i})=-Mt/a$ which is right (up to an overall sign which we do not
keep track of).

On the other hand, it is possible to fix the time coordinate
$\tau$ in Weyl coordinates for this superextremal S-dihole II and
calculate the S-charge.  In this case $A|_{\rho=0}=0$ and
$A|_{\rho\to\infty}=-2 M^2/a$ so the S-charge is $Q_s=M^2/a$. This
result is independent of $\tau$ and so the S-dihole II has a
`conserved' charged in a limited sense.

\begin{figure}[htb]
\begin{center}
\epsfxsize=3.5in\leavevmode\epsfbox{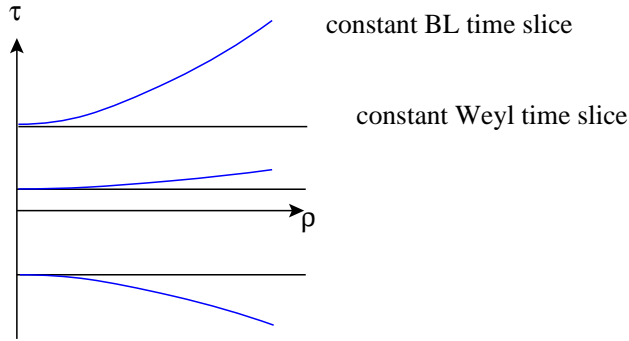}
\caption{Calculation of S-charge on Boyer-Lindquist versus Weyl
time slices can lead to different results.} \label{Scharge}
\end{center}
\end{figure}

Compared to the S-charge in BL coordinates, the S-charges are the
same for $t=M$ (or $\tau=0$) which is where the solution
experiences a `bounce.'  The difference between the BL and Weyl
S-charges can be seen from looking at the surfaces ${\cal S}$ in
Weyl coordinates:  The BL constant-$t$ slice asymptotes to a
finite, nonzero slope at large values of $\theta$ as shown in
Fig.~\ref{Scharge}.

Examining the EM field strength for S-dihole II in the ${\cal E}$
universe, we notice the following fact.  On the horizontal III
card there is an electric field in the $t$ direction and for large
values of $\theta$ $F_{\phi t}=2M/a$ is constant. One can then
interpret this as an background electric field which is related to
the two dimensional object along the ergosphere. As time passes
(we eventually go up the vertical II$_\pm$ cards) the electric
field eventually goes to zero so this gives support for the
interpretation of this S-dihole II as the creation of a localized
two dimensional unstable object. The vertical II$_\pm$ cards are
entirely accessible to timelike observers who start on the
horizontal III card, and yet they have spacelike Killing
directions which yield spacetime points not accessible to any one
observer. In contrast S-dihole I is the formation and decay of a
localized fluxbrane, which would be a one dimensional object.
The S-charge is finite and roughly corresponds to the fluxbrane
charge which is conserved. The main electro-gravitational wave
activity of S-dihole I is concentrated along the light cone.

The subextremal case is less amenable to S-charge.  The noncompact
wedge universes which are regions I$_\pm$ and II$_\pm$ have finite
and nonconstant S-charge as we compute along a constant-time (say
BL time $t$) slice out to the boundary.  However, these surfaces
${\cal S}_t$ asymptote to the conformal infinity ${\cal I}^\pm$,
not to $i^0$. Any attempt to find a surfaces to $i^0$ gives an
infinite S-charge.

On the other hand, the compact wedge universes have a clear $i^0$
on the card diagram.  S-charges are conserved and finite; one
evaluates $A_\phi$ at $i^0$ and subtracts $A_\phi$ evaluated
anywhere on the $\rho=0$ boundary.  The S-charges are
$Q_s^-={M\over a}(M-\sqrt{M^2-a^2})$, $Q_s=a$, and $Q_s^+={M\over
a}(M+\sqrt{M^2-a^2})$ for regions IV$_-$, IV, and IV$_+$.
Evidently, one can define an S-charge in the middle of the ${\cal
U}$ universes (between the horizons) but not in the future or past
regions. The results for $Q_s^+$ and $Q_s^-$ match the S-charge
for the $r\geq r_+$ and $r\leq r_-$ S-dihole I, where we naively
continue $M\to i M$.

\section{Infinite alternating array}

In the previous section we discussed some of the simplest S-brane
solutions sourced by black holes.  In this section we find black
hole and S-brane solutions that generalize the infinite
alternating array of \cite{JonesRG}.  The motivation for studying
such arrays arises from the rolling tachyon solution \cite{roll}
which was obtained by Wick rotating a periodic array of
codimension one branes and anti-branes.  These array solutions
should also exist for any codimension, however, including the
codimension three case which are Israel-Khan arrays black holes in
four dimensions.  Analytic continuation then produces S0-brane
solutions in four dimensions or this configuration may be lifted
to S6-branes in ten dimensions.

More specifically the array solutions were Israel-Khan arrays of
black holes and in Weyl coordinates they are rods of length $2M$
whose midpoints are separated by a distance $2k$. When
$\alpha=M/k<1$, the black holes rods do not overlap and upon
twisted KK reduction, give alternating-charge extremal black
holes.  The conical singularities between the rods and on the rods
can be removed by appropriately choosing constants $\gamma_{\rm
divergent}$ and $U_{\rm divergent}$, while a dimensionful constant
$R$ can be fixed to scale the black hole charges.

The solution with dilaton has a vector potential $A={\rho^2 R\over
2(\rho^2+R^2 e^{4U})}d\phi$.  When $\alpha<1/2$ the rods occupy
less than half of the $z$-axis, and we will take a symmetric setup
where no rods cross $z=0$.  In the limit $\rho\to\infty$, we have
$U\sim \alpha\log\rho+{\rm consant}$, so $A\to Rd\phi/2$ and the
S-charge (or rightwards flux through $z=0$) is $R/4$. The black
hole charge is still $R/4$ as in \cite{JonesRG}, so that means
that the rightwards flux through $z=-k$ vanishes.  We see that the
symmetrical set-up of $\alpha=1/2$ in \cite{JonesRG} where
S-charge flux alternates from pointing right to left with
magnitude $R/8$, upon slightest deformation to $\alpha<1/2$,
changes discontinuously so S-charge vanishes between close
neighbor pairs and is all lumped between far neighbor pairs.

The case $\alpha>1/2$ yields a spacetime physically equivalent to
a spacetime with parameter $1-\alpha$.  In particular the S-charge
never points the `wrong way' as it does in the $a<0$ `anomalous'
region of the dihole-in-Melvin of \cite{Emparan:1999au}. The
solutions we have are analogous to the $a=0$ case of
dihole-in-Melvin.

We can continue $z\to i\tau$ to get a generalized alternating
S-brane solution with a doubled S-charge $R/4$. Alternatively if a
rod had crossed $z=0$ we would get a solution with S-charge $0$.
We write the solution as

\begin{eqnarray}
ds^2& =& (e^{2U} + e^{-2U} \rho^2/R^2)^{1/2}((dx^4)^2 + e^{-2U}
e^{2\gamma} (d\rho^2-d\tau^2)) + \frac{\rho^2}{e^{2U}+e^{-2U} \rho^2/R^2}
d\phi^2 \nn \\
A &=& \frac{\rho^2 R}{2\rho^2+ 2R^2 e^{4U}}
d\phi
\nn \\
e^{-4\Phi/\sqrt{3}}&=& e^{2U} + e^{-2U} \rho^2/R^2 \nn \\
e^{2U}&=&e^{-2U_{\rm div}} \prod_{p=-\infty}^{\infty} \frac{r_p
+
\tilde{r}_p-2M_p}{r_p + \tilde{r}_p +2M_p} \nn \\
e^{2\gamma}&=&e^{-2\gamma_{\rm div}}
\prod_{p,q=-\infty}^{\infty} \frac{\tilde{r}_q r_p - (\tau+iz_q +
iM_q)(\tau+iz_p -iM_p) +\rho^2}{\tilde{r}_q \tilde{r}_p
-(\tau+iz_q +i M_q)(\tau+iz_p +iM_p) +\rho^2} \ \times \nn \\
&&  \frac{r_q \tilde{r}_p - (\tau+iz_q-i M_q)(\tau+iz_p+iM_p)
+\rho^2}{r_qr_p-(\tau+iz_q -i M_q)(\tau+iz_p -iM_p) +\rho^2}  \nn \\
r_p &=&\sqrt{\rho^2- (\tau+iz_p+iM_p)^2} ,\hspace{.2in}
\tilde{r}_p=\sqrt{\rho^2-(\tau+iz_p-iM_p)^2}
\end{eqnarray}
where $z_p=(2p+1)k$, $M_p=M$, tilde means to the left end of the
rod \cite{Myers:rx}, and there is no problem taking
the standard branch cut for square roots.
Using $r_p^*=-\tilde{r}_{-p-1}$, it is easy to see that $U$ is real,
where every rod matches up with its image under ${\rm Im}\,\tau\to -{\rm
Im}\,\tau$.
The reality of $\gamma$ follows from its PDE.

As described in \cite{Gaiotto:2003rm}, we can deform the black
hole locations while keeping the real-$\tau$ geometry real, as
long as they obey the symmetry of charge conjugation times ${\rm
Im}\,\tau\to -{\rm Im}\,\tau$. This means we may perform some
further operations to deform our S-brane geometry.  For example we
may change the individual masses $M_p\to \lambda_p M_p$ as long as
$\lambda_p^*=\lambda_{-p-1}$.  We may also move the positions
$z_p$ in the complex $z$-plane as long as ${\rm Re}\,z_p= -{\rm
Re}\,z_{-p-1}$. Although naively the rules prohibit black holes
from being on the real $\tau$-axis, in fact one or more rods may
cross the real $\tau$-axis.  Under these deformations, the
original $z$-geometry is not real, or if it were, it has conical
singularities; the $\tau$-geometry is however real and smooth.
Note that non-finite deformations can require an infinite shift in
$U_{\rm div}$, $\gamma_{\rm div}$.

From the S-dihole II geometry we known that we can place black
holes on the real $\tau$-axis.  This breaks the rule of
\cite{Gaiotto:2003rm} but seemingly restricts us to wedge
universes instead of a full vertical card. Take for the present
scenario a rod setup where there is a central rod $p=0$ that
crosses $z=0$. This rod has no pair, and the only option for
rotation is by $90^\circ$, $M_0\to iM_0$.  The function
$f_0={r_0+\tilde{r}_0+2iM\over r_0+\tilde{r}_0-2iM}$ is now real
but only in the Weyl regions $\rho\leq |\tau\pm M|$. This
comprises two noncompact wedge cards and one compact 45-45-90
wedge card.  Let us call the Weyl wedges $W_1$, $W_2$, $W_3$ in
chronological order.  The solution is different from S-dihole II
in Melvin in that there are the other rods at complex time $\tau$,
which we anticipate smoothly deform the solution.

As discussed in Sec.~\ref{2rodgravwavesec}, there are actually different
branch
choices for the pairs of rods at complex $\tau$.  We will not
focus on this because any good branch choice for these complex
sources will be constant over the connected spacetime.

The rule for branches for $p=0$ is as follows.  We may pick any set of
branches in wedge $W_1$. Then when we pass through the dS$_2$ horizon
to wedge $W_2$, we must change the branch for $\tilde{r}_0$.  Now
we are in wedge $W_2$. When we pass through the dS$_2$ horizon to
wedge $W_3$, we must change the branch for $r_0$.  This gives
three distinct non-singular ${\cal U}$ type universes, where the
branch choices are $(-,-)$, $(+,-)$, $(+,+)$ which is like ${\cal
U}$ (or its time-reverse); $(+,-)$, $(-,-)$, $(-,+)$ which is like
${\cal U}_-$, and $(-,+)$, $(+,+)$, $(+,-)$ which is like ${\cal
U}_+$.

One could take any of these wedges with a branch choice and turn the
wedge on its side, yielding an ${\cal E}$ universe.
Looking at the pure Einstein-Maxwell solution,
\begin{eqnarray*}
ds^2&=&(e^{2U}+e^{-2U}\rho^2/R^2)^2((dx^4)^2+e^{-8U}e^{8\gamma}(-
d\tau^2+d\rho^2)
+(e^{2U}+e^{-2U}\rho^2/R^2)^{-2}\rho^2 d\phi^2\\
A&=&{\rho^2 R\over \rho^2+R^2 e^{4U}}d\phi,
\end{eqnarray*}
we note that putting $\rho\to i\rho'$ gives us possible zeroes for
$e^{2U}-e^{-2U}\rho'^2/R^2$.  Comparing to the form of S-dihole
II, this is precisely an ergosphere singularity.  Its locus may be
hard to describe but we expect for this single rod that the locus
is qualitatively similar to that of S-dihole II.  This solution is
constituted in the same manner as the rolling tachyon\cite{roll}
in that they are Wick rotations of an infinite array of branes and
anti-branes.

One can put more than one vertical rod on the real $\tau$-axis.  To
put an even number, start with rods none of which cross $z=0$, and
move an even number to real $\tau$.  To put an odd number, start
with rods one of which crosses $z=0$.  In any case, the geometry
will be real for $\rho\leq |\tau-\tau_i|$ for all foci $\tau_i$.
If we put $n$ rods, then we have an initial noncompact wedge
followed by $2n-1$ compact wedges and then one final noncompact
wedge.  We may pick any choice for initial branches, and as we
pass through each dS$_2$ horizon at all the $\tau_i$, we change
the branch of $R_i$.

If the divergent constant renormalization procedure does not break
down, then it is possible to put an infinite number of rods along
the real $\tau$-axis.  Then we can have a universe which is
periodic in time and has an infinite number of dS$_2$ horizons. To
get periodicity in time, the only acceptable scheme of branches is
to go, across a given focus, from $(\cdots ,-,-,+,+,+,\cdots)$ to
$(\cdots,-,-,-,+,+,\cdots)$, which is the same.

\section{Card diagrams}

In this section we construct the card diagrams for a wide
assortment of solutions including black holes and S-branes.  The
card diagrams are shown to be useful in representing global
spacetime structure such as how Reissner-Nordstr\o m black holes
change as we take their chargeless and extremal limits.  For the
superextremal black holes we discuss for the first time the
technical issue of how to deal with the appearance of branch
points in Weyl coordinates.  The card diagram also clearly
represents the Kerr ring singularity and how it is possible to
traverse the ring into a second asymptotic spacetime. In addition
the five dimensional black ring solution, which is a black hole
with non-spherical horizon topology, is analyzed.  Many other
examples are presented.

\subsection{Black holes}

\subsubsection{Reissner-Nordstr\o m}
\label{RNsubsub}

\begin{figure}[tp]
\begin{minipage}{70mm}
\begin{center}
\includegraphics[width=6.5cm]{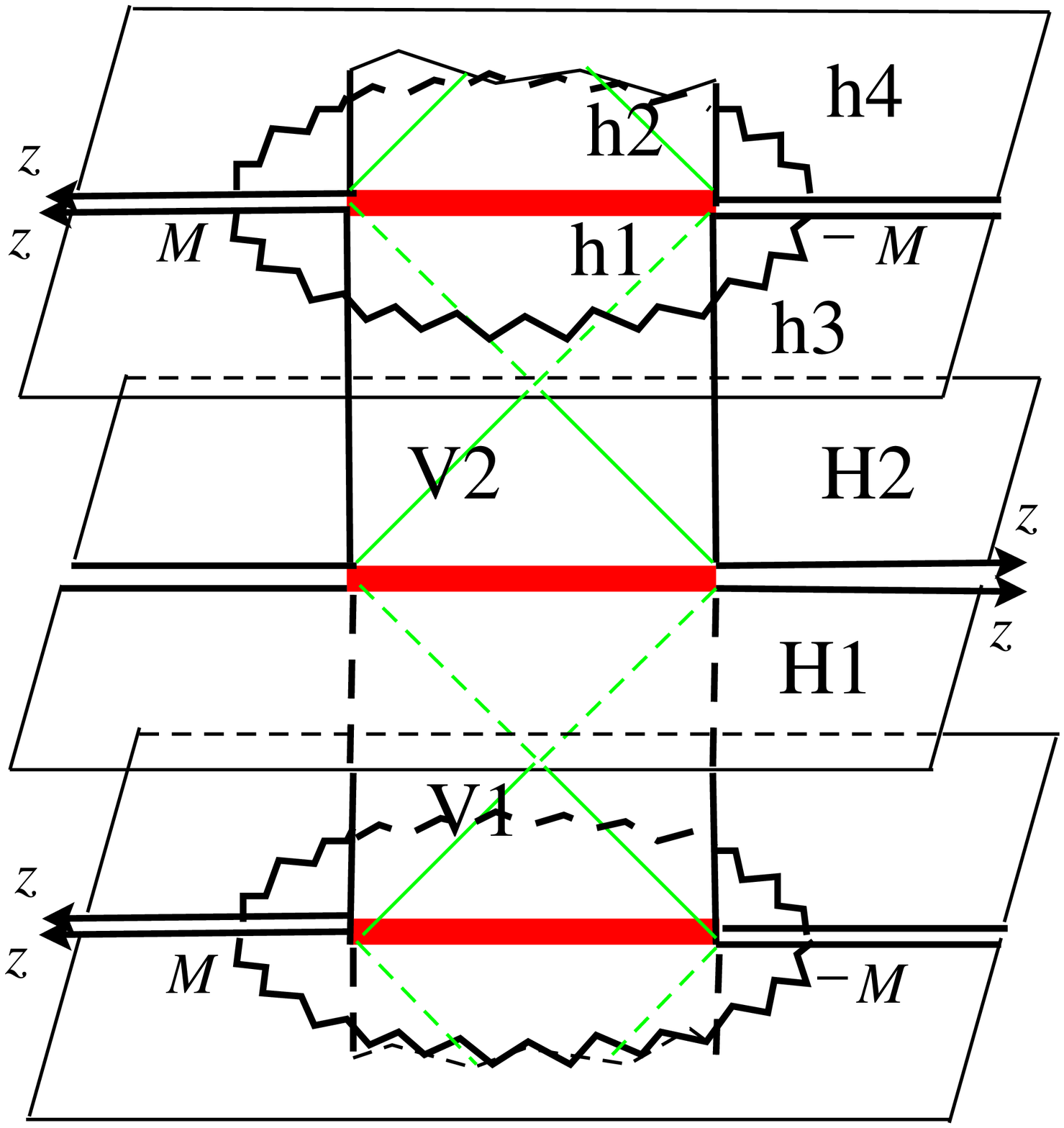}
\caption{The subextremal Reissner-Nordstr\o m card diagram.  The
ellipse singularity has semimajor axes $z=\pm M$ and $\rho=Q$, and
the rod endpoints are the foci on the $z$-axis at
$z=\pm\sqrt{M^2-Q^2}$.} \label{RN-card}
\end{center}
\end{minipage}
\hspace*{15mm}
\begin{minipage}{70mm}
\begin{center}
\includegraphics[width=6cm]{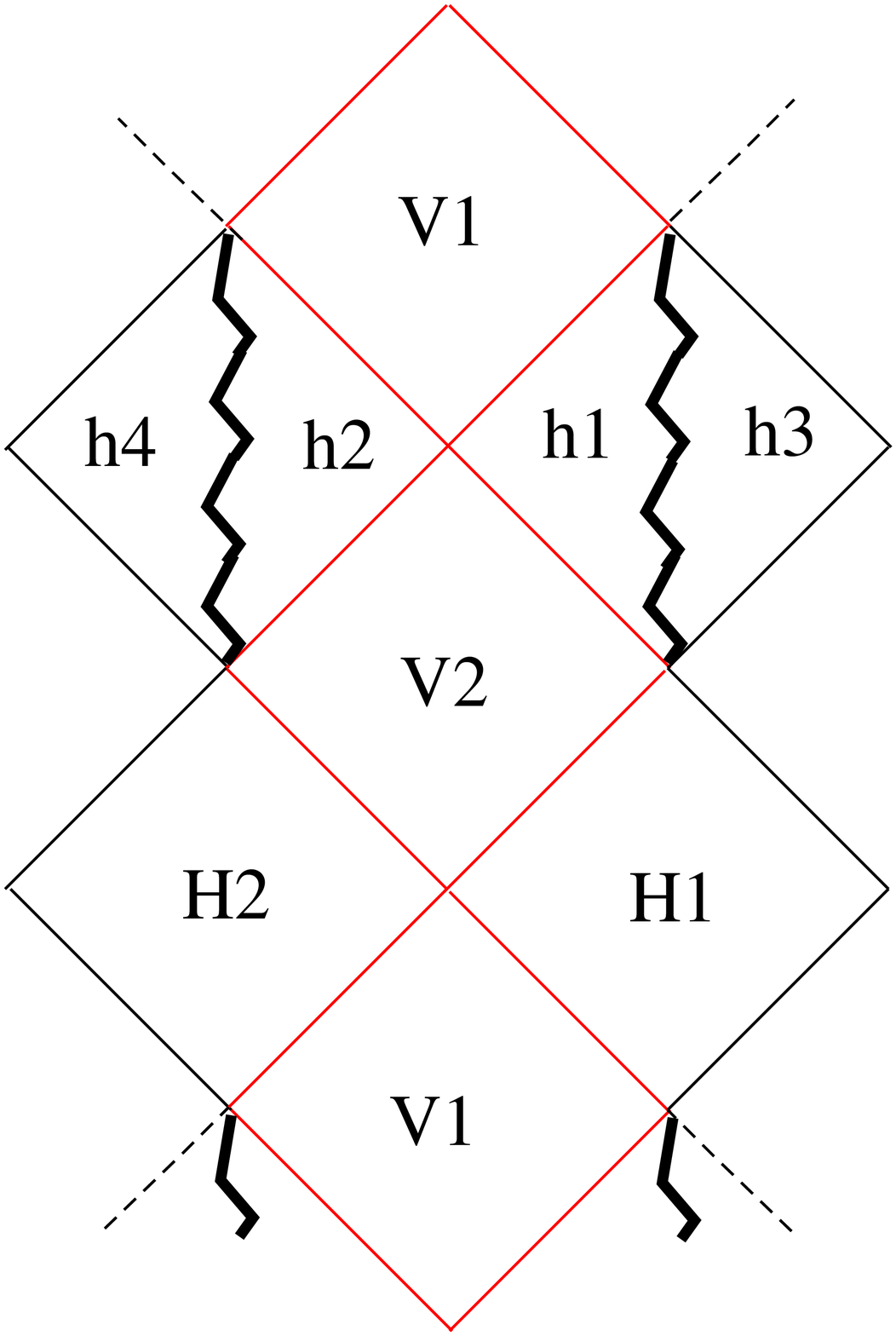}
\caption{The extended Reissner-Nordstr\o m Penrose diagram, including
negative mass universes.}
\label{RN-Penrose}
\end{center}
\end{minipage}
\hspace*{5mm}
\end{figure}

In Weyl coordinates the Reissner-Nordstr\o m black hole is
\begin{eqnarray} \rho&=&\sqrt{r^2-2Mr+Q^2}\sin\theta, \hspace{.3in} z=(r-M)
\cos\theta \label{RNsoln} \\
\nonumber
R_{\pm}&=& \sqrt{\rho^2 + (z\pm \sqrt{M^2-Q^2})^2}= r-M \pm
\sqrt{M^2-Q^2} \cos\theta \\
\nonumber
f &=& \frac{(R_+ + R_-)^2 - 4(M^2-Q^2)}{(R_+ + R_- +2M)^2} \\
\nonumber
e^{2\gamma}&=&\frac{(R_+ + R_-)^2 - 4(M^2-Q^2)}{4 R_+ R_-} \\
\nonumber
A&=& -\frac{2Qdt}{R_+ + R_- +2M} \end{eqnarray} and the card
diagram for $|Q|<|M|$ is shown in Fig.~\ref{RN-card}.  The
construction of the card diagram proceeds along similar lines to
the Schwarzschild card diagram.  There are two adjacent
horizontal half-planes, H1 and H2, which represent the positive
mass asymptotically flat regions. The outer horizon is a rod which
lies on the $z$-axis for $-\sqrt{M^2-Q^2}<z<\sqrt{M^2-Q^2}$.  The
vertical cards, V1 and V2, are squares of length $2\sqrt{M^2-Q^2}$
and the diagonal lines connecting opposite corners of the square
are special null lines. In the Schwarzschild case, the black hole
singularity is at the top of the vertical card V1 and the bottom
of the card V2.  When the black hole is charged, however, these
vertical cards end in horizons with four card junctions.  (In
going past the two special null lines, the orientation of the
$z$-axis has flipped direction.)

The vertical card ends at the inner horizon and connects to
another vertical card as well as two horizontal half plane cards
each containing a region h1 and h2 where $0<r<r_-$, and also a
negative-mass universe. The black hole singularity lies on the
boundary of the h1 and h2 regions and takes the form of a
semi-ellipse $\rho^2/Q^2+z^2/M^2=1$.   The black hole singularity
is typically where we stop the space. For completeness though we
extend past the singularity on the card diagram.
For the case of charged black holes, negative values of $r$
lead to a nakedly singular spacetime.  In fact this
singular spacetime is exactly what fills out the rest of the
horizontal card with the regions h3 and h4.
The card diagram correctly captures the
fact that negative mass universes do not have horizons and that
their singularity is timelike.

Finally at each horizon, the card diagram should be continued
vertically through horizons to obtain a tower of cards extending
infinitely up and down in the vertical direction.

Let us compare this Weyl card diagram to the Reissner-Nordstr\o m
Penrose diagram.   The Penrose diagram for all black holes with
$M>|Q|$ are the same (see Figure~\ref{RN-Penrose}), so the causal
structure is the same for all such black holes.  When $Q=0$
however spacetime is described by the Schwarzschild Penrose
diagram of Fig.~\ref{Penrose-Weyl-comparison} which is very
different. At once this tells us both that there is a
discontinuous jump in the causal structure, but also that this
discontinuity is not easy to understand from the Penrose diagram.

On the other hand the limit $Q\to 0$ is easy to understand by
examining the Weyl card diagram in Figure~\ref{RN-card}. In this
case the vertical cards expands to a $2M\times 2M$ square and the
singularity, which formed a semi-ellipse, degenerates to a line
segment covering the inner horizon.  This singularity now sits on
and seals the horizons and the multiple asymptotic regions from
each other.  So while in the presence of charge the singularity
used to be timelike and avoidable, it now becomes spacelike as
seen from the vertical card $0<r<2M$.  Therefore a nice feature of
the card diagram is that it shows how regions h1, h2 collapse and
the singularity smoothly falls onto the vertical cards V1 and V2. An
explorer on the vertical card who could continue up forever in the
case of a charged black hole, is now doomed at the singularity.

\subsubsection{Extremal Reissner-Nordstr\o m}
\begin{figure}[htb]
\begin{center}
\epsfxsize=3.5in\leavevmode\epsfbox{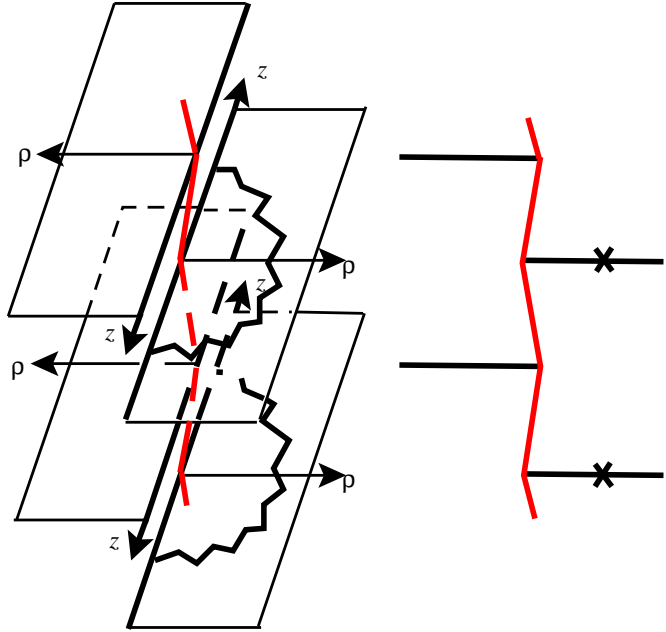}
\caption{Extremal RN horizontal cards are connected in an AdS$_2$
fashion at their origins.  A side-view $z=0$ cross section is also
shown.} \label{RNextremalfig}
\end{center}
\end{figure}

Starting from the above card diagram we now examine the extremal
limit $Q\to \pm M$. In this case the vertical cards which
represent the regions between the two horizons get smaller and
disappear.  When $Q=M$, the horizontal cards are now only attached
at point-like extremal-horizons and only half of the horizontal
cards remain connected, see Fig.~\ref{RNextremalfig}. The region
near the point-horizons are anti-de Sitter throats although cards
themselves cannot adequately depict the throat region. The
throat is a `connected' sequence of points on vertically adjacent
horizontal cards.

To understand how to go beyond the horizon, it is important to
remember that there were special null lines inside the black hole.
In the extremal case the foci for the null lines become
degenerate.  In passing a special null line, we use the opposite
branch of the square root for the Minkowski distance to the focus.
For extremal black holes, both foci degenerate, and so we change
the sign of all occurrences of the square root when we pass
through the throat. This is a result of the simple fact that for
extremal Reissner-Nordstr\o m the functions $R=R_+=R_-=r-M$ both
become negative inside the horizon. Hence we
change the sign of $R$ to continue past the point
$\rho=z=0$ onto either of the two adjacent horizontal cards, which
have a singularity in the form of a half circle.

For Majumdar-Papapetrou solutions, this `sign change rule' agrees
with that in \cite{Myers:rx}.  For an axisymmetric array, one can
then easily draw the interior regions on horizontal cards, and
find negative-mass-object universes on their complements.  The
work of \cite{Myers:rx} of course holds for an arbitrary array.

\subsubsection{Superextremal black holes}

The superextremal $|Q|>|M|$ Reissner-Nordstr\o m black hole does
not have horizons or vertical cards.  Its card diagram consists of
two horizontal cards, connected along the segment
$0\leq\rho\leq\sqrt{Q^2-M^2}$, $z=0$.  One card has a semi-ellipse
singularity passing through the points $(\rho=0,z=\pm M)$ and
$(\rho=Q, z=0)$.

These two horizontal cards are connected in the same sense as a
branched Riemann sheet. By choosing Weyl's canonical coordinates
(meaning $Z=\rho+iz$ with
$-($Coef$\,dt^2)($Coef$\,d\phi^2)=($Re$\,Z)^2$),  the solution is
no longer accurately represented on the horizontal card.  This can
be seen by examining the coordinate transformation (\ref{RNsoln})
from Schwarzschild coordinates to Weyl coordinates.  The
coordinate transformation from $(r,\theta)$ double covers the Weyl
half plane $(\rho\geq0,z)$.  For fixed radius and varying
$\theta$, the coordinates from $M<r<\infty$ cover the Weyl plane
in semi-ellipses which degenerate to the segment $(0\leq\rho\leq
\sqrt{Q^2-M^2}, z=0)$, which serves as a branch cut. The range
$-\infty<r<M$ again covers the half-plane with $r=0$ forming an
ellipse singularity. Therefore the point $(\rho=\sqrt{Q^2-M^2},
z=0)$ acts as a branch point with a branch cut segment running to
the $z$-axis.  In summary while for the subextremal black hole the
cards are joined by a four card horizon along a line segment of
length $2\sqrt{M^2-Q^2}$, superextremal black holes involve a
horizontal card branch cut along a line segment of length equal to
the `rod length' $2\sqrt{Q^2-M^2}$.

\begin{figure}[htb]
\begin{center}
\epsfxsize=2.5in\leavevmode\epsfbox{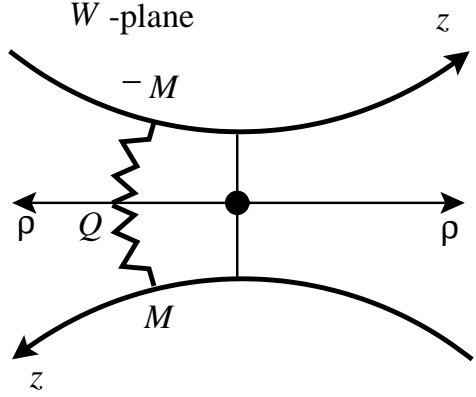} \caption{The
superextremal RN black hole, after conformal transformation
to the $W$-plane.  This is a branched, static horizontal card,
with two boundaries and one branch point.} \label{superRNfig}
\end{center}
\end{figure}

This double cover of the Weyl plane can be mapped onto a singly
covered space by taking the appropriate square root.   By choosing
a new coordinate $W=\sqrt{Z-\sqrt{Q^2-M^2}}$, we map both the
positive and negative-mass universes into the region $({\rm
Im}\,W)^2-({\rm Re}\,W)^2\leq\sqrt{Q^2-M^2}$ (see Fig.
\ref{superRNfig}). The image of the $z$-axis boundary is a
hyperbola where the $\phi$ circle vanishes.  The origin $W=0$ is
the image of the branch point and the image of the line segment
$(0\leq\rho\leq \sqrt{Q^2+M^2}, z=0)$ is a line connecting the two
hyperbolas and intersecting the origin of the $W$-plane. In the
$W$-plane the $z$-axis and $\rho$-axis are no longer orthogonal.
Larger values of $\rho$ past the branch point are mapped to the
real $W$-axis. The singular nature of $e^{2\gamma}\propto 1/R_+
R_-\propto1/|\Delta Z|\propto 1/|W|^2$ has been fixed by
$e^{2\gamma}dZ d\overline{Z}=4|W|^2 e^{2\gamma}dW d\overline{W}$.
Finally the black hole singularity is mapped to a curved segment
stretching from one hyperbola line to the other, to the left of
the branch cut.

Alternatively one can use a Schwarz-Christoffel transformation to
map the two universes onto the strip $|{\rm Im}\,W|\leq W_0$. This
is useful when the horizontal card boundaries are horizons as it
allows for multiple horizontal cards to be placed adjacent to each
other at the horizons (Fig.~\ref{SRN2fig}).   This technique of
fixing a horizontal card with a branch point will also be used in
more complicated geometries such as the hyperbolic representation
of S-RN and multi-rod solutions in four and five dimensions.

Card diagrams distinguish between spacetimes which are `truly'
superextremal, $|Q|>|M|$, from those simply with $M<|Q|$. The
superextremal card diagram of this section for example is
different from a negative mass Schwarzschild card diagram which is
only region h3, for example,  of the subextremal card diagram of
Fig.~\ref{RN-card}.

\subsubsection{Kerr} \label{Kerrsubsec}

Written in Weyl-Papapetrou form, the Kerr black hole is
\begin{eqnarray*}
ds^2&=&-f(dt-\omega
d\phi)^2+f^{-1}(e^{2\gamma}(d\rho^2+dz^2)+\rho^2d\phi^2),\\
f&=&\frac{(R_++R_-)^2-4M^2+{a^2\over M^2-a^2}(R_+-R_-)^2}
{(R_++R_-+2M)^2+{a^2\over M^2-a^2}(R_+-R_-)^2},\\
e^{2\gamma}&=&\frac{(R_++R_-)^2-4M^2+{a^2\over
M^2-a^2}(R_+-R_-)^2}
{4R_+R_-},\\
\omega&=&\frac{2aM(M+{R_++R_-\over 2})(1-{(R_+-R_-)^2\over
4(M^2-a^2)})} {{1\over 4}(R_++R_-)^2-M^2+a^2{(R_+-R_-)^2\over
4(M^2-a^2)})},
\end{eqnarray*}
where $R_\pm=\sqrt{\rho^2+(z\pm \sqrt{M^2-a^2})^2}=r-M \pm
\sqrt{M^2-a^2} \cos\theta$. The transformation to Boyer-Lindquist
coordinates is $\rho=\sqrt{r^2-2Mr+a^2}\sin\theta$,
$z=(r-M)\cos\theta$.

\begin{figure}[htb]
\begin{center}
\epsfxsize=3.5in\leavevmode\epsfbox{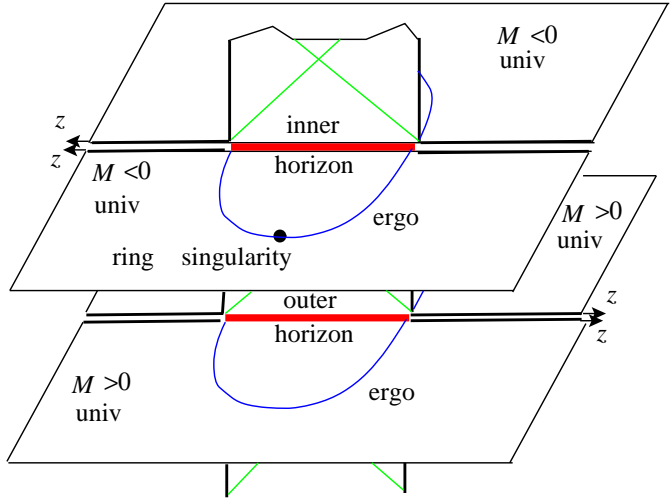} \caption{The
Kerr card diagram.} \label{Kerrcard}
\end{center}
\end{figure}

For $|a|<|M|$ the Kerr black hole card diagram (see
Figure~\ref{Kerrcard}) is similar to Reissner-Nordstr\o m except
that the singularity is a point and lies at $\rho=a$, $z=0$ on
each negative-mass card. The outer and inner ergospheres lie on
the positive-and negative-mass cards and are both described by the
curve $z^2=\alpha^2-(\alpha^2/a^2-1)\rho^2-\rho^4/a^2$ where
$\alpha^2=M^2-a^2$.  The boundary of the region with closed
timelike curves is also described by a quartic polynomial in Weyl
coordinates.  On the vertical card, which has length
$2\sqrt{M^2-a^2}$, there are two special null lines.

The $r=0$ surface in BL coordinates is a semi-ellipse
$\rho^2/a^2+z^2/M^2=1$ on the negative-mass card; but it is not
a distinguished locus on the card diagram.  Attempting to
make one loop around the ring in BL coordinates clearly does not
make a loop in Weyl space.  When a traveller passes `through' the
ring at $r=0$, he merely traverses this non-singular semi-ellipse.
For example if we start at the point $(r=0,\theta)$ and then
attempt to loop around the ring in BL coordinates we arrive at
$(r=0,\pi-\theta)$. In Weyl space this trajectory only connects
two points on the non-singular semi-ellipse by a segment and
so does not enclose the singularity.  The question as to what
happens when we go through the $r=0$ region can be answered simply
in Weyl coordinates.  At minimum it is clear that we must fill out
the rest of the horizontal plane.  In our analysis of the
Reissner-Nordstr\o m black hole we extended past $r=0$ by simply
going to negative values of $r$ which filled out the rest of
the horizontal half plane.  The key difference is that for Kerr
the singularity is point-like and so it is clear that it is
physically possible for an observer to enter this second
asymptotically flat universe.  Finally the question as to whether
we identify universes gotten to by going through the disk from
above or below, now becomes the question of do we identify parts
of the negative-mass card, as we circle the ring singularity
point. A card diagram for a charged Kerr-Newman solution can
similarly be constructed.

In examining the Reissner-Nordstr\o m solution it was found that
the Weyl card diagram smoothly interpolates between the different
sub/super-extremal parameter ranges.  We now check the card
diagram for Kerr for various parameter values.  For fixed mass, if
the angular momentum of the black hole is increased, the horizon
shrinks to a point at the extremal limit and disappears when we go
above the extremal limit.  Extending past the pointlike horizon in
the extremal case is again accomplished by changing the sign of
both $R_\pm=R$.  At the same time the ring singularity at $\rho=a$
moves farther from the origin.  Therefore when the horizon
disappears, the singularity is separated by a finite distance in
Weyl space.  The superextremal Kerr solution is similar to the
superextremal Reissner-Nordstrom black hole except that the
ellipse singularity is replaced by a point, and the `ergospheres'
map to an $\infty$-looking locus centered at $W=0$.
Decreasing the angular momentum to zero, the singularity moves
towards the horizon which increases in length. However this limit
is not smooth since when the angular momentum reaches exactly
zero, the curvature singularity jumps from being a point to a
segment on the card diagram, to fully cover the horizon.

The superextremal Kerr solution is similar to the superextremal
RN, except that the ellipse singularity is replaced by a
point ($z=0$), and the ergospheres map to an $\infty$-looking
locus centered at $W=0$.

\subsubsection{The Black Ring}

The 5d black ring solution of \cite{EmparanWK} is
\begin{eqnarray}\label{blackringsol}
&ds^2&=-{F(x)\over F(y)}dt^2\\
&+&{1\over A^2(x-y)^2}\left[F(x)\Big((y^2-1)d\psi^2 +{F(y)\over
y^2-1}dy^2\Big)+F(y)^2\Big({dx^2\over 1-x^2}+{1-x^2\over
F(x)}d\phi^2\Big)\right]\nonumber
\end{eqnarray}
where $F(x)=1-\mu x$, $F(y)=1-\mu y$, and $0\leq\mu\leq 1$. The
coordinates $x$, $y$ are 4-focus coordinates (in the sense that
polar coordinates have two foci and prolate coordinates have three;
we are counting the focus at infinity) that parametrize
a half-plane of Weyl space $\rho\geq0$, $-\infty<z<\infty$:
\begin{eqnarray*}
\rho&=&{1\over A(x-y)^2}\sqrt{F(x)F(y)(1-x^2)(1-y^2)}\\
z&=&{(1-xy)(F(x)-F(y))\over 2A(x-y)^2} \ .
\end{eqnarray*}
The foci are on the $z$-axis at $z=\pm\mu/2A$ and $z=1/2A$. The
black ring horizon is also on the $z$-axis along $-\mu/2A\leq
z\leq \mu/2A$. The $\phi$-circle vanishes along $z\leq-\mu/2A$ and
$\mu/2A\leq z\leq1/2A$, and the $\psi$-circle vanishes along
$z\geq1/2A$. Curves of constant $y$ value degenerate to the
horizon line segment as $y\to -\infty$, and degenerate to the
$(1/2A,\infty)$ line segment (better pictured with a conformally
equivalent disk) for $y\to -1$.  Curves of $x=$constant degenerate
to the vanishing $\phi$-circle line segment for $x\to 1$ and to
the left semi-line $-\infty<z<-\mu/2A$ for $x\to -1$.

\begin{figure}[htb]
\begin{center}
\epsfxsize=5in\leavevmode\epsfbox{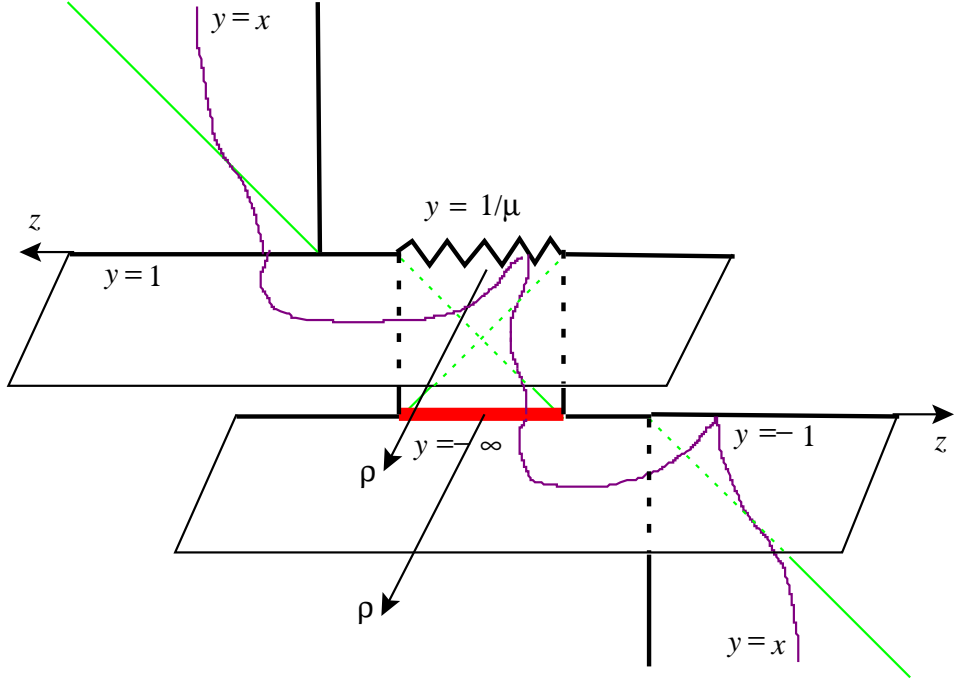} \caption{An
extended card diagram for the black ring, where we have continued
past $y=$constant boundaries for $-1\leq x\leq 1$.  A $y$-orbit is
drawn curving through several cards.  Only one of the vertical and
horizontal cards are drawn at each four card V-H-H-V card junction
to avoid too many overlapping figures.} \label{blackringfig}
\end{center}
\end{figure}

The card diagram is easy to construct and is not much different
from the four dimensional Schwarzschild case.  Past $y=-\infty$ we
can go to $y=+\infty$ and hence imaginary $\rho=i\rho'$, and move
up a $\mu/A\times\mu/A$ square with two special null lines.  At
the top of the square, at $y=1/\mu$ we have the curvature
singularity. Continuing again to real $\rho$ and running $y$ down
to $1$, we map out a (negative-mass) horizontal card.  The locus
$y=1$ is the semi-line $z>1/2A$.  The space closes off here as the
$\psi$-circle vanishes, but one may formally continue to make an
`extended card diagram' which is useful in several applications.
Past $y=1$, we see that for fixed $x$, reducing $y$ down to $x$
makes a topological half-line in a vertical card with a special
null line like the elliptic card representations of
S-Schwarzschild. Then for $-1<y<x$, we traverse another vertical
card, which we could attach to our original positive-mass
horizontal card along $z>1/2A$ (see Fig. \ref{blackringfig}). Thus
when $y$ runs down from $x$ on the extended real line, it
traverses a topological line in a V(ertical)-H(orizontal)-V-H-V
sequence of cards.  These noncompact vertical cards would be
physically relevant if $\psi$ were continued to be time.

The above assumed $-1\leq x\leq 1$.  One can instead fix
$-\infty<y<-1$ and continue $x$ past $1$ to find a square card above the
line segment $\mu/2A<z<1/2A$. The top of this square is $x=1/\mu$
and if $\phi$ were time, this would be the curvature singularity
of a black hole (in an expanding KK bubble). For $1/\mu<x<\infty$
we sweep out a negative mass card, for $-\infty<x<y$ we sweep up a
noncompact vertical card, and for $y<x<-1$ we sweep down a
noncompact vertical card to attach to our original card at
$z<-\mu/2A$.

Note that when passing through the black ring horizon at
$y=-\infty$, the Weyl conformal factor \cite{EmparanWK}
$$e^{2\nu}={1+\mu\over 4A}{Y_{23}\over R_1 R_2 R_3}\sqrt{Y_{12}\over
Y_{13}}\sqrt{R_2-\zeta_2\over R_3-\zeta_3},$$ stays real;
$R_3-\zeta_3$ and $Y_{13}$ go negative.  As we pass the special
null lines, explicit appearances of $R_1$ and $R_2$ in the Weyl
functions $e^{2U_i}$, $e^{2\nu}$ change sign.

\subsection{Spacelike Branes}

\subsubsection{S-Reissner Nordstrom: elliptic diagram}

\begin{figure}[htb]
\begin{center}
\epsfxsize=4in\leavevmode\epsfbox{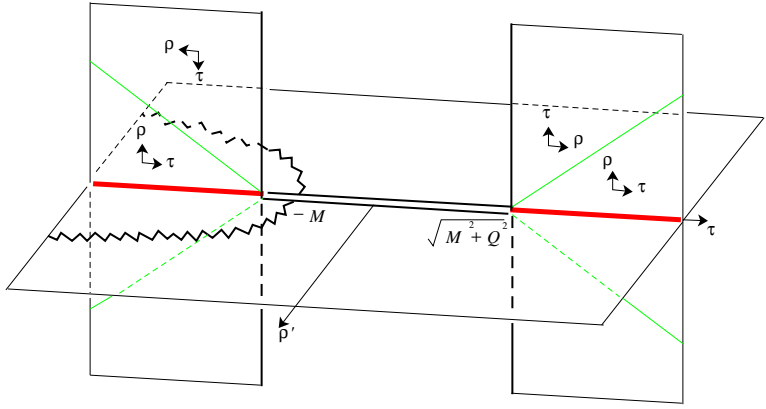}
\caption{The elliptic S-RN card diagram.  The singularity is a
hyperbola which does not intersect a horizon.}
\label{chargedSbranecard}
\end{center}
\end{figure}

Adding charge to the Schwarzschild solution moves the singularity
from a four card junction and leaves a horizon.  Adding charge has
a similar effect on the Schwarzschild S-brane.  This has already
been described previously in Sec.~\ref{thirdcard-Poincare} in the
case of the parabolic card representation.  Turning to the S-RN
elliptic card diagram, as shown in Figure~\ref{chargedSbranecard},
the S-RN elliptic card diagram is similar to the elliptic
Schwarzschild S-brane except the singularity is now a hyperbola
$(\rho',\tau)=(|Q|\sinh\theta,-M\cosh\theta)$, and the negative-mass
connected universe has a 4-card horizon qualitatively similar to
the positive-mass connected universe.  As we increase the charge
the singularity moves further out from the $\tau$-axis but any
$Q\neq 0$ gives the same qualitative diagram.

\subsubsection{S-Reissner-Nordstr\o m: hyperbolic card diagram and
branch points} \label{SRN2sec}

The Witten bubble in four dimensions has three card diagram
representations as a result of its large $SO(2,1)$ symmetry group.
There are likewise three card diagrams for S-Reissner-Nordstr\o m
which is equivalent to the statement that there are three ways to
Killing-parametrize the hyperboloid $Z^2-X^2-Y^2=1$ embedded in
${\bf R}^{2,1}$ where $dZ$ is timelike.  The more well known ways
to parametrize the hyperboloid are elliptic, where we group terms
as $Z^2-(X^2+Y^2)=1$ and this leads to two Weyl foci being real,
and hyperbolic where $(Z^2-Y^2)-X^2=1$ and the Weyl foci are both
imaginary.

First let us recap in Boyer-Lindquist coordinates the
Reissner-Nordstr\o m solution is commonly written as
\begin{eqnarray*}
ds^2&=&-(1-2M/r+Q^2/r^2)dt^2+(1-2M/r+Q^2/r^2)^{-1}dr^2+r^2(d\theta^2+
\sin^2\theta d\phi^2),\\
A&=&{Qdt\over r} \ .
\end{eqnarray*}
If we send $t\to i x^4$, $M\to i M$, $r\to i t$, $\theta\to
\pi/2+i\theta$, and $\phi\to i\phi$ then we get the solution
\begin{eqnarray*}
ds^2&=&(1-2M/t-Q^2/t^2)(dx^4)^2-(1-2M/t-Q^2/t^2)^{-1}dt^2+t^2(d\theta^
2
+\cosh^2\theta d\phi^2),\\
A&=&{Qdx^4\over t}.
\end{eqnarray*}
Here we leave $\phi$ noncompact so the ranges on the coordinates
are $-\infty<\tilde\theta<\infty$, $-\infty<\phi<\infty$ and take
the following embedding
$$X=\sinh\theta,\qquad Y=\cosh\theta\sinh\phi,
\qquad Z=\cosh\theta \cosh\phi \ .$$ We see that the metric
$d\theta^2+\cosh^2\theta d\phi^2$ has a 1-1 map onto the
hyperboloid $Z^2-X^2-Y^2=1$ which is ${\bf H}_2$.  Therefore this
Wick rotation is the entire S-RN solution but just in a different
coordinate system.  In this solution it possible to compactify the
$\phi$ direction to produce  an orbifold of ${\bf H}_2$ and
therefore the S-RN solutions as well.  Euclideanized AdS$_2$ is
just ${\bf H}_2$, so this orbifold puts it at a finite
temperature.  In these coordinates, there is no longer a Killing
circle vanishing at $X=Y=0$.  This means that there is a card
diagram for S-RN that has no such boundaries.

Now let us turn to the construction of the hyperbolic card diagram
representation for S-RN which has a branch point on the horizontal
card.  In Weyl coordinates the metric and gauge field for
subextremal Reissner-Nordstr\o m are given in (\ref{RNsoln}).
To obtain the hyperbolic S-RN analytically continue $M\to iM$,
$t\to ix^4$ and $\phi\to i\phi$ and change branches $R_-\to -R_-$.
In this case the special null lines, $R_\pm=\sqrt{\rho^2+(z\pm
i\sqrt{M^2+Q^2})^2}=0$, will intersect the real manifold at only
one point and so their zero will correspond to a branch point.
{}From the definition of the functions $R_\pm$ these S-branes can be
thought of as being sourced by rods which have been turned on
their side in the imaginary $z$-direction.  In this analytic
continuation it is also necessary to replace all instances of
$R_-$ by $-R_-$ in the functions $f$, $e^{2\gamma}$ and $A$ for
reality as in the superextremal S-dihole II solution of Subsection
\ref{supersdiholeii}.  We now have
$R_-=\overline{R_+}\equiv\overline{R}$; issues of branches are
discussed shortly. The metric can be written
\begin{eqnarray}\label{srn2}
ds^2&=&-f(dx^4)^2+f^{-1}(e^{2\gamma}(d\rho^2+dz^2)+\rho^2d\phi^2),\\
f&=&{M^2+Q^2-({\rm Im}\,R)^2\over ({\rm Im}\,R+M)^2},\nonumber\\
e^{2\gamma}&=&{M^2+Q^2-({\rm Im}\,R)^2\over |R|^2},\nonumber\\
A&=&{Q\,dx^4\over {\rm Im}\,R+M}.\nonumber
\end{eqnarray}
In this case, on the horizontal card, $x^4$ is a timelike
coordinate and $\phi$ is spacelike.

To understand how the branch point arises, let us examine the
coordinate transformation from Schwarzschild to Weyl for the
range $M\leq t\leq M+\sqrt{M^2+Q^2}$.  Although we have Wick
rotated the Schwarzschild coordinates to obtain the S-RN solution,
the Weyl coordinates $\rho$ and $z$ are
not Wick rotated.  At a fixed value of $\theta>0$, the $t$-orbits
are ellipses and there is a degenerate limit as $\theta\to 0$.
In this limit the ellipse becomes the segment $z=0$,
$0\leq\rho\leq\sqrt{Q^2+M^2}$ traced back and forth.  We take
this segment as the branch cut.  Ellipses for $\theta<0$
fill out another copy of the horizontal half-plane.

The locus $t=M+\sqrt{M^2+Q^2}$ maps to the entire $z$-axis
where there is a horizon which we call $\cal{H}_+$.  Here it is
possible to analytically continue $\rho \to i\rho'$ and we find
the usual four-card junction.  The vertical cards are in the
`positive mass' S-RN universe.  On the vertical cards, $f$ and
$e^{2\gamma}$ are strictly negative and well-behaved; the
coordinate $\rho'$ is timelike while $x^4$, $z$ and $\phi$ are
spacelike.  These `positive-mass' vertical cards have no
singularities, no special null lines and no boundaries so the
$\phi$-direction does not vanish and we may leave $\phi$
noncompact.

Using the same techniques as in the superextremal RN or Kerr cases
it is possible to obtain a singly covered card by a conformal
transformation.  Henceforth the card diagram for S-RN will be
drawn in the conformally-fixed coordinate $W$ but we will continue
to label points using the $(\rho,z)$ coordinates.
In the $W$-plane the branch cut maps to a line segment connecting
the two horizons ${\cal H_\pm}$ and passing through the branch point.
The $z$ and large-$\rho$ axes no longer appear orthogonal.
A curvature singularity
divides the horizontal card at $\rho^2=Q^2+{Q^2z^2\over M^2}$ or
${\rm Im}\,R+M=0$.  Extending past the singularity we reach a
second horizon ${\cal H}_-$ (at $z<0$ with the standard branch
prescription).  The vertical cards here are in the `negative-mass'
universe.  These are qualitatively similar to the positive mass
ones and have no singularities, no boundaries, and no special null
lines.  The difference between going towards ${\cal H}_+$ and
${\cal H}_-$ is we must replace $R\to -R$ in (\ref{srn2}) which
affects the denominators of $f$ and $A$, and is equivalent to
$M\to -M$ and $A\to -A$.

It is easy to take the chargeless limit
$Q\to 0$ and we get the S-Schwarzschild solution where the
curvature singularity coincides with the horizon ${\cal H}_-$.

\begin{figure}[htb]
\begin{center}
\epsfxsize=3in\leavevmode\epsfbox{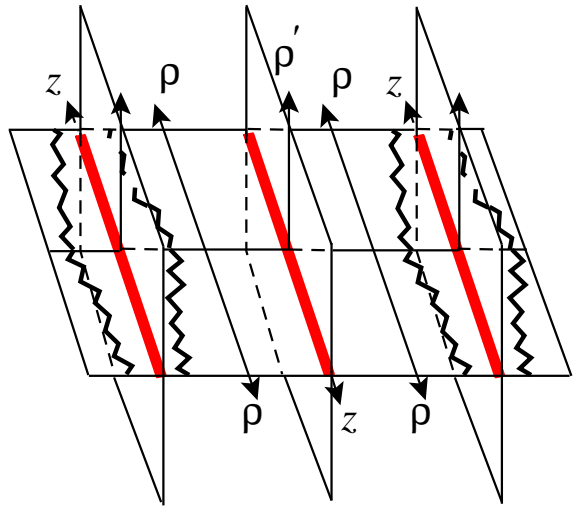} \caption{Hyperbolic
S-RN card diagram after transformation to the $W$-plane.  There
are no boundaries where spacelike Killing circles vanish.  One may
identify every other horizontal card.  The singularities are
closer to the ${\cal H}_-$ horizon and are farther from the ${\cal
H}_+$ horizon.}\label{SRN2fig}
\end{center}
\end{figure}

The full S-RN card diagram is drawn in Fig.~\ref{SRN2fig}. The images
of $\rho=0$ become two horizons ${\cal H}_+$ and ${\cal H}_-$,
each being a copy of $-\infty<z<\infty$, $\rho=0$.  The surface
gravities are $\kappa_\pm=\sqrt{M^2+Q^2}/(\sqrt{M^2+Q^2}\pm M)^2$
at ${\cal H}_\pm$.

Comparing the elliptic and hyperbolic representations of S-RN, the
horizon to the positive-mass quarter-plane vertical card of the
elliptic is the same as the horizon to the positive-mass
half-plane vertical card of the hyperbolic; the left-edge boundary
of the horizon of the elliptic is the center point of that of the
hyperbolic but only at $\phi_{\rm hyp}=0$; both of these points
are $X=Y=0$ of $Z^2-X^2-Y^2=1$.  The line segment where the
$\phi_{\rm ell}$ circle vanishes in the elliptic corresponds to
the branch cut of the hyperbolic, but only at $\phi_{\rm hyp}=0$.

We point out two ways to get the S-Schwarzschild from the Witten
bubble: One can either take $M\to i M$ which sends elliptic
(hyperbolic) Witten $\to $ hyperbolic (elliptic) S-Schwarzschild
respectively.  Alternatively, one can turn a vertical card on its
side which preserves elliptic/hyperbolicity.

Just as we took the hyperbolic Witten bubble on its side and got
hyperbolic S-RN, we can take the vertical half-plane card diagrams
for the Kerr bubble, S-dihole I, and superextremal S-Kerr and
S-dihole II and apply the $\gamma$-flip to yield new spacetimes.
These solutions are not identical to any previously described
solutions and will be described in \cite{joneswangfuture}.

To achieve the E-RN solution, Wick rotate S-RN by $Q\to i Q$;
we will discuss the superextremal E-RN card diagrams below.

\subsubsection{S-Kerr}

\begin{figure}[htb]
\begin{center}
\epsfxsize=2.5in\leavevmode\epsfbox{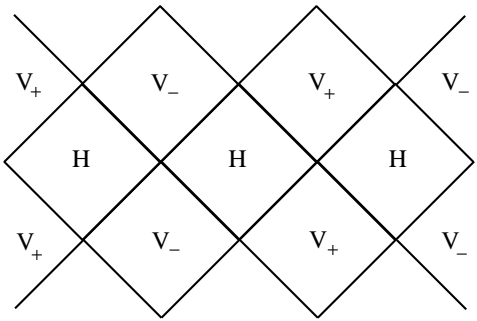}
\caption{Subextremal $|a|<|M| $ S-Kerr Penrose diagram.  V$_\pm$
map to vertical cards of positive and negative `mass,' while all H
diamonds give identical horizontal cards.  It is possible to
identify cards (say, every other H diamond) so there are only
a finite number of regions
in the spacetime.} \label{SKerrPenrosefig}
\end{center}
\end{figure}

The twisted S-brane~\cite{Wang:2004by}, see also
\cite{regSbraneQuevedo}, is also known as S-Kerr, and is another
example of a nonsingular time-dependent solution.  It can be
obtained from the Kerr black hole by analytically continuing the
Boyer-Lindquist coordinates $r,t,\theta$ and the parameters $a,M$.
For the parameter range $|a|<|M|$ there are horizons and the
global spacetime has a unique representation as a series of cards
connected in the same manner as the elliptic bubble (see
Fig.~\ref{bubblecard1}) except the vertices ending the horizons
are at $\tau=\pm\sqrt{M^2-a^2}$. The image of the ergosphere lies
on the horizontal card and has the same qualitative shape as it
does for the Kerr black hole diagram. The whole spacetime is
connected and each horizon is a 4-card junction which we can
extend to obtain an infinite number of distinct cards; it is
possible to truncate to a finite number of distinct cards by
suitably identifying cards. In comparison the Penrose diagram of
the $z,t$ plane is in Fig.~\ref{SKerrPenrosefig}.

In the extremal limit $|a|\to |M|$, $\theta$-orbits on the
vertical card shift up relative to the special null line, and any
fixed ($t$, $\theta$) point is sent above the null line.  The
region below the null line disappears in this limit and the
horizontal card collapses to a point. Furthermore, those geodesics
in the upper-right card can only reach the lower-left card (and
the same with upper-left and lower-right), splitting the universe
into connected $45^\circ$ wedges just like the parabolic
representation for the Witten bubble (Fig. \ref{bubblecard3}),
where the connections for dS$_2$ were added for clarity.

The case $|a|>|M|$ for S-Kerr does not have horizons and can be
represented as a single vertical half plane card with $\rho\geq 0$
and with no special null lines.  As we send $|a|\to |M|$, the
$\theta$-orbits pull away from $\tau=0$ and the spacetime gets put
into two wedges, $\rho<|\tau|$ with $d\tau$ timelike; this again
gives the extremal card diagram discussed above.
Finally in the limit where $|a|\to\infty$, the $\theta$-orbits
flatten out and the solution becomes flat space.

\subsubsection{5d Schwarzschild and Reissner-Nordstr\o m, and
S-variants}\label{5dcard-subsec1}

The Weyl Ansatz has been extended to higher dimensions.  In this
and the next subsection we discuss charged black holes, Witten
bubbles and S-Schwarzschild solutions in five dimensions. Although
most of the properties will be the same there will be some
important differences regarding the card diagrams.  A discussion
on a charged Weyl Ansatz in higher dimensions is included in the
appendix.

To begin with, the five dimensional Schwarzschild black hole
can be written in the generalized Weyl Ansatz \cite{EmparanWK}
and has a card diagram which is similar to the connected
positive-mass universe of the four dimensional case.  To fit the
Weyl Ansatz the three sphere is parametrized as $d\theta^2
+\sin^2\theta d\phi^2+\cos^2\theta d\psi^2$.  There are two
half-plane horizontal cards representing asymptotically flat
regions and two vertical square cards $-\mu/4\leq z\leq\mu/4$
inside the horizon. Even though the card diagram is similar to the
four dimensional case, however, the $z$-axis plays a different
role.   In the five dimensional case the boundary the $z$-axis
where $z>\mu/4$ and the right boundary of the vertical square
which it connects to, are both where the $\phi$-circle vanishes.
The $z$-axis where $z<-\mu/4$ and the left boundary of the square
is where the $\psi$-circle vanishes.

As discussed in the appendix,
the Weyl formalism also holds in the presence of an electrostatic
potential, so the charged five dimensional black hole of
\cite{Krori:iz}
$$ds^2=-\Big(1-{\mu\over r^2}+{Q^2\over 4r^4}\Big)dt^2
+\Big(1-{\mu\over r^2}+{Q^2\over
4r^4}\Big)^{-1}dr^2+r^2d\Omega_3^2$$ can be put into Weyl's
coordinates and a card diagram can be drawn.  In this choice of
units $Q=\mu$ is extremal.  Now the coordinate transformation from
the above Schwarzschild coordinates to Weyl coordinates is
$$\rho={1\over 2}\big(1-{\mu\over r^2}+{Q^2\over 4r^4}\big)^{1/2}r^2\sin
2\theta,\qquad z={1\over 2}(r^2-{\mu\over 2})\cos 2\theta.$$ As in
the four dimensional black hole solutions, the charge $Q$ does not
enter in the definition of the $z$-coordinate.

The card diagram is similar to that of the four dimensional
Reissner-Nordstr\o m in Fig.~\ref{RN-card} except the foci are at
$z={1\over 4}\sqrt{\mu-Q^2}$ and the singularity is a semi-ellipse
passing through the points $(\rho,z)=(0,\pm\mu/4)$ and
$(\rho,z)=(|Q|/4,0)$ on a horizontal card.  It is again possible to
attach a negative mass universe to the singularity to fill out the
rest of the horizontal plane.  Since Weyl coordinates depend only
on $r^2$, filling out the rest of the horizontal card
unambiguously take us to imaginary values of $r$ and not negative
values of $r$.

To obtain the charged five dimensional Witten bubble, analytically
continue the solution with $t\to ix^5$, $\phi\to i\phi$, $Q\to iQ$.
This analytic continuation gave the elliptic representation
of the Witten bubble in four dimensions.
There is a quarter-plane vertical card above
the $\phi$ ray with an $x^5$ vertical boundary; this boundary
continues on a segment of the nonsingular horizontal card, and
then there is a $\psi$ boundary.

To obtain the five dimensional S-RN, take the vertical card from
the Witten bubble and turn it on its side using our flip
$\gamma\to\gamma+i\pi$.  This vertical card now has a $\phi$
boundary and an $x^5$ horizon.  Continuing onto the horizontal
card, the $\phi$ boundary terminates after a segment and at $r=0$,
there is the S-brane singularity which maps to a hyperbola on the
card.   Past the singularity we complete the horizontal card with
a negative mass universe.  The rest of the $z$-axis is filled out
by an $x^5$ horizon which is a four card junction.  In short the
card diagram looks just like the four dimensional case, except
there is a nonvanishing spacelike $\psi$-direction which is of
course not evident on the diagram.

\subsubsection{Multiple representations in 5d and ${\bf H}_3$}
\label{5dcard-subsec2}

As discussed in Section~\ref{bubble-sbrane-subsec} the four
dimensional S-RN and charged Witten bubble have three card
representations which corresponds to the three difference ways of
putting a Killing congruence on ${\bf H}_2$. In contrast, the five
dimensional S-RN and charged Witten bubbles have 2 representations
because there are only two ways to put two orthogonal Killing
congruences on ${\bf H}_3$. The elliptic and parabolic card
diagrams in four dimensions will be analogous to what will call
elliptic-hyperbolic and parabolic-parabolic diagrams in five
dimensions.

The standard representation of five dimensional S-RN involves
parametrizing ${\bf H}_3$ by $ds^2=d\theta^2+\cosh^2\theta
d\psi^2+\sinh^2\theta d\phi^2$ or
$$W'=\cosh\theta\cosh\psi,\qquad Z=\cosh\theta\sinh\psi,\qquad
X=\sinh\theta\cos\phi,\qquad Y=\sinh\theta\sin\phi,$$
where $-W'^2+Z^2+X^2+Y^2=1$ and $dW'$ is timelike.

It is a fact, which we will not prove here, that
the isometries of ${\bf H}_3$ are precisely the conformal isometries
of its conformal infinity, the Riemann sphere $S^2$ \cite{Matsuzaki}.
This is the complex M\"obius group $PSL(2,{\bf C})$.  To see the
relation between ${\bf H}_3$ and $S^2$ in the above representation,
send $\theta\to\infty$ to achieve $-W'^2+Z^2+X^2+Y^2=0$
modulo rescaling.  Then set $W'=1$ to get $Z^2+X^2+Y^2=1$.  A conformal
isometry of $S^2$ induces an isometry of ${\bf H}_3$ and in simple
cases it is easy to guess which M\"obius transformation corresponds
to which isometry.  In the present
example, $Z=\tanh\psi$, $X={\rm sech}\,\psi \cos\phi$, and $Y={\rm
sech}\,\psi \sin\phi$.
So the hyperbolic $\psi$ boost isometry corresponds to $z\to
(1+\epsilon)z$
on $S^2$, whereas the elliptic $\phi$ azimuthal rotation isometry
corresponds to $z\to e^{i\epsilon}z$ on $S^2$.  It is easy to see that the
only other
pair-type of orthogonal M\"obius transformations is perpendicular parabolic
ones, $z\to z+\epsilon$ and $z\to z+i\epsilon$, and that this gives
the only other pair-type of orthogonal Killing congruences on ${\bf H}_3$.
For the half-3-space representation of ${\bf H}_3$ with coordinates
$(x,y,1/\sigma>0)$ and metric
$d\sigma^2/\sigma^2+\sigma^2(dx^2+dy^2)$, these are just translations in $x$
and $y$.

The parabolic-parabolic representation of five dimensional S-RN is
\begin{eqnarray*}
ds^2&=&f_5 (dx^5)^2+f_x (dx)^2+f_y(dy)^2+e^{2\nu}(-d\rho'^2+dz^2),\\
e^{2\nu}&=&t^2/(\mu^2-Q^2)\sigma^2,\\
f_5&=&1-\mu/t^2+Q^2/4t^4,\\
f_x=f_y&=&t^2\sigma^2,
\end{eqnarray*}
where $\rho'=\sigma^2\sqrt{\mu^2-Q^2}(\sinh\zeta)/2$,
$z=\sigma^2\sqrt{\mu^2-Q^2}(\cosh\zeta)/2$, and
$t^2-\mu=\sqrt{\mu^2-Q^2}(\cosh\zeta)/2$, written for $t\geq t_+$.
The card diagram looks like the parabolic representation of 4d
S-RN.

Applying the $\gamma$-flip on the vertical wedge card of
S-Schwarzschild we get the parabolic-parabolic Witten bubble.  At
the pointlike tip, we attach one downward-facing copies of
that wedge in a dS$_3$ fashion.

There are no five dimensional analogues of the hyperbolic S-RN and
hyperbolic superextremal E-RN (see next subsection) card diagrams.
For all these solutions, the lack of a third card diagram in the
five dimensional case can be understood from two different
perspectives: the lack of a $z\to i\tau$ continuation in the Weyl
functions $f_x$, $f_y$; or the restrictions on orthogonal
congruences of ${\bf H}_3$.

\subsubsection{Three representations for superextremal E-RN}

Although it is not clear if E-brane solutions \cite{Hull} should
exist in a stable theory, they are interesting simple examples of
new card diagrams.  We examine the superextremal E-RN,
$Q^2-M^2>0$, which can be obtained by starting from S-RN and
continuing the charge $Q\to iQ$.

The elliptic representation can also be directly obtained from the
RN black hole by continuing $t\to ix^4$, $r\to it$, $M\to iM$,
$Q\to iQ$, and $\theta\to i\theta$. The elliptic representation
for the superextremal E-brane Reissner-Nordstr\o m solution is a
vertical half-plane card with $\rho=\sqrt{Q^2-M^2}\cosh\zeta
\sinh\theta$, $\tau=\sqrt{Q^2-M^2}\sinh\zeta\cosh\theta$, where
$\zeta$ is defined as $t-M=\sqrt{Q^2-M^2}\sinh\zeta$. It has no
special null lines and a singularity along a $\theta$-orbit at
$t=0$ which forms a hyperbola in Weyl space.

\begin{figure}[htb]
\begin{center}
\epsfxsize=5in\leavevmode\epsfbox{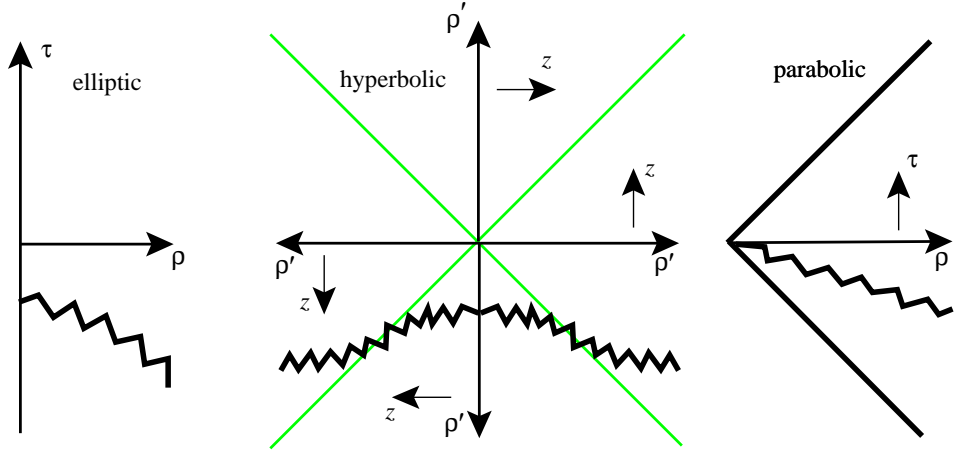} \caption{Three
representation for superextremal E-RN.  The hyperbolic is a whole
plane vertical card; there are two special null lines which meet
at a point.  This spacetime has the indicated singularity.}
\label{ERNfig}
\end{center}
\end{figure}

The hyperbolic representation, achieved by instead sending
$\theta\to\pi/2+i\theta$, and $\phi\to i\phi$ in BL coordinates
gives an interesting card.  The Weyl coordinates are
$\rho'=\sqrt{Q^2-M^2}\cosh\zeta\cosh\theta$,
$z=\sqrt{Q^2-M^2}\sinh\zeta\sinh\theta$, so we have the relation
$\rho'\geq |z|+\sqrt{Q^2-M^2}$.  Although $\rho'$ is positive, the
$z$ coordinate can take either sign; neither the $x^4$- nor the
$\phi$-circle closes anywhere.  The coordinates cover a vertical
quarter-plane wedge with $\rho'$ timelike.  Extending past the
null lines by reflection to make four wedges, the card becomes an
entire vertical plane card.  The two special null lines that meet
at $\rho'=\sqrt{Q^2-M^2}$ (see Figure \ref{ERNfig}).

This card diagram can be turned on its side to obtain a new
solution the superextremal Witten E-bubble.  By continuing
$\tau\to iz$, the intersection of special null lines becomes the
branch point of superextremal RN's horizontal card.  By
considering complex $z$ and real $\rho$, one sees that quite
generally, branch points on horizontal cards and intersecting
special null lines on vertical cards are really the same thing.

There is also a parabolic representation which is a noncompact
vertical $45^\circ$ card, with $\rho>|\tau|$.  Constant-$t$ loci
are straight rays $\tau/\rho=$constant.

The superextremal Kerr-Newman E-Bubble with $Q^2-a^2-M^2>0$ has a
card diagram structure like hyperbolic superextremal E-RN, except
the naked singularity is replaced by an instantaneous S1-brane
ring singularity, which appears as a point on the vertical card
diagram, on the left wedge.

The five dimensional superextremal E-RN has two card diagram
representations which look like the four dimensional elliptic and
parabolic ones.

For the hyperbolic subextremal E-RN the card diagram is similar to
the hyperbolic card diagram for S-RN except the singularity lies
on the negative-mass vertical cards at $\rho'^2=Q^2+{Q^2z^2\over
M^2}$. For the extremal case $Q=M$, The horizontal card totally
collapses and we take ourselves on the positive-mass vertical
card.  Now $R^2=z^2-\rho'^2$, and the numerator of $f$ vanishes
outside $\rho'>|z|$; only consider this region.  The spacetime is
complete like extremal RN.

\subsubsection{S-Black Rings}

Starting from the black ring, one may try to continue $x\to ix$,
$y\to iy$, and $\mu\to -i\mu$ in (\ref{blackringsol}), in the
hopes of finding a new time dependent S-Black Ring solution.
Unfortunately there is a singularity which rides up the boundary
of the vertical card and so the solution is nakedly singular. The
problems seems to be that the horizon was adjacent to a singular
boundary on the horizontal card.  If we can find spacetimes where
the horizon is adjacent to a regular boundary where a circle
closes, then the vertical boundary on the vertical quarter-plane
card must be nonsingular; that same circle will close with the
same periodicity.  To find such spacetimes, we eschew $xy$
coordinates and use Weyl methods.

One can define an S-black ring by taking the black ring's two
bubble solutions, with vertical quarter-plane cards either at
$z<-\mu/2A$ or $z>1/2A$, and turn them on their sides.  One finds
four new extended card diagrams three of which are not nakedly
singular.  With the help of our description of the $xy$
parametrization of these cards, we can fill out the new card
diagram structure which is the same as the extended black ring
structure.  The boundaries are labelled by the order of vanishing
of $x^5$, $\phi$, $\psi$ as follows:  $y=\pm 1$ is $(0,0,2)$,
$x=\pm1$ is $(0,2,0)$, $y=\infty$ is $(2,0,0)$, $x=\infty$ is
$(-2,2,2)$, $y=1/\mu$ is $(-2,4,0)$, $x=1/\mu$ is $(2,-2,2)$.
However all these spacetimes can be constructed directly using the
procedure of \cite{EmparanWK} by prescribing singularities,
boundaries, and acceleration/black hole horizons, and hence are
not really new.

It is possible that some electrification or spinning
generalization would yield less singular spacetimes with the same
card diagram structure. If canonical card diagrams (or even
more general card diagrams) do not apply,
the card diagram structure of the vacuum Weyl solution is still
useful as it gives us an approximation of the new global
structure.

A quite interesting new five dimensional solution built using Weyl
methods will be described in Section \ref{interesting5d}.

\subsection{Israel-Khan solutions}

Israel-Khan solutions\cite{IsraelKhan} are arrays of black holes
held apart by conical singularities or struts of pressure.  Their
Weyl representation is a series of rods along the $z$-axis. Their
card representation is obtained by simply treating each separate
rod as a horizon with a four card junction with two horizontal and
two vertical cards.
In particular special null lines still emanate from the foci of
Weyl space, and a triangle is unfolded four times to give a square
for each inner-horizon region.  Because the function $f$ is
multiplicative for each source of an Israel-Khan solution and
because the singularities occur when $f=\infty$, the
singularities lie at the top of the vertical square
cards just as in the isolated black hole case.

Over connected parts of the boundary the conical excess angle is
constant.  In the case of two black holes physically this means
that inside the left black hole's horizon, two-spheres have
conical excess singularities at their north poles, and inside the
right black hole's horizon, two-spheres have conical excess
singularities at their south poles.

By taking the distance between them to zero, black holes can be
made to merge in a discontinuous fashion.  Adjacent square
vertical cards go from having size $2m_1\times 2m_1$ and
$2m_2\times 2m_2$ to size $2(m_1+m_2)\times 2(m_1+m_2)$.  This
parametric merging is not a physical process.

\subsubsection{A 2-rod gravitational wave example}
\label{2rodgravwavesec}

As an application of our Weyl techniques we examine the case of
two rods neither of which will cross the real-$\tau$ line, similar
to a situation considered in \cite{JonesRG}.  The rods have linear
density $\lambda$ which need not be $1/2$, are centered at $z=\pm
a$ and have length $2M$.  The new step is to rotate the right rod
counterclockwise in the complex $z$-plane and the left one
clockwise.  We rotate them by $90^\circ$.\footnote{Rotating by a
nonorthogonal angle would give a real solution $U$ that is neither
even nor odd in time $\tau$.}
Using the standard branch cut the distances from the
rod endpoints are
\begin{eqnarray*}
\tilde r_1&=&\sqrt{\rho^2+a^2-(\tau-M)^2+2ia(\tau-M)};{\rm\
distance\
to\ }z=-a+ib\\
r_1&=&\sqrt{\rho^2+a^2-(\tau+M)^2+2ia(\tau+M)};{\rm\ distance\
to\ }z=-a-ib\\
\tilde r_2&=&\sqrt{\rho^2+a^2-(\tau+M)^2-2ia(\tau+M)};{\rm\
distance\
to\ }z=a-ib\\
r_2&=&\sqrt{\rho^2+a^2-(\tau-M)^2-2ia(\tau-M)};{\rm\ distance\ to\
}z=a+ib.
\end{eqnarray*}
Note that $\tilde r_1\leftrightarrow r_2$ and $\tilde
r_2\leftrightarrow r_1$ under complex conjugation; this exchanges
a point with its mirror across the $\tau$-axis.  The Weyl function
$f=e^{2U}$ is in this case
\begin{equation}
U_{\rm SLO}=\lambda\log {r_1+\tilde r_1+2iM\over r_1+\tilde
r_1-2iM} \cdot{r_2+\tilde r_2-2iM\over r_2+\tilde r_2+2iM}
\end{equation}
where both the numerator and denominator are real.  The notation
`S' stands for `sum' since the distances are summed $r_1+\tilde
r_1$ and $r_2+\tilde r_2$, `L' stands for `log,' and `O' stands
for `odd' since the solution has $U$ odd in $\tau$.  This is one
branch choice but it is not the only good choice.

Alternatively we can change the branch of $r_2$ and $\tilde r_2$,
or equivalently the orientation of the rod at $z>0$.  (If we had
kept $M_1$ and $M_2$ distinct, we could send $M_2\to -M_2$
everywhere it appears.  We deduce that the equivalence must also
hold for $\gamma$ from its PDEs sourced by $U$.) This makes the
complex logarithm purely imaginary, so to obtain a sensible metric
we send $\lambda\to -i\lambda$ and get
\begin{equation}
U_{\rm SAE}=\lambda\arg {r_1+\tilde r_1+2iM\over r_1+\tilde
r_1-2iM} \cdot{r_2+\tilde r_2+2iM\over r_2+\tilde r_2-2iM}
\end{equation}
where the product of fractions is unimodular.  This solution has
sums, the angular argument function and is even in $\tau$.  This
branch choice is seemingly good and produces a different even
wave.

Thirdly we can have
\begin{equation}
U_{\rm DAO}=\lambda\arg {r_1-\tilde r_1+2iM\over r_1-\tilde
r_1-2iM} \cdot{r_2-\tilde r_2-2iM\over r_2-\tilde r_2+2iM},
\end{equation}
which involves differences, arg, and is odd in $\tau$.  If one is
careful to take a continuous branch of the argument, the function
$U$ is smooth.

Lastly we can have
\begin{equation}
U_{\rm DLE}=\lambda\log {r_1-\tilde r_1+2iM\over r_1-\tilde
r_1-2iM} \cdot{r_2-\tilde r_2+2iM\over r_2-\tilde r_2-2iM};
\end{equation}
which involves differences, log, and is even in $\tau$.  This
function is problematic in that it is not asymptotically flat as
$\tau\to\pm\infty$, but is nonsingular.

The function $e^{2\gamma}$ can be gotten from
\cite{Myers:rx}\cite{IsraelKhan} by multiplying by $4\lambda^2$
and analytically continuing.  It is smooth but does not have good
asymptotic properties.  Recall that $\gamma=\gamma_{11}+
2\gamma_{12}+\gamma_{22}$,\footnote{These $\gamma_{11}$ and
$\gamma_{22}$ were accidentally left out of \cite{JonesRG}; they
are present in the horizontal rod case as well.  Also there are
four choices for branches, leading to SLE, SAO, DAO, and DLE.
Here, SLE is problematic in that $U$ it is not asymptotically flat
as $\tau\to\pm\infty$.} where $\gamma_{ij}$ is the interaction
of the $i$'th rod with the $j$'th rod.

\subsubsection{The C-Metric}

\begin{figure}[htb] \begin{center}
\epsfxsize=3in\leavevmode\epsfbox{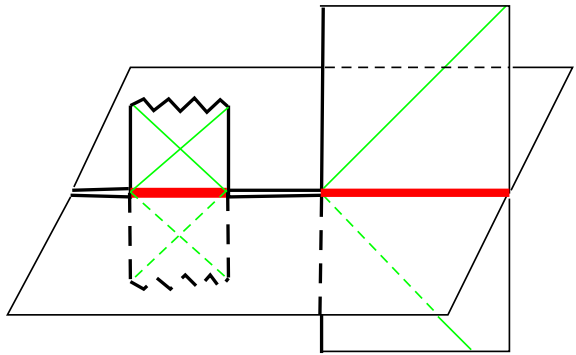} \caption{The card
diagram for the C-metric.  It is quite natural to identify the two
black holes.} \label{cmetricfig}
\end{center}
\end{figure}

A half-plane horizontal card for the C-metric has one finite rod
to give a black hole, and one semi-infinite rod to represent an
acceleration horizon. The full card diagram is then shown in
Fig.~\ref{cmetricfig} which shows that it is quite natural to
identify the two black holes that are accelerating apart from
``each other,'' and that there is an Einstein-Rosen bridge from
one horizontal card to the other \cite{Hawking:1994ii}.

One can similarly draw card diagrams for spacetimes with two
acceleration horizons and black holes in between; these are $\phi
\to it$, $t\to i\phi$ Wick rotations of Israel-Khan solutions.

\section{2-Rod solutions without $z\to i\tau$}

With a few techniques, one can find new S-brane solutions which
are like deformations of the hyperbolic representation of S-RN by
multiple rods and this will break the ${\bf H}_2$ symmetry of the
geometry. The ease of constructing such solutions is the advantage
of Weyl coordinates.  In this analytic continuation we will stay
on a horizontal card and we will not continue $z\to i\tau$.

\subsection{One vertical rod, one horizontal rod}

One can find a solution which is somewhat like the elliptic Witten
bubble and somewhat like hyperbolic S-Schwarzschild by turning one
rod and leaving the other one alone.  This solution shows how
Weyl solutions are specified by their rod sources and boundary
information on a horizontal card.

Here we begin with a horizontal half-plane card.  Let us label the
Killing directions as $x^4$ (time), $\phi$ (space) in that order
and label a boundary by $(m,n)$ where as $\rho\to 0$, the metric
component $g_{44}=f\sim\rho^m$ and $g_{\phi \phi}\equiv \rho^2
f^{-1}\sim \rho^n$. By definition we have the relation $m+n=2$. In
this scheme a horizon is $(m,n)=(2,0)$, an ordinary boundary where
the Weyl card ends is $(0,2)$, and a Schwarzschild black hole
singularity is $(-2,4)$.

Next we specify a background boundary which is of type $(0,2)$
along the $z$-axis of the Weyl card. On top of this place a rod
from $-M\leq z\leq M$ of Schwarzschild type $(+2,-2)$ and a second
rod of type $(+2,-2)$ from $a-\mu\leq z\leq a+\mu$. These sources
generate
$$f_1={R_++R_--2M\over R_++R_-+2M},\qquad f_2={r_++r_--2\mu\over
r_++r_-+2\mu}$$ where $R_\pm=\sqrt{\rho^2+(z\pm M)^2}$ and
$r_\pm=\sqrt{\rho^2+(z-a\pm\mu)^2}$.  Turn the second rod on its
side by the analytic continuation $\mu\to i\mu$, $r_-\to -r_-$,
$x^4\to ix^4$, $\phi\to i\phi$ which is like the one used for the
hyperbolic representation of S-RN.  The function $f=f_1 f_2$ is
negative, the coordinate $x^4$ is timelike and the solution is
ready to become a charged S-brane upon imaginary-$C$
electrification, see Appendix~\ref{electrify-app}. Denoting
$r=r_+=\overline{r_-}$, the solution has
\begin{eqnarray*}
e^{2\gamma_{11}}&=&\frac{(R_++R_-)^2-4M^2}{4 R_+ R_-},\\
e^{2\gamma_{22}}&=&\frac{\mu^2-({\rm Im}\,r)^2}{|r|^2},\\
e^{4\gamma_{12}}&=&\frac{R_-r_++(z-M)(z-a+i\mu)+\rho^2}
{-R_-r_-+(z-M)(z-a-i\mu)+\rho^2}\,
\frac{-R_+r_-+(z+m)(z-a-i\mu)+\rho^2}{R_+r_++(z+M)(z-a+i\mu)+\rho^2}
\end{eqnarray*}
The function $\gamma_{12}$ is real as can be checked.

The effect of turning the rod on its side is to induce a branch
point at $(\rho=\mu,z=a)$.  Fixing this branch point by taking the
appropriate square root, we find two
boundaries, labelled $+$ and $-$, which are the images of
the $z$-axis on the horizontal card. The $+$ boundary is uniformly
affected by $(+2,-2)$ and the $-$ boundary is uniformly affected
by $(-2,+2)$.  This makes the $+$ boundary a sequence, of
horizon $(2,0)$, singularity $(4,-2)$, and horizon $(2,0)$ from
left to right. The $-$ boundary is $(-2,4)$, $(0,2)$, $(-2,4)$.

Let us recall two solution generating techniques.  There is a
`sign flip' procedure which is applicable to four dimensional
vacuum Weyl solutions where
\begin{equation}\label{signflip}
U\to -U+\log\rho,\qquad \gamma\to \gamma-2U+\log\rho.
\end{equation}
In Israel-Khan solutions, this replaces rods with space and space
with rods.  When rods are rotated parallel to imaginary $z$, the
interpretation is less clear.  The sign flip sends $(m,n)\to (2-m,2-n)$.

Another technique is Harrison's third theorem \cite{Harrison} which sends
\begin{equation}
2U\to 2U-\mu_H\log\rho,\qquad 2\gamma\to 2\gamma-\mu_H 2U+{\mu_H^2\over
2}\log
\rho,\label{harrisonthird}
\end{equation}
where $\mu_H$ is real.  This changes the order of $f$'s
vanishing near $\rho=0$ uniformly at all boundaries.

\begin{figure}[htb] \begin{center}
\epsfxsize=3.5in\leavevmode\epsfbox{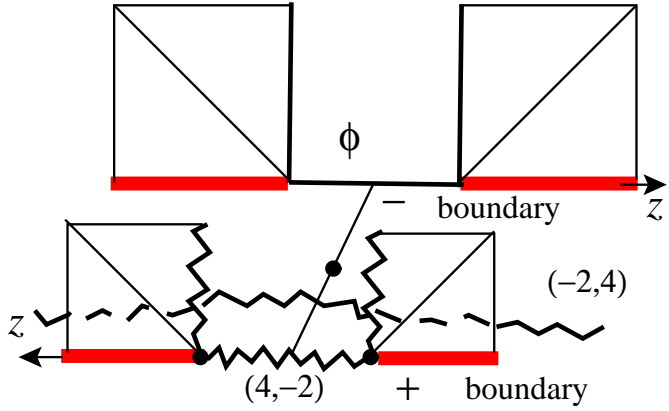} \caption{4d
electrified solution; two rods, one turned on its side.}
\label{4D-2rods}
\end{center}
\end{figure}

Performing the sign flip (\ref{signflip}) and then Harrison's
third theorem with $\mu_H=2$ on our solution gives a $-$ boundary
of orders $(2,0)$, $(0,2)$, $(2,0)$, and a $+$ boundary of orders
$(-2,4)$, $(-4,6)$, $(-2,4)$.  The $+$ boundary has two
Schwarzschild singularities and a bad singularity where $f\propto
\rho^{-4}$.  In the case of the Schwarzschild and S-Schwarzschild
solutions, we saw that singularities can be moved/created by
adding charge.  Here we change the
singularities by imaginary-$C$ electrification. Setting
$f=-e^{2\overline{U}}$, we get a RN singularity wherever
$\tanh\overline{U}={C\over\sqrt{1+C^2}}$. For large enough
positive $C$, the RN singularity then hugs the $+$ boundary and
leaves the rest of the universe (with Witten bubble horizons on
the $-$ boundary) connected.  The card diagram is shown in
Figure~\ref{4D-2rods}.

\subsection{Two vertical or imaginary-displaced rods}

There are three basic rod configurations which give
topological-line horizons and topological-line singularities.  All
of these configurations involve branch cuts from the boundary of
the $\rho- z$ half-plane out to a branch point in the interior of
the card.

First, place a horizontal rod from $z=-M+ib$ to $z=M+ib$, and
place a mirror rod from $z=-M-ib$ to $z=M-ib$.  Branch points now
occur (for the distances $r_\pm=\sqrt{\rho^2+(z\pm M-ib)^2}$ to
the upper rod endpoints, and their conjugates) at $z=\pm M$,
$\rho=b$, and the standard square root branch cut gives us branch
cuts along $z=\pm M$, $0\leq\rho<b$.  Undoing the two branch cuts
gives $2^2=4$ boundaries (which may be horizons, singularities, or
true boundaries where a circle vanishes).

Second, place a vertical rod from $z=i(b-M)$ to $z=i(b+M)$ and a
mirror rod from $z=-i(b+M)$ to $z=-i(b-M)$.  There are branch
points at $z=0$, $\rho=b-M$ and $\rho=b+M$.  One branch cut runs
out to $\rho=b-M$ and an independent branch cut runs out to
$\rho=b+M$. Again undoing the branch cuts gives four boundaries.

Third, place a vertical rod from $z=-iM$ to $z=iM$ like the second
representation of S-Schwarzschild.  If we place two such rods, one
displaced from along real $z$, we get two branch cuts and four
boundaries.

The orientations of rods (reversing the sign of some $M$'s) and
branches to all endpoints can be considered; there will always be
some solution for $U$ which is real and involves the real
logarithm.  For brevity this will not be spelled out.

In all cases, rods must be given density $1/2$ (sourcing $U$) to
generate a $\gamma$ such that $e^{2\gamma}$ blows up to order 1 at
the branch point.  This is necessary to get a nonsingular
spacetime after conformally fixing the branches on the horizontal
card.  For the hyperbolic S-Schwarzschild, we know
that a single rod generates a horizon (where $f=e^{2U}$ vanishes
$\propto \rho^2$), and a singularity (where $f$ blows up $\propto
1/\rho^2$).  For that case we had $ds^2\supset f
(dx^4)^2-f^{-1}\rho^2 d\phi^2$ where $f<0$.

For the cases of two rods, each boundary has a $\pm$ status
relative to each rod.  The $++$ boundary has $f\propto \rho^4$, a
$+-$ or $-+$ boundary has $f\propto 1$, and the $--$ boundary has
$f\propto 1/\rho^4$.  So for the third case of two vertical-in-$z$ rods
crossing the
$z$-axis, not continuing Killing directions, we have $ds^2\supset -f_1 f_2
dt^2+f_1^{-1}f_2^{-1}\rho^2 d\phi^2$ where each $f_1$, $f_2$
is negative.

We perform the sign flip (\ref{signflip}).  Here, this gives
$f\propto \rho^{-2},\rho^2,\rho^2, \rho^6$ on the $++$, $+-$,
$-+$, $--$ boundaries.  This is ready to be turned into a charged
S-brane solution via a real-$C$ electrification (see the next
subsection).

In this case, writing $f\to e^{2\overline{U}}$ for real
$\overline{U}$ on the horizontal card, the electrified solution
goes singular at the locus
$\coth\overline{U}={C\over\sqrt{C^2-1}}$.  Picking $C$
sufficiently negative, there are S-Schwarzschild singularities
hugging around the $+-$ and $-+$ boundaries which are still
horizons.  Each has an associated connected 4-card spacetime which
generalizes S-RN.

It is possible to get three S-Schwarzschild singularities and one
horizon: Instead of the sign flip, we can utilize Harrison's
third theorem (\ref{harrisonthird}).  Picking
$\mu_H=2$, we get $f\propto \rho^2,\rho^{-2},\rho^{-2}, \rho^{-6}$
on the $++$, $+-$, $-+$, $--$ boundaries.  This is
ready to be turned into a charged S-brane solution via a real-$C$
electrification.

In this case we can pick $C$ sufficiently positive so there are
S-Schwarzschild singularities hugging the $+-$, $-+$, and $--$
boundaries.  The $++$ boundary is still a horizon.  So this is
like S-RN but the horizontal card has three disjoint
singularities.

\subsection{5d solution with two rods, one turned on its side}
\label{interesting5d}

\begin{figure}[htb] \begin{center}
\epsfxsize=3.5in\leavevmode\epsfbox{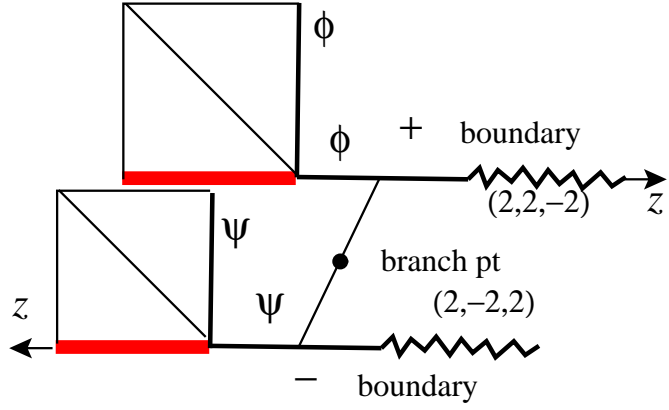} \caption{5D solution;
two rods, one turned on its side.} \label{5Dringcard}
\end{center}
\end{figure}

We now give a prescription for a non-nakedly singular 5d spacetime
using two rods, one turned on its side. Write the Killing
coordinates $x^5$, $\phi$, $\psi$ in that order. Say that part of
the boundary has order $(m,n,p)$ if $f_5$, $f_\phi$, and $f_\psi$
vanish to orders $\rho^m$, $\rho^n$, and $\rho^p$.  Of course
$m+n+p=2$. Take Killing orders $(2,-2,2)$, $(0,0,2)$, and
$(2,0,0)$ for $z<-M$, $-M<z<M$, and $M<z$.  Add in a $(+0,+2,-2)$
rod of length $\mu$ centered at $z=a$ and turn it on its side.  We
then get (see Figure~\ref{5Dringcard}) a $+$ boundary of
$(2,0,0)$, $(0,2,0)$, and $(2,2,-2)$. Note that the horizon is
adjacent to a $\phi$-circle closing and its quarter-plane vertical
card will be nonsingular.  The $-$ boundary is $(2,-2,2)$,
$(0,0,2)$, and $(2,0,0)$.  Again its horizon is adjacent to the
$\psi$-circle closing and its quarter-plane vertical card will be
nonsingular.  The singularities in this spacetime are confined to
the horizontal card and they are related by trivial Killing
rotation to $(-2,2,2)$ boundaries which are ordinary 5d black hole
ones; so they are no more singular than the 5d S-Schwarzschild.

\subsection{Appendix: Weyl Electrification} \label{electrify-app}

Weyl's electrification procedure \cite{Fairhurst:2000xh} can be
applied to any Israel-Khan solution or Wick rotation thereof.
Calling our old gravitational potential $\overline{U}$, the new
one is given by
\begin{eqnarray}\label{electrify}
e^{-U}&=&{e^{-v}\over
2}\left(\Big(1+{C\over\sqrt{C^2-1}}\Big)e^{-\overline{U}}
-\Big({C\over\sqrt{C^2-1}}-1\Big)e^{\overline{U}}\right),\\
e^{2U}&=&e^{2v}-2C e^{v}A+A^2,\nonumber
\end{eqnarray}
where $C\geq 1$ and $v$ are constants and $e^{2\gamma}$ does not
change. Using the conformal symmetry of the Einstein-Maxwell
system and the freedom to shift $\gamma$ by a constant we may take
$v=0$.  As applied to Schwarzschild of mass $M_0$, this generates
Reissner-Nordstr\o m of $C=M/Q$ where $M_0^2=M^2-Q^2$.  As applied
to S-Schwarzschild, since we had sent $t\to i x^4$ this generates
an E-RN.

The point of (\ref{electrify}) is that the Weyl cards of the
spacetime do not change, but singularities move.  Where
$\overline{f}$ used to behave like $\rho^\alpha$, $f$ behaves like
$\rho^{|\alpha|}$.  So horizons stay horizons and Schwarzschild
singularities become horizons.  New singularities are generated
where the quantity in parentheses vanishes, and generically this
is to first order, generating Schwarzschild singularities.

For imaginary $C$,
\begin{eqnarray}\label{electrifyimag}
e^{-U}&=&{e^{-v}\over
2}\left(\Big(1+{C\over\sqrt{C^2+1}}\Big)e^{-\overline{U}}
+\Big(1-{C\over\sqrt{1+C^2}}\Big)e^{\overline{U}}\right),\\
e^{2U}&=&1-2iCA+A^2,\nonumber
\end{eqnarray}
will turn the S-Schwarzschild into an S-RN.  Any value of $C$ is
then allowed. This transformation also charges the Witten bubble.

These electrifications seemingly do not generate superextremal
solutions, which require a change in the card structure.  (They
have branch cuts instead of line segments.)

\section{Discussion}

In this paper we have examined the utility of the Weyl Ansatz and
constructed an associated card diagram.  The card diagram
conveniently captures most of the interesting properties of a
spacetime including its singularities, horizons, null infinity and
some of its causal structure.  The main technical details that one
has to deal with regarding card diagrams seem to be analytic
continuation of the coordinates, special null lines and the choice
of the branch of a square root, and branch points.

Here we give a summary of the solutions we have discussed in this
paper.  The card diagrams correctly capture the different regions
of the charged Reissner-Nordstr\o m black hole and its various
charged and chargeless limits, and its negative mass complement.
We also analyzed the Kerr black
hole and its singularity structure; the passage to the second
asymptotic universe through $r=0$ is quite clearly depicted.

The Witten bubble and S-brane had three card diagrams since there
were three choices of coordinates which were compatible with the
Weyl Ansatz.  The hyperbolic representations had no foci on the
card diagram.  In the case of the bubble the card diagram is a
half-plane vertical card, while the S-Schwarzschild was more
technically involved and had a branch point which we fixed with a
conformal mapping. The elliptic representations were very similar
and consisted of two foci and six individual cards.  The main
difference is that for the S-Schwarzschild, the left 4-card
junction is not a horizon but a singularity.  Finally the parabolic
representation of the bubble was an infinite array of triangles
connected pointwise while the S-Schwarzschild had a 6-card butterfly
shape.  This third representation has a useful representation of
their null infinity.

The S-Schwarzschild can be obtained from the bubble in two ways.
One may start with the bubble and analytically continue $M\to i M$
in Weyl coordinates.  In this case the
hyperbolic/elliptic/parabolic representation of the bubble maps to
the elliptic/hyperbolic/parabolic representation of the S-brane.
There was also a second way to relate these two solutions which we
termed the $\gamma$-flip and which was conveniently visualized as
a flip of the associated cards about a null line.  This
procedure maintains the number of Weyl foci on the card and so
maps all bubbles to similarly labelled S-branes.  The
$\gamma$-flip provides a very simple and interesting way to relate
Schwarzschild with the bubble and the S-brane.  In fact all
spacetimes related in this way by $\gamma$-flips can be
simultaneously drawn together in a complexified Weyl spacetime
diagram.

There are six S-dihole universes which we called $\cal{U}$,
$\cal{U}_\pm$ and $\cal{E}$, $\cal{E}_\pm$.  The three $\cal{U}$
universes were non-singular and had interesting near horizon
scaling limits which we named $\cal{W}$.  The $\cal{E}$ universes
were singular on a so called ergosphere, represented the decay of
two dimensional unstable objects, and ${\cal E}$ itself had an
S-Melvin universe scaling limits. It was also previously noted
that both the dihole and the S-dihole I solutions had Melvin
scaling limits.  The reason for this is that all these Melvin
limits come from the same neighborhood of the complexified Weyl
space.  These solutions were then generalized to include new
arrays as compared to those mentioned in \cite{Gaiotto:2003rm}.
The generalization includes allowing for black holes on the real
time axis which gives dS$_2$ horizons with ${\cal W}$-type scaling
limits, and we were even able to construct a periodic universe.

We also detailed multiple rod solutions such as black rings,
S-black rings, the C-metric.  New electrified
4d solutions and a 2-horizon vacuum rod 5d solutions were given.
Gravitational wave solutions were shown to have several choices
for branches.

Finally we turn to some comments on future work.  As for the
S-brane solutions we have presented here, it seems possible that
the relationship between the general array solution and the Melvin
scaling limit may have some connection to the discussion on the
decay of unstable branes.  In particular while the S-dihole is
generally unstable and will decay, the decay time can be made as
long as we wish by appropriately scaling the parameters.  Also
while we have now found an infinite array solution with horizons,
there is still potentially much to understand regarding this
solution and its exact relation to the rolling tachyon.  Further
investigation of this decay process, its relation to the work in
Ref.~\cite{Strotalk} and the possibility of an open-closed string
duality could prove interesting.

We hope the techniques herein and this method of visualization are
helpful for keeping track of the numerous Wick rotations and
mentally partitioning complicated spacetimes into simpler regions.
Aside from being a useful way to describe spacetime structures,
and follow their Wick rotations, card diagrams may have
alternative uses. As discussed in \cite{EmparanWK}, the higher
dimensional Schwarzschild solutions cannot be written in Weyl form
and so do not have Weyl card diagrams.  However, card diagrams are
more general than as described in this paper; the recently
developed Weyl-Papapetrou formalism \cite{Harmark:2004rm} for
$D\geq 5$ will yield card diagrams. Furthermore card diagrams do
not really require Weyl's canonical coordinates. Spacetimes with
Weyl-type symmetry and yet where Weyl's procedure fails
algebraically can still admit card diagrams.  An example is the
inclusion of a nonzero cosmological constant $\Lambda$, where a
$\gamma$-flip changes the sign of $\Lambda$.  We also hope that
these methods, or their further generalizations, have even greater
applicability than to the multitude of spacetimes already
discussed.

\section*{Acknowledgements}
We thank D.~Jatkar, A. Maloney, W. G. Ritter, A. Strominger,
T.~Wiseman and X. Yin for valuable discussions and comments.
G.~C.~J.~would like to thank the NSF for funding.  J.\ E.\ W.\ is
supported in part by the National Science Council, the Center for
Theoretical Physics at National Taiwan University, the National
Center for Theoretical Sciences and would like to thank the
organizers of Strings 2004 for support and a wonderful conference
where part of this research was conducted.


\appendix
\section{Appendix: Electrostatic Weyl Formalism} \label{EWeylapp}

The formalism of \cite{EmparanWK} can be extended for general $D$
to include an electrostatic potential.  This is somewhat
surprising since the electromagnetic energy-momentum tensor
$$T_{\mu\nu}=F_{\mu\rho}F_\nu{}^\rho-{1\over 4}g_{\mu\nu}F^2$$
is traceless only in $D=4$ and so Einstein's equations are more
complicated. Nevertheless, a cancellation does occur and one may
sum the diagonal Killing frame components of the Ricci tensor to
achieve a harmonic condition.

Follow the notation of \cite{EmparanWK} and add a 1-form potential
$A(Z,\Zbar)dt$ where $t=x^1$ is timelike ($\epsilon_1=-1$) and all
other $x^i$, $i=2,\ldots,D-2$ are spacelike (with
$\epsilon_i=+1$).  The metric takes the form
$$ds^2=-e^{2U_1}dt^2+\sum_{i=2}^{D-2}e^{2U_i}(dx^i)^2+e^{2C}dZd\Zbar,$$
from which we extract the frame metric
$$g_{\hat\mu\hat\nu}={\rm
diag}(-1,+1,\ldots,+1)\oplus\left[\begin{array}{cc}0&1/2\\
1/2&0\end{array}\right].$$ For $F=dA$ we have $F_{\hat Z\hat
t}=-F_{\hat t\hat Z}=\partial A\, e^{-U_1-C}$ and
$F_{\hat{\Zbar}\hat t}=-F_{\hat t\hat{\Zbar}}=\overline{\partial}
A\, e^{-U_1-C}$, all other components vanishing.  We compute
$F^2=-8\partial A\overline{\partial}A\,e^{-2U_1-2C}$ and
\begin{eqnarray*}
T_{\hat t\hat t}&=&2\partial A\overline{\partial}A\,e^{-2U_1-2C}\\
T_{\hat i\hat i}&=&2\partial
A\overline{\partial}A\,e^{-2U_1-2C}\quad(i\neq
1)\\
T_{\hat Z\hat Z}&=&-(\partial A)^2 e^{-2U_1-2C}\\
T_{\hat{\Zbar}\hat{\Zbar}}&=&\overline{T_{\hat Z\hat Z}}\\
T_{\hat Z\hat{\Zbar}}&=&0.
\end{eqnarray*}
The field equations are $R_{\hat\mu\hat\nu}-{1\over
2}g_{\hat\mu\hat\nu}R=T_{\hat\mu\hat\nu}$; taking the trace, we
get
$$R=-{4(D-4)\over D-2}\partial A\overline{\partial}A\,e^{-2U_1-2C}$$
and Einstein's equations are then
\begin{equation}\label{neweinstein}
R_{\hat\mu\hat\nu}=T_{\hat\mu\hat\nu}-{2(D-4)\over
D-2}g_{\hat\mu\hat\nu}
\partial A\overline{\partial}A\,e^{-2U_1-2C}.
\end{equation}
Form the sum $\sum_{i=1}^{D-2} R_{\hat i\hat i}\epsilon_i$; the
right side of (\ref{neweinstein}) gives
$$(D-4)2\partial A\overline{\partial}A\,e^{-2U_1-2C}-{2(D-4)\over D-2}(D-2)
\partial A\overline{\partial}A\,e^{-2U_1-2C}=0.$$
Hence (following (2.4)-(2.5) of \cite{EmparanWK}) we get
$$\partial\overline{\partial}\exp\left(\sum_{i=1}^{D-2}U_i\right)=0,$$
the Weyl harmonic condition.

One can add magnetostatic potentials along spatial Killing
directions as well.  We skip remaining details and give the
equations.  Let us assume $x^1$ is timelike and $x^i$ are
spacelike for $i=2,\ldots,D-2$, the potential is $A_{\it
1}=\sum_{i=1}^{D-2}A_i dx^i$, and the metric is
$ds^2=-e^{2U_1}(dx^1)^2+\sum_{i=2}^{D-2}e^{-2U_i}(dx^i)^2
+e^{2\nu}(d\rho^2+dz^2)$ and $w=\rho+iz$, $\partial_w={1\over
2}(\partial_\rho-i\partial_z)$. Einstein's equations are
\begin{eqnarray*}
\Delta U_1&=&{1\over 2}\Big(\sum_{i=1}^{D-2}(\nabla
A_i)^2e^{-2U_i}
+{D-4\over D-2}\sum_{i=1}^{D-2}(\nabla A_j)^2 e^{-2U_i},\\
\Delta U_k&=&{1\over 2}\Big(-(\nabla A_1)^2e^{-2U_1}-(\nabla
A_k)^2
e^{-2U_k}+\sum_{i\neq k,1}(\nabla A_i)^2 e^{-2U_i}\\
&&\qquad-{D-4\over D-2}\sum_{i\neq 1}(\nabla A_i)^2 e^{-2U_i}
+{D-4\over D-2}(\nabla A_1)^2 e^{-2U_1}\Big),
\end{eqnarray*}
and
\begin{eqnarray*}
\partial_w \sum_{i=1}^{D-2}\nu =-2\rho(\sum_{i<j}\partial_w U_i \partial_w
U_j +{(\partial_w A_1)^2 e^{-2U_1}\over
2}-\sum_{i=2}^{D-2}{(\partial_w A_i)^2\over 2}e^{-2U_i}).
\end{eqnarray*}
Maxwell's equations are
$$\nabla\cdot\big(\nabla A_i e^{-2U_i}\big).$$
All Laplacians and divergences are with respect to a flat 3d
axisymmetric auxiliary space with coordinates $\rho,z$.\\
\\

\end{document}